\documentclass[a4paper,12pt]{article}
\pdfoutput=1 % if your are submitting a pdflatex (i.e. if you have
\usepackage{jheppub} % for details on the use of the package, please
\usepackage[T1]{fontenc}
\usepackage{bodegraph}

\usetikzlibrary{intersections}
\usetikzlibrary{calc}
\usetikzlibrary{positioning}
\usetikzlibrary{shapes, arrows, shadows} 
\usepackage[english]{babel}
\usepackage{hyperref}
\usepackage{ifpdf}
\usepackage{mathtools}
\usepackage{slashed}
\usepackage{array}
\newcolumntype{P}[1]{>{\centering\arraybackslash}p{#1}}
\usepackage{subfigure}
\usepackage{amssymb}
\usepackage{amsfonts}
\usepackage{epsf}
\usepackage{rotating}
\usepackage{graphicx}
\usepackage{float}
\usepackage{amsmath}
\usepackage{fancyhdr}
\usepackage{babel}
\usepackage{graphics}
\usepackage{pstricks}
\usepackage{color}
\usepackage{multirow}
\usepackage{float}
\usepackage[utf8]{inputenc}

\newcommand{\nn}{\nonumber}
\newcommand{\lsim}{\mathrel{\mathop{\kern 0pt \rlap
  {\raise.2ex\hbox{$<$}}}
  \lower.9ex\hbox{\kern-.190em $\sim$}}}
\newcommand{\gsim}{\mathrel{\mathop{\kern 0pt \rlap
  {\raise.2ex\hbox{$>$}}}
  \lower.9ex\hbox{\kern-.190em $\sim$}}}

\newcommand{\be}{\begin{equation}}
\newcommand{\ee}{\end{equation}}
\newcommand{\bea}{\begin{eqnarray}}
\newcommand{\eea}{\end{eqnarray}}

\title{\boldmath Vacuum Stability in Inert Higgs Doublet Model with Right-handed Neutrinos }
\preprint{ IITH-PH-0001/20}
\author[a]{Shilpa Jangid,}
\author[a]{Priyotosh Bandyopadhyay,}
\author[b]{P. S. Bhupal Dev,}
\author[c]{Arjun Kumar}

\affiliation[a]{
	Indian Institute of Technology Hyderabad, Kandi,  Sangareddy-502287, Telengana, India}
\affiliation[b]{Department of Physics and McDonnell Center for the Space Sciences, Washington University, St. Louis, MO 63130, USA}
\affiliation[c]{
	Indian Institute of Technology Delhi, Hauzkhas,  New Delhi-110016, Delhi, India}
\emailAdd{bpriyo@iith.ac.in, bdev@wustl.edu, ph19resch02006@iith.ac.in, Arjun.Kumar@physics.iitd.ac.in}

\abstract{We analyze the vacuum stability in the inert Higgs doublet extension of the Standard Model (SM), augmented by right-handed neutrinos (RHNs) to explain neutrino masses at tree level by the seesaw mechanism. We make a comparative study of the high- and low-scale seesaw scenarios and  the effect of the Dirac neutrino Yukawa couplings on the stability of the Higgs potential.  Bounds on the scalar quartic couplings and Dirac Yukawa couplings are obtained from vacuum stability and perturbativity considerations.  These bounds are found to be relevant only for low-scale seesaw scenarios with relatively large Yukawa couplings. The regions corresponding to stability, metastability and instability of the electroweak vacuum are identified.  These theoretical constraints give a very predictive parameter space for the couplings and  masses of the new scalars and RHNs which can be tested at the LHC and future colliders. The lightest non-SM neutral CP-even/odd scalar can be a good dark matter candidate and the corresponding collider signatures are also predicted for the model. }

\keywords{Beyond Standard Model, Extended Higgs Sector, Vacuum Stability, Dark Matter, Large Hadron Collider}

\begin{document}
\maketitle
\flushbottom
\section{Introduction}
The last missing piece of the Standard Model (SM) particle spectrum was found in 2012 with the discovery of a SM-like Higgs boson with a mass of about 125 GeV at the Large Hadron Collider (LHC)~\cite{Aad:2012tfa, Chatrchyan:2012xdj}, followed by increasingly-precise measurements~\cite{Aad:2013xqa, Khachatryan:2014kca, Sirunyan:2018koj, Aad:2019mbh} on its spin, parity, and couplings to SM particles, all of which are consistent within the uncertainties with those expected in the SM~\cite{Djouadi:2005gi}.  On the other hand, there are ample experimental evidences, ranging from observed dark matter (DM) relic density and matter-antimatter asymmetry in the universe to nonzero neutrino masses, that necessitate an extension of the SM, often involving the scalar sector. Moreover, from the theoretical viewpoint, it is known that the SM by itself cannot ensure the absolute stability of the electroweak (EW) vacuum up to the Planck scale~\cite{Isidori:2001bm, Bezrukov:2012sa, Degrassi:2012ry, Buttazzo:2013uya}.\footnote{This is not a problem per se, as for the current best-fit values of the SM Higgs and top-quark masses~\cite{Tanabashi:2018oca}, the EW vacuum is metastable in the SM with a lifetime much longer than the age of the universe~\cite{Markkanen:2018pdo}. However, absolute stability is desired, for instance, for the success of minimal Higgs inflation~\cite{Bezrukov:2007ep} (see Ref.~\cite{Bezrukov:2014ipa} for a way around, though). Moreover, Planck-scale higher-dimensional operators can have a large effect to render the metastability prediction unreliable in the SM~\cite{Branchina:2013jra, Lalak:2014qua, Branchina:2014rva}.} An extended scalar sector with additional bosonic degrees of freedom can alleviate the stability issue, by compensating for the destabilizing effect of the top-quark Yukawa coupling on the renormalization group (RG) evolution of the SM Higgs quartic coupling. The issue of vacuum stability in presence of additional scalars has been extensively studied in the literature. An incomplete list of models include SM-singlet scalar models~\cite{Gonderinger:2009jp, Gonderinger:2012rd, Lebedev:2012zw, EliasMiro:2012ay, Balazs:2016tbi, Athron:2018ipf, Dev:2019njv}, Two-Higgs doublet models (2HDM)~\cite{Ferreira:2004yd,  Maniatis:2006fs, Barroso:2006pa, Battye:2011jj, Kannike:2016fmd, Xu:2017vpq}, type-II seesaw models with $SU(2)_L$-triplet scalars~\cite{Gogoladze:2008gf, Chun:2012jw, Dev:2013ff, Kobakhidze:2013pya, Bonilla:2015eha, Haba:2016zbu, Dev:2017ouk}, $U(1)$ extensions~\cite{Datta:2013mta, Chakrabortty:2013zja, Coriano:2014mpa, Haba:2015rha, Oda:2015gna, Das:2015nwk, Das:2016zue}, left-right symmetric models~\cite{Mohapatra:1986pj, Dev:2018foq, Chauhan:2019fji},  universal seesaw models~\cite{Mohapatra:2014qva, Dev:2015vjd}, Zee-Babu model~\cite{Chao:2012xt, Babu:2016gpg}, models with Majorons~\cite{Sirkka:1994np, Bonilla:2015kna},  axions~\cite{EliasMiro:2012ay, Masoumi:2016eqo}, moduli~\cite{Rummel:2013yta, Ema:2016ehh}, scalar leptoquarks~\cite{Bandyopadhyay:2016oif} or higher color-multiplet scalars~\cite{He:2013tla, Heikinheimo:2017nth}, as well as various supersymmetric models~\cite{Curtright:1975yf, Gabrielli:2001py, Datta:2004td, Evans:2008zx, Giudice:2011cg, Basso:2015pka, Bagnaschi:2015pwa, Mummidi:2018nph, Camargo-Molina:2013sta, Staub:2018vux, Ahmed:2019xon}. In contrast, additional fermions typically aggravate the EW vacuum stability, as shown e.g.~in type-I~\cite{Casas:1999cd, EliasMiro:2011aa, Rodejohann:2012px, Masina:2012tz, Farina:2013mla, Ng:2015eia, Bambhaniya:2016rbb}, III~\cite{Gogoladze:2008ak, Chen:2012faa, Lindner:2015qva, Goswami:2018jar}, linear~\cite{Khan:2012zw} and inverse~\cite{Rose:2015fua, Das:2019pua} seesaw scenarios, fermionic EW-multiplet DM models~\cite{Baek:2012uj, Lindner:2016kqk, DuttaBanik:2018emv, Wang:2018lhk}, or models with vectorlike fermions~\cite{Xiao:2014kba, Gopalakrishna:2018uxn}.  

As alluded to above, nonzero neutrino masses provide a strong motivation for beyond the SM physics. Arguably, the simplest paradigm to account for tiny neutrino masses is the so-called type-I seesaw mechanism with additional right-handed heavy Majorana neutrinos~\cite{Minkowski:1977sc, Mohapatra:1979ia, Yanagida:1979as, GellMann:1980vs, Schechter:1980gr}. However, it comes with the additional Dirac Yukawa couplings which contribute negatively to the RG running of the SM Higgs quartic coupling, thus aggravating the vacuum stability problem. One way to alleviate the situation is by adding extra scalars~\cite{Ghosh:2017fmr, Garg:2017iva, Bhattacharya:2019fgs,Chakrabarty:2015yia,Chakrabarty:2014aya, Bhattacharya:2019tqq} which compensate for the destabilizing effect of the right-handed neutrinos (RHNs). Following this approach, we consider in this paper an inert 2HDM~\cite{Deshpande:1977rw, Barbieri:2006dq} with the addition of RHNs for seesaw mechanism. The lighest of the $Z_2$ doublet is stable and we choose the parameter space in such a way that the neutral $Z_2$ odd component of the inert doublet comes out to be lightest and therefore, can be identified as the DM candidate~\cite{Barbieri:2006dq, LopezHonorez:2006gr, Dolle:2009fn, Honorez:2010re, LopezHonorez:2010tb, Goudelis:2013uca, Arhrib:2013ela, Belyaev:2016lok}.\footnote{A variant of this model with an additional scalar singlet was considered in Refs.~\cite{Bhattacharya:2019fgs, Bhattacharya:2019tqq} to obtain a multi-component DM scenario.} Though the second Higgs doublet remains inert as far as the EW symmetry breaking is concerned, it plays an important role in deciding the stability of the EW minimum for given Dirac neutrino Yukawa couplings. For sizable quartic couplings in the 2HDM sector, we find that the effect of large Dirac Yukawa couplings from the RHN sector can be compensated to keep the EW vacuum stable all the way up to the Planck scale. It should be emphasized here that the effect of the RHNs on vacuum stability is only relevant in the low-scale seesaw scenarios with relatively large Dirac Yukawa couplings, which can be realized either via cancellations in the type-I seesaw matrix or via some form of inverse seesaw mechanism (see Section~\ref{sec:2.2} for details). We also discuss the collider phenomenology of this model, and in particular, new exotic decay modes of the RHNs involving the heavy Higgs bosons (see Section~\ref{pheno}).     

          The rest of this article is organized as follows: In Section~\ref{model} we briefly review the inert 2HDM with RHNs. 
%model with electro-weak symmetry breaking conditions with various extensions of type-I 2HDM with right-handed neutrinos. 
In Section~\ref{loops}, the RG running effects are discussed in the context of perturbativity. In Section~\ref{stability}, the stability of the EW vacuum has been studied in detail as a function of the Yukawa couplings. Some LHC phenomenology is touched upon in Section~\ref{pheno}. Our conclusions are given in Section~\ref{concl}. For completeness, we give the expressions for two-loop beta functions used in our analysis in Appendix~\ref{betaf}. 
%In Appendix~\ref{metaf}, we give the expression for the tunnelling probability.  %\BD{May not need Appendix B.}
%%%%%%%%%%%%%%%%%%%%%%%%%%%%%%%%%%%%%%%%%%%%%%%%%%%%
\section{The Model}\label{model}
We extend the SM by adding another $SU(2)_L$-doublet scalar field and three RHNs which are singlets under the SM gauge group. The scalar sector of the model is discussed in Section~\ref{sec:2.1}. For the vacuum stability analysis, we consider  two different scenarios for the RHNs, viz., a canonical type-I seesaw with small Yukawa couplings and an inverse seesaw with large Yukawa couplings, which are discussed in Section~\ref{sec:2.2}.  We consider the SM gauge-singlet RHNs which are even under $Z_2$ symmetry and thus generate small neeutrino masses via type-I seesaw mechanism, while the lightest component of the $Z_2$-odd inert doublet is the DM candidate.\footnote{This is different from the scotogenic model~\cite{Ma:2006km}, where the RHNs are also $Z_2$-odd and the Dirac neutrino masses are forbidden. The observed neutrino masses in this model are obtained via one-loop radiative effects.}

\subsection{The Scalar Sector} \label{sec:2.1}
The scalar sector of this model consists of two $SU(2)_L$-doublet scalars $\Phi_1$ and $\Phi_2$ with the same hypercharge $1/2$: 
\begin{align}
\Phi_1
	\ = \ \left(\begin{array}{c}
	G^+   \\
    h+ i G^0   \end{array}\right) \, , \qquad
		\Phi_2
	\ = \ \left(\begin{array}{c}
	H^+   \\
	H+ i A   \end{array}\right) \, .
\end{align}

%\end{document}

The tree-level  Higgs potential symmetric under the SM gauge group  $SU(2)_L \times U(1)_Y$ is given by~\cite{Branco:2011iw} 
\begin{align}
  V_{\rm scalar} \ &= \ m_{11}^2\Phi_1^\dagger \Phi_1 + m_{22}^2\Phi_2^\dagger\Phi_2-(m_{12}^2\Phi_1^\dagger \Phi_2+{\rm H.c}) \nonumber \\
  & \qquad +\lambda_1(\Phi_1^\dagger \Phi_1)^2+\lambda_2(\Phi_2^\dagger \Phi_2)^2+
 \lambda_3(\Phi_1^\dagger \Phi_1)(\Phi_2^\dagger \Phi_2) +\lambda_4(\Phi_1^\dagger \Phi_2)(\Phi_2^\dagger \Phi_1) \nonumber \\ 
 & \qquad +\big[\lambda_5(\Phi_1^\dagger \Phi_2)^2+\lambda_6(\Phi_1^\dagger \Phi_1)(\Phi_1^\dagger \Phi_2) +   \lambda_7(\Phi_2^\dagger \Phi_2)(\Phi_1^\dagger \Phi_2) +{\rm H.c}\big], \label{eq:2.2}
\end{align}
where the mass terms $ m_{11}^2, m_{22}^2 $ and the quartic couplings $\lambda_{1,2,3,4}$ are all real, whereas $m_{12}^2$ and the $\lambda_{5,6,7}$ couplings are in general complex. To  avoid the dangerous flavor changing neutral currents at tree-level and to make $\Phi_2$ inert for getting a DM candidate,  we impose an additional $Z_2$ symmetry under which $\Phi_2$ is odd and $\Phi_1$ is even. This removes the $m_{12}$, $\lambda_6$ and $\lambda_7$ terms from the potential and Eq.~\eqref{eq:2.2} reduces to 
\begin{align}
V_{\rm scalar} & \ =  \ m_{11}^2\Phi_1^\dagger \Phi_1 + m_{22}^2\Phi_2^\dagger\Phi_2 + \lambda_1(\Phi_1^\dagger \Phi_1)^2 + \lambda_2(\Phi_2^\dagger \Phi_2)^2  \nonumber \\ 
& \qquad +
\lambda_3(\Phi_1^\dagger \Phi_1)(\Phi_2^\dagger \Phi_2)+ \lambda_4(\Phi_1^\dagger \Phi_2)(\Phi_2^\dagger \Phi_1) + \big[\lambda_5(\Phi_1^\dagger \Phi_2)^2 + {\rm H.c}\big].  
\label{eq:2.3}
\end{align}

The EW symmetry breaking is achieved by giving real vacuum expectation value (VEV) to the first Higgs doublet, i.e 
\begin{equation}
	\langle \Phi _1 \rangle \ = \ \frac{1}{\sqrt 2}\left(
\begin{array}{c}
	0 \\
	v \\
\end{array}
\right) \, ,
\end{equation}
with $v\simeq 246$ GeV, whereas the second Higgs doublet, being $Z_2$-odd, does not take part in symmetry breaking (hence the name `inert 2HDM').

Using minimization conditions, we express the mass parameter $m_{11}$ in terms of other parameters as follows: 
\begin{align}
m_{11}^2 \ = \ -\lambda _1v^2 \, ,
\end{align}
whereas the physical scalar masses are given by 
\begin{eqnarray}\label{mass}
M_{h}^2 & \ = \ & 2\lambda_1 v^2  \, , \nn\\
M_{H}^2 &\ = \ & \frac{1}{2}[2m_{22}^2+v^2(\lambda_3+ \lambda_4+2\lambda_5)] \, , \nn\\
M_{A}^2 &\ = \ & \frac{1}{2}[2m_{22}^2+v^2(\lambda_3+\lambda_4-2\lambda_5)] \, , \nn\\
M_{H^\pm}^2 &\ = \ & m_{22}^2+\frac{1}{2}v^2 \lambda_3 \, .
\end{eqnarray} 
Here we get one $CP$-even neutral Higgs boson $h$ which is identified as the SM-like Higgs boson of mass 125 GeV discovered at the LHC. We also get two heavy neutral Higgs bosons $H$ and $A$ with opposite $CP$ parities and a pair of charged Higgs bosons $H^\pm$. Notice from Eq.~\eqref{mass} that the heavy Higgs bosons $H$, $A$ and $H^\pm$ are nearly degenerate. Depending upon the sign of $\lambda_5$ one of scalars between $H$ and $A$ can be a cold DM candidate. Since all the physical Higgs bosons except $h$ are $\Phi_2$-type, i.e., $Z_2$-odd, this also restricts their decay modes. Since $\Phi_2$ is inert, there is no mixing between $\Phi_1$ and $\Phi_2$ and the gauge eigenstates are same as the mass eigenstates for the Higgs bosons. The $Z_2$-symmetry prevents any such mixing through the Higgs portal. In this scenario, the second Higgs doublet does not couple to fermions.

To ensure that the tree-level potential~\eqref{eq:2.3} is bounded from below in all the directions, the quartic couplings must satisfy the tree-level stability conditions~\cite{Branco:2011iw}
	\begin{align} \label{stabTHDM1}
	\lambda_1 > \ 0 \, , \quad \lambda_2 \ > \ 0 \, , \quad 
	2\sqrt{\lambda_1 \lambda_2} + \lambda_3 \ > \  0\, ,\quad 2\sqrt{\lambda_1 \lambda_2}+\lambda_3+\lambda_4- 2|\lambda_5| \ > \  0 \, .
	\end{align}
	Similarly, a neutral, charge-conserving vacuum can be ensured by demanding that  
	\begin{align}
	\lambda_4-2|\lambda_5| < 0 \, ,
	\end{align}
	which is a sufficient but not necessary condition.

Another constraint comes from the fact that the scalar potential~\eqref{eq:2.3} can have two minima at different depths~\cite{Belyaev:2016lok, Branco:2011iw, Barroso:2013awa, Chakrabarty:2016smc, Chakrabarty:2017qkh, Branchina:2018qlf}. In order to avoid the possibility of having a pseudo-inert vacuum as the global minimum, the following constraints must be satisfied~\cite{Belyaev:2016lok}, along with $m^2_{11} <0$: 
	\bea\label{cons2}
	m^2_{22} \ > \ \left\{\begin{array}{ll}R \sqrt{\frac{\lambda_2}{\lambda_1}}m^2_{11} & \quad  {\rm for}  \quad |R| \ < \ 1 \, , \\ 
		\sqrt{\frac{\lambda_2}{\lambda_1}}m^2_{11} & \quad {\rm for}  \quad R \ > \ 1 \, ,
	\end{array}\right.
	\eea
	where $\lambda_{345}=\lambda_3+\lambda_{4}+2|\lambda_{5}|$ and $R=\lambda_{345}/2\sqrt{\lambda_1\lambda_2}$.  
	Such constraints will  affect the RG-evolution of the dimensionless couplings, depending on their values at the electroweak scale.  In our case, $\lambda_i \geq 0.03$ (for $i=2,\cdots, 5$) corresponds to $R>1$ and $\lambda_i < 0.03$ corresponds to $|R|<1$ at the electroweak scale. Demanding $R>1$ turns out to be a stronger constraint than Eq.~\eqref{cons2},  as we test the perturbativity and stability profiles. The values of $m^2_{11,22}$ are taken suitably  at the electroweak scale in order to avoid the pseudo-inert vacuum for the RG-evolution in Section~\ref{loops}, as well as for the benchmark points discussed in the Section~\ref{pheno}.

\subsection{The Fermion Sector} \label{sec:2.2}
In the fermion sector, we just add SM gauge-singlet RHNs which are $Z_2$ even, to the SM particle content to generate tree-level neutrino mass via seesaw mechanism. In the canonical type-I seesaw, we just add three RHNs $N_{R_i}$, where $i=1,2,3$ and the relevant part of the Yukawa Lagrangian is given by  \begin{eqnarray}
{\cal L}_{\rm I} \ = \ i \overline{N}_{R_i} \slashed{\partial} N_{R_i}- \left(Y_{N_{ij} } \overline{L}_i \widetilde{\Phi}_1 N_{R_j} + \frac{1}{2}\overline{N}_{R_i}^c M_{R_i} N_{R_i} + {\rm H.c.} \right)\, ,
\end{eqnarray}
where  $L\equiv \left(\nu, \ell \right)_{L}$ is the SM lepton doublet, $\widetilde{\Phi}_1=i \sigma_2 \Phi_1^\star$ (with $\sigma_2$ being the second Pauli matrix), $N_R^c \equiv N_R^\intercal C^{-1}$ (with $C$ being the charge conjugation matrix), $Y_{N}$ is the 3$\times$3 Yukawa matrix and $M_R$ is the 3$\times$3 diagonal mass matrix for RHNs.

After EW symmetry breaking by the VEV of $\Phi_1$, the $Y_{N}$ couplings generate the Dirac mass terms for the neutrinos: 
\begin{align}
M_D \ = \ \frac{v}{\sqrt 2}Y_N \, ,
\end{align}
which mix the left- and right-handed neutrinos. This leads to the full neutrino mass matrix 

\begin{equation}\label{numass}
\mathcal{M}_\nu \ = 
\ 
\left( {\begin{array}{cc}
	0 & M_D \\
	M_D^\intercal & M_R \\
	\end{array} } \right) \, .
\end{equation}
After block diagonalization and in the seesaw limit $||M_D||\ll ||M_R||$, we obtain the mass eigenvalues for the light neutrinos as
\begin{eqnarray}\label{megst}
m_{\nu} & \ \simeq & -M_D M_R^{-1} M_D^\intercal \,  ,
\end{eqnarray}
whereas the RHN mass eigenstates have masses of order $M_R$. From Eq.~\eqref{megst}, it is clear that in order to have the correct order of magnitude of light neutrino mass $m_\nu\lesssim 0.1$ eV, as required by oscillation data as well as cosmological constraints, the Yukawa couplings in the canonical seesaw have to be very small, unless the RHNs are super heavy. For instance, for $M_R\sim {\cal O}(100~{\rm GeV})$, we require $Y_N \lesssim \mathcal{O}(10^{-6})$. We will see later that these coupling values are too small to have any impact in the RG evolution of other couplings, and thus, the RHNs in the canonical seesaw have effectively no contribution to the vacuum stability in this model. 

However, most of the experimental tests of RHNs in the minimal seesaw rely upon larger Yukawa couplings~\cite{Atre:2009rg, Deppisch:2015qwa}.  There are various ways to achieve this theoretically, even for a ${\cal O}(100~{\rm GeV})$-scale RHN mass.  One possibility is to arrange special textures of $M_D$ and $M_R$ matrices and invoke cancellations among the different elements in Eq.~\eqref{megst} to obtain a light neutrino mass~\cite{Kersten:2007vk, He:2009ua, Adhikari:2010yt, Ibarra:2010xw, Mitra:2011qr, Dev:2013oxa, Chattopadhyay:2017zvs, CarcamoHernandez:2019kjy}. Another possibility is the so-called inverse seesaw mechanism~\cite{Mohapatra:1986aw, Mohapatra:1986bd}, where one introduces another set of fermion singlets $S_i$ (with $i=1,2,3$), along with the RHNs $N_{R_i}$. The corresponding Yukawa Lagrangian is given by 
	\begin{align}\label{lmass}
	{\cal L}_{\rm ISS} \ = \ i \overline{N}_R \slashed{\partial} N_R + i \overline{S} \slashed{\partial} S-\left(     Y_{N_{ij}} \overline{L}_i \widetilde{\Phi}_1  N_{R_j} + \overline{N}_{R_i} M_{R_{ij}}  S_j + \frac{1}{2} \overline{S}^c_i \mu_{S_{ij}} S_j + {\rm H.c.}   \right),
	\end{align}
	where  $M_R$ is a 3$\times$3 Dirac mass matrix in the singlet sector and $\mu_S$ is the small lepton number breaking mass term for the $S$-fields.  In the basis of $\{\nu^c_L, N_R, S\}$, the full $9 \times 9$ neutrino mass matrix takes the form
	\be\label{massm}
	\mathcal{M}_{\nu} \ = \ \begin{pmatrix}
		0 \, \quad M_D\, \quad 0 \\
		M^\intercal_D\, \quad 0\, \quad M_R\\
		0\,\quad   M^\intercal_R\, \quad  \mu_S 
	\end{pmatrix}.
	\ee
After diagonalization of the mass matrix Eq.~\eqref{massm} we get the three light neutrino masses 
\bea\label{neueigen}
	m_{\nu} \ \simeq \ M_D M^{-1}_R\mu_S (M^\intercal_R)^{-1}M^\intercal_D \, , 
	\eea
	whereas the remaining six mass eigenstates are mostly sterile states with masses given by $M_R\pm \mu_S/2$. The key point here is that the presence of additional fermionic singlet and the extra mass term $\mu_S$ give us the freedom to accommodate any $M_R$ values while having sizable Yukawa couplings. 
	
Irrespective of the underlying model framework, if we take large $Y_N \sim \mathcal{O}\left(1\right)$, it will have a significant negative contribution to the running of quartic couplings via the RHN loop at scales $\mu>M_R$ \cite{Ipek:2018sai}. This must be taken into account in the study of vacuum stability in low-scale seesaw scenarios, as we show below. 
%In order to keep our results generically applicable for low-scale type-I as well as inverse seesaw scenarios, we will use the Casa-Ibarra paramterisation for the Yukawa couplings~\cite{Casas:2001sr}
%\bea\label{casapm}
%Y_N \ = \ \frac{\sqrt{2}}{v} V^\star \sqrt{\hat{M}}R\sqrt{\hat{m_\nu}} U_{\rm PMNS}^\dagger \, ,
%\eea
%where $\hat{m_\nu}$ is the diagonal matrix with eigenvalues corresponding to the three light neutrino masses, $M \equiv M_R$ for type-I seesaw and  $M\equiv M_R \mu_S^{-1}M_R^\intercal$ for inverse seesaw~\cite{Rose:2015fua}, which is diagonalized by matrix $V$ such that $\hat{M}= V^\intercal MV$, $R$ is an arbitrary $3\times3$ matrix parameterized by three complex angles and $U_{\rm PMNS}$ is the unitary Pontecorvo-Maki-Nakagawa-Sakata lepton mixing matrix. 

%\BD{You took Yukawa couplings to be diagonal and degenerate, I believe. So why need Casas-Ibarra discussion, if we aren't fitting neutrino masses?}
		
%%%%%%%%%%%%%%%%%%%%		
\section{RG Evolution of the Scalar Quartic Couplings}\label{loops}
To study the RG evolution of the couplings, the inert 2HDM+RHN scenario was implemented in {\tt SARAH 4.13.0}~\cite{Staub:2013tta}  and the $\beta$-functions for various gauge, quartic and Yukawa couplings in the model are evaluated up to two-loop level. The explicit expressions for the two-loop $\beta$-functions can be found in Appendix~\ref{betaf}, and are used in our numerical analysis of vacuum stability in the next section. To illustrate the effect of the Yukawa and additional scalar quartic couplings on the RG evolution of the SM Higgs quartic coupling $\lambda_1$ in the scalar potential~\eqref{eq:2.3}, let us first look at the one-loop $\beta$-functions. At the one-loop level, the $\beta$-function for the SM Higgs quartic coupling $\lambda_h$ (which is equal to $\lambda_1$ at tree level) in this model receives three different contributions: one from the SM gauge, Yukawa and quartic interactions, the second from the RHN Yukawa couplings and the third from the inert scalar sector as shown in Eq.~\eqref{lfull}. 
\begin{align}\label{lfull}
\beta_{\lambda_h } & \ = \ \beta_{\lambda _1}^{\rm SM} +\beta_{\lambda _{1}}^{\rm RHN} + \beta_{\lambda _1}^{\rm inert} \, ,
\end{align}
with  
		\begin{eqnarray}\label{b1}
		\beta_{\lambda_1}^{\rm SM} & \ = \ & \frac{1}{16\pi^2}\Bigg[
		\frac{27}{200} g_{1}^{4} +\frac{9}{20} g_{1}^{2} g_{2}^{2} +\frac{9}{8} g_{2}^{4} -\frac{9}{5} g_{1}^{2} \lambda_1 -9 g_{2}^{2} \lambda_1 +24 \lambda_1^{2}\nonumber \\
		&& \qquad  +12 \lambda_1 \mbox{Tr}\Big({Y_u  Y_{u}^{\dagger}}\Big) +12 \lambda_1 {\rm Tr}\Big({Y_d  Y_{d}^{\dagger}}\Big) +4 \lambda_1 \mbox{Tr}\Big({Y_e  Y_{e}^{\dagger}}\Big)  \nonumber\\
		&& \qquad  
		-6 \mbox{Tr}\Big({Y_u  Y_{u}^{\dagger}  Y_u  Y_{u}^{\dagger}}\Big)-6 \mbox{Tr}\Big({Y_d  Y_{d}^{\dagger}  Y_d  Y_{d}^{\dagger}}\Big) -2 \mbox{Tr}\Big({Y_e  Y_{e}^{\dagger}  Y_e  Y_{e}^{\dagger}}\Big) \Bigg], \label{eq:3.2} \\
				\beta_{\lambda_1}^{\rm RHN} & \ = \ & \frac{1}{16 \pi^2}\Big[4 \lambda_1 \mbox{Tr}\Big({Y_N  Y_{N}^{\dagger}}\Big)- 2 \mbox{Tr}\Big({Y_N  Y_{N}^{\dagger}  Y_N  Y_{N}^{\dagger}}\Big)\Big], \label{eq:3.3} \\
				\beta_{\lambda_1}^{\rm inert} & \ = \ & \frac{1}{16 \pi^2}\Big[ 2 \lambda_{3}^{2} +2 \lambda_3 \lambda_4 +\lambda_{4}^{2}+4 \lambda_{5}^{2}\Big]. \label{eq:3.4}
		\end{eqnarray}		
		Here $g_1,g_2$ are respectively the $U(1)_Y$, $SU(2)_L$ gauge couplings, and $Y_u, Y_d, Y_e$ are respectively the up, down and electron-type Yukawa coupling matrices in the SM. We use the SM input values for these parameters at the EW scale~\cite{Tanabashi:2018oca}: $\lambda_1=0.1264$, $g_1=0.3583$, $g_2=0.6478$, $y_t=0.9511(0.9369)$ at one (two) loop, while other Yukawa couplings are neglected~\cite{Buttazzo:2013uya}. It is important to note that the RHN contribution to the RG evolution of $\lambda_1$ is applicable only above the threshold of $M_R$. 
		
\begin{figure}[t!] 
		\hspace*{-0.9cm}
		\mbox
		{
			\subfigure[$Y_N=0.4$]
			{
				\includegraphics[width=0.33\linewidth,angle=-0]{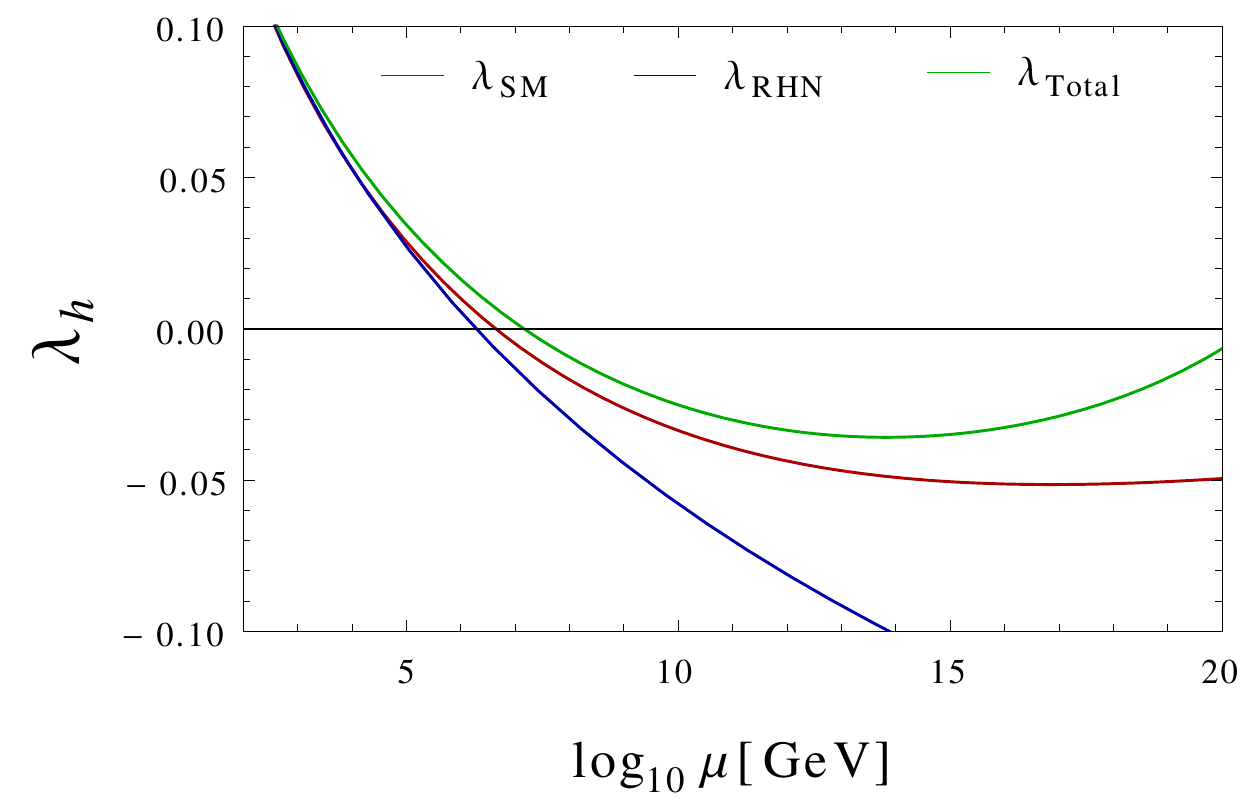}}
			\subfigure[$Y_N=0.01$]
			{
				\includegraphics[width=0.33\linewidth,angle=-0]{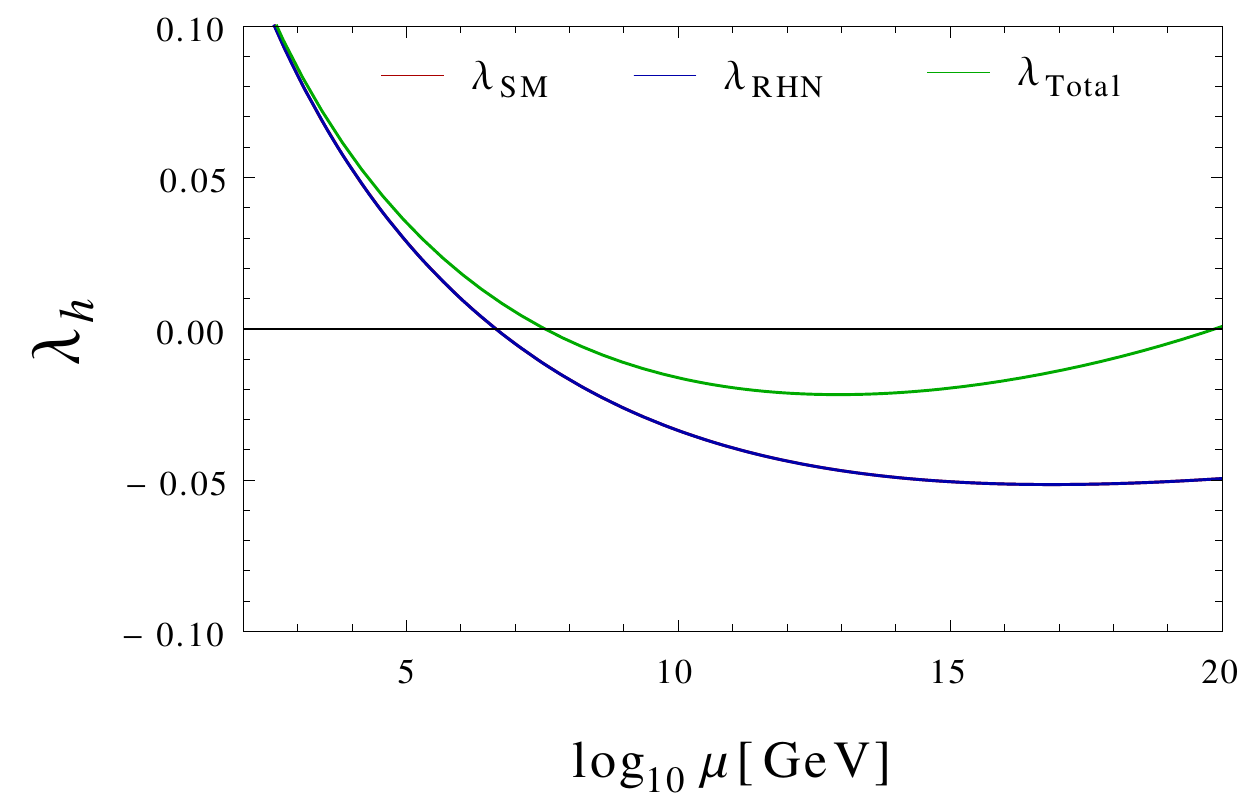}
			}
			\subfigure[$Y_N=10^{-7}$]
			{
				\includegraphics[width=0.33\linewidth,angle=-0]{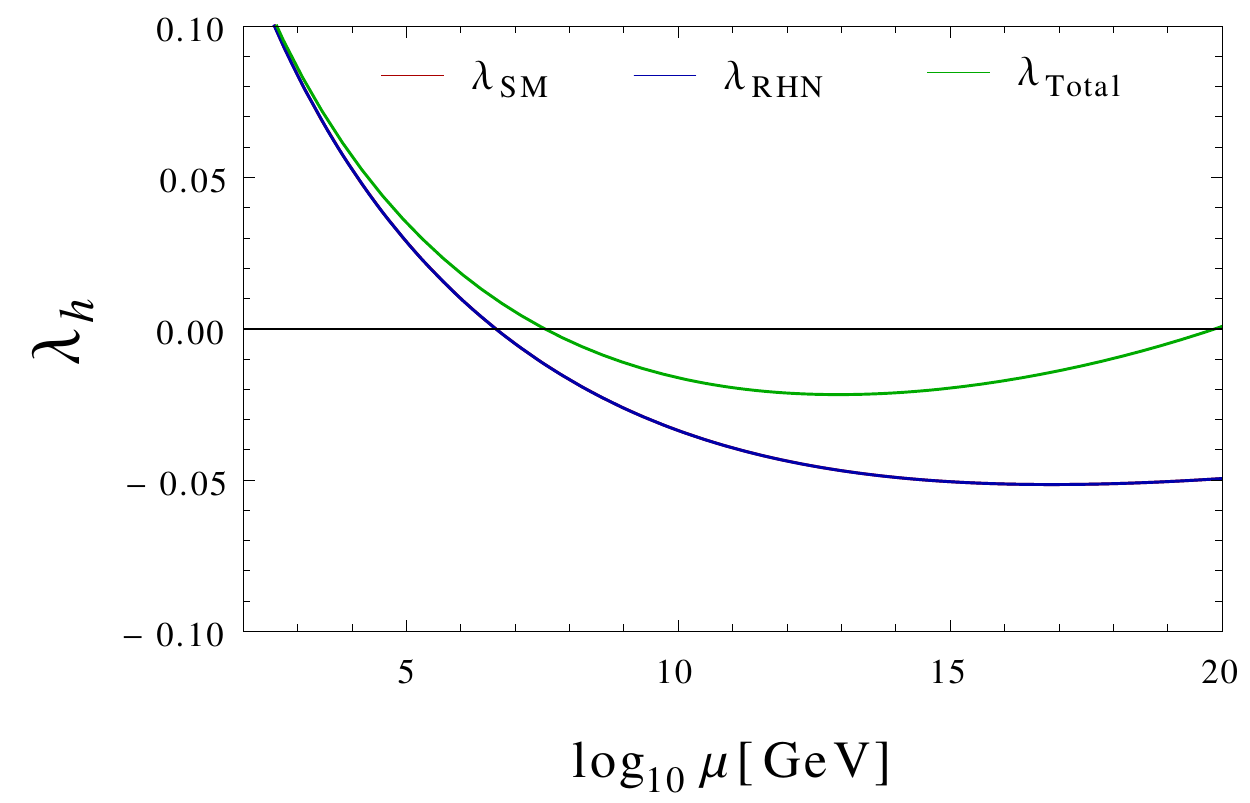}}
		}
		\caption{One-loop running of  the Higgs quartic coupling $\lambda_h$ as a function of the energy scale $\mu$ for three benchmark values of the Yukawa coupling $Y_N$. Here we have taken $M_R$=100 GeV and set $\lambda_{i=2,3,4,5}=0.1,\, y_t=0.9511$ for the other quartic couplings at the EW scale. The red, blue, and green curves respectively correspond to the $\beta$-functions in the SM, including the RHN contribution and the total contribution including both RHNs and inert scalars to the SM. The horizontal line corresponds to $\lambda_h=0$, which is the stability line.} \label{fig1l}
		
	\end{figure}

		For illustration, we assume $M_R=100$ GeV and fix all other quartic coupling values to $\lambda_{i}=0.1$  (with $i=2,3,4,5$) with $y_t=0.9369$ at the EW scale. The added effects of these new contributions in Eq.~\eqref{lfull} on the RG evolution of the SM Higgs quartic coupling $\lambda_h$ as a function of the energy scale $\mu$ are shown in Figure~\ref{fig1l}. Here the red curve shows the RG evolution of  $\lambda_h$ using $\beta_{\lambda_1}^{\rm SM}$ only [cf.~Eq.~\eqref{eq:3.2}], while the blue curve shows the evolution using $\beta_{\lambda_1}^{\rm SM}+\beta_{\lambda_1}^{\rm RHN}$, and finally the green curve shows the full evolution using $\beta_{\lambda_h}\equiv \beta_{\lambda_1}^{\rm SM}+\beta_{\lambda_1}^{\rm RHN}+\beta_{\lambda_1}^{\rm inert}$ [cf.~Eq.~\eqref{lfull}]. The three panels correspond to three benchmark values for the diagonal and degenerate Yukawa coupling values  $Y_N=0.4$ (left), 0.01 (middle), and $10^{-7}$ (right). As shown in the left panel of Figure~\ref{fig1l}, for large $Y_N=0.4$, the negative RHN contribution to the $\beta$-function in Eq.~\eqref{eq:3.3} brings down the stability scale (below which $\lambda_{h}\geq 0$) from $10^{6.6}$ GeV in the SM (at one-loop level) to $10^{6.2}$ GeV, which is then neutralized by the positive inert scalar contribution [cf.~Eq.~\eqref{eq:3.4}], that pushes the stability scale back to $10^{7.2}$ GeV and makes $\lambda_h>0$ again near the Planck scale. As shown in the middle and right panels, for smaller $Y_N$ values, the RHN contribution to the running of $\lambda_h$ is negligible, and therefore, the red and blue curves almost coincide. In these cases, the addition of inert scalar contribution pushes the stability scale up to $10^{7.6}$ GeV, and then $\lambda_h$ again becomes positive at $\sim 10^{19.6}$ GeV. 
		
		For completeness, we show the full two-loop evolution using the $\beta$-functions given in Appendix~\ref{betaf} in Figure~\ref{fig2l}. In this case, the stability scale in the SM is $10^{9.5}$ GeV, whereas including the inert scalar contribution always leads to a stable vacuum all the way up to the Planck scale, even for the case when the Yukawa coupling is chosen to be large, $Y_N=0.4$ (left panel). From this illustration, we conclude that although large Yukawa couplings involving RHNs in low-scale seesaw models tend to destabilize the vacuum at energy scales lower than that in the SM, the additional scalar contributions in the inert 2HDM extension under consideration here have the neutralizing effect of bringing back (or even enhancing) the stability up to higher scales, and in the particular example shown above, all the way up to the Planck scale\cite{Plascencia:2015xwa}.

	%%%%%%%%%%%%% beta functions at one-loop %%%%%%%%%%
	%%%%%%%%%%%%%%%%%%%%%%%%%%%
	%%%%%%%%% beta functions at two loop %%%%%%%%%%%%%%%
	\begin{figure}[t!]
		\hspace*{-0.9cm}
		\mbox
		{
			\subfigure[$Y_N=0.4$]
			{
				\includegraphics[width=0.33\linewidth,angle=-0]{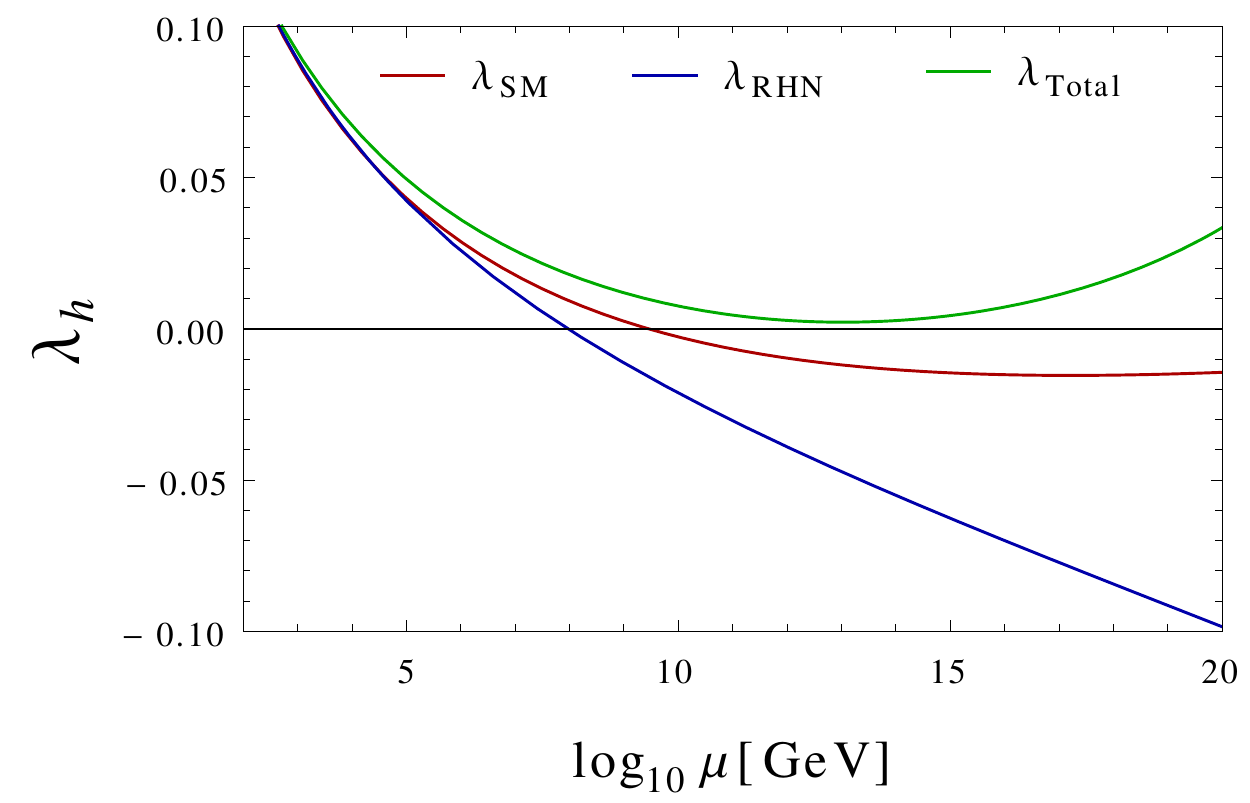}\label{pf1}}
			\subfigure[$Y_N=0.01$]
			{
				\includegraphics[width=0.33\linewidth,angle=-0]{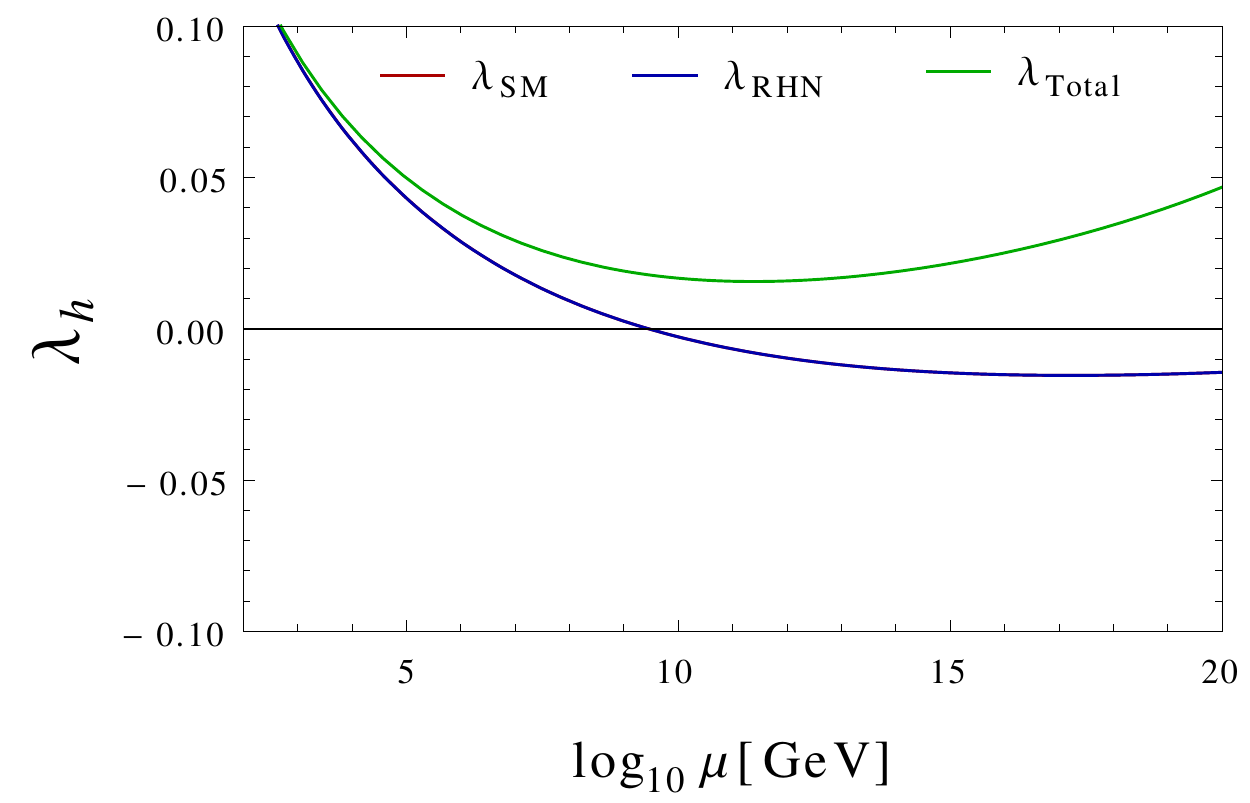}\label{pf2}
			}
			\subfigure[$Y_N=10^{-7}$]
			{
				\includegraphics[width=0.33\linewidth,angle=-0]{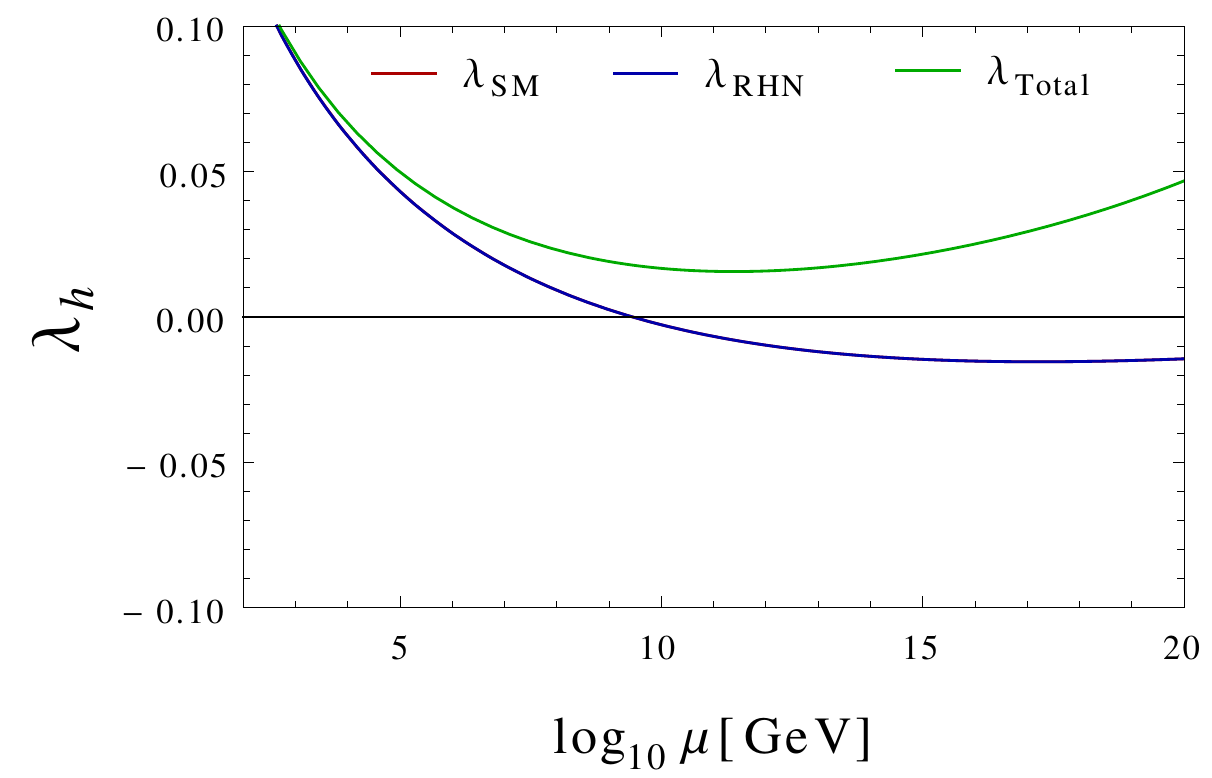}\label{pf3}}
		}
		\caption{Two-loop running of  the Higgs quartic coupling $\lambda_h$ as a function of energy for three benchmark values of the Yukawa coupling $Y_N$. Here we have taken $M_R$=100 GeV and $\lambda_i=0.1$ for the values of the quartic couplings $\lambda_{2,3,4,5}$ at the EW  scale. For the top Yukawa coupling, we use the two-loop value $y_t=0.9369$ at the EW  scale. The red, blue, and green curves respectively correspond to the $\beta$-functions in the SM, including the RHN contribution and the total contribution including both RHNs and inert scalars to the SM.} \label{fig2l}
	\end{figure}
%%%%%%%%%%%%%%%%%%%%%%%%%%%%

	%%%%%%%%%%%%% Stability Bound plots %%%%%%%%%%%%%
	\subsection{Stability Bound} 
	
\begin{figure}[t!]
		\begin{center}
			\includegraphics[width=0.6\linewidth,angle=-0]{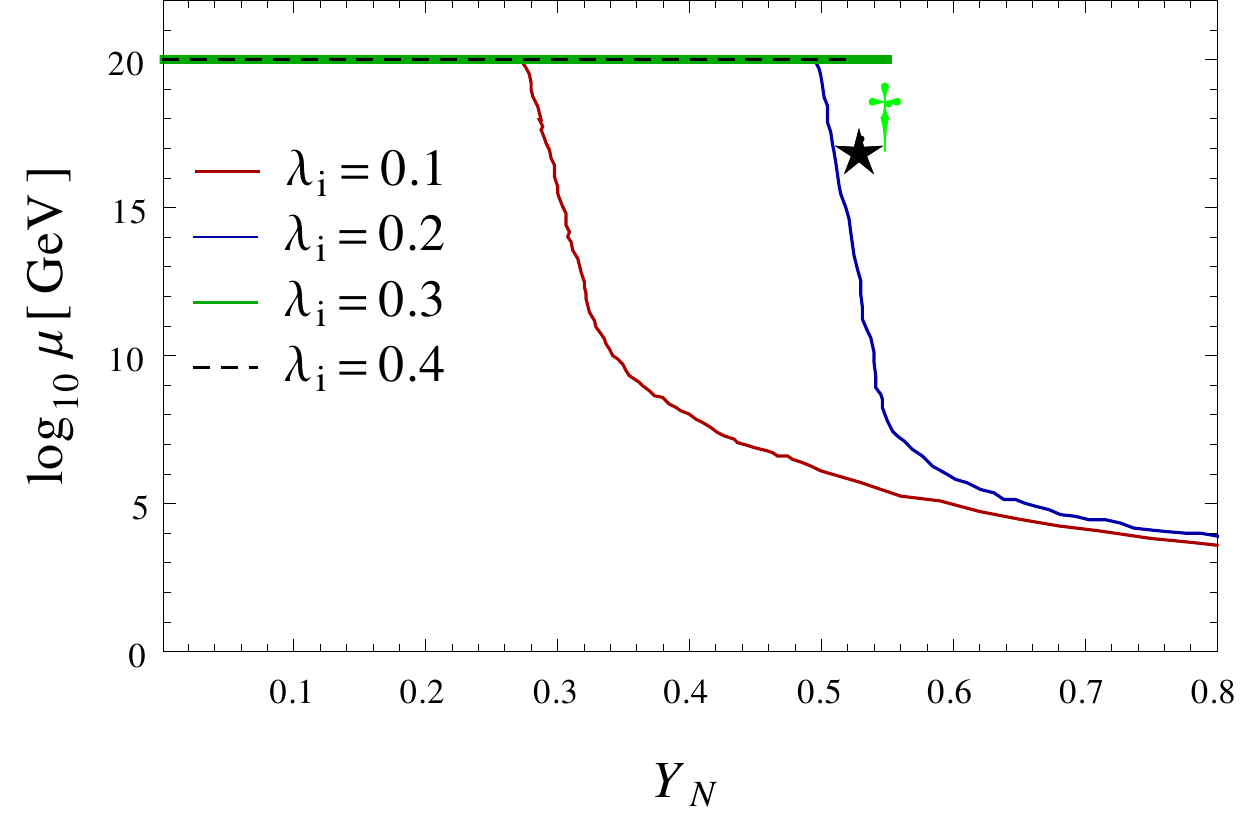}
			\caption{Effect of Yukawa coupling on the stability bound for different  values of $\lambda_i$ and $y_t=0.9369$. Here, the red curve corresponds to  $\lambda_i=0.10$ which gives stability till the Planck scale for $Y_N \leq 0.30$. The blue curve corresponds to $\lambda_i$=0.2 which gives stability till the Plank scale for $Y_N\leq 0.50$. The green curve corresponds to $\lambda_i$=0.3 which hits Landau pole at $Y_N$=0.58 and $\mu=10^{18.5}$ GeV (as shown by the dagger).  The purple curve corresponds to  $\lambda_i$=0.4 which hits Landau pole at $Y_N$= 0.55 and  $\mu=10^{17.8}$ GeV(as shown by the star). Otherwise, the green and purple curves almost coincide.}\label{fig3l}
		\end{center}
	\end{figure}

	The variation of the stability scale with the size of $Y_N$ and $\lambda_i$ is depicted in Figure~\ref{fig3l} for the choice of $y_t=0.9369$ at EW scale. For smaller values of $\lambda_i$, say 0.1 (red curve), the stability can be ensured up to the Planck scale only for $Y_N\leq 0.30$, beyond which the negative contribution from the RHNs take over and pull $\lambda_h$ to negative values at scales below the Planck scale. As we increase the $\lambda_i$ values, the compensating effect from the scalar sector gets enhanced and stability can be ensured up to the Planck scale for higher values of $Y_N$. This is illustrated by the blue curve corresponding to $\lambda_i=0.2$, for which $Y_N\leq 0.50$ is allowed. However, arbitrarily increasing  $\lambda_i$ does not help, as the theory encounters a Landau pole below the Planck scale. For instance, with $\lambda_i=0.3$ (green curve), a Landau pole is developed at $Y_N=0.58$ and $\mu=10^{18.5}$ GeV(dagger). Similarly, with $\lambda_i=0.4$ (purple curve), a Landau pole is developed at $Y_N=0.55$ and $\mu=10^{17.8}$ GeV(star). This leads us to the discussion of the perturbativity bound below.

	%%%%%%%%%%%%%%%%%%%%%%%%%%%%%%%%%%%%%%%%%%%%%%%%%%%%%%%

	\subsection{Perturbativity Bound}
	Apart from the stability constraints on the model parameter space, we also need to consider the perturbativity behaviour of the dimensionless couplings as we increase the validity scale of the theory.  We impose the condition that all dimensionless couplings of the
	model must remain perturbative for a given value of the energy scale $\mu$, i.e. the couplings must satisfy the following constraints:
	\begin{align}
	\left|\lambda_i\right|  \ \leq \ 4 \pi, \qquad
		\left|g_j\right| \ \leq \ 4 \pi, \qquad \left|Y_k\right|  \ \leq \ \sqrt{4\pi} \, ,
	\end{align}
	where $\lambda_i$ with $i=1,2,3,4,5$ are all scalar quartic couplings, $g_j$ with $j=1,2$ are EW gauge couplings,\footnote{The running of the strong coupling $g_3$ is same as in the SM, so we do not show it here.}  and $Y_k$ with $k=u,d,e,N$ are all Yukawa couplings.

\begin{figure}[t!]
		\begin{center}
			\includegraphics[width=0.6\linewidth,angle=-0]{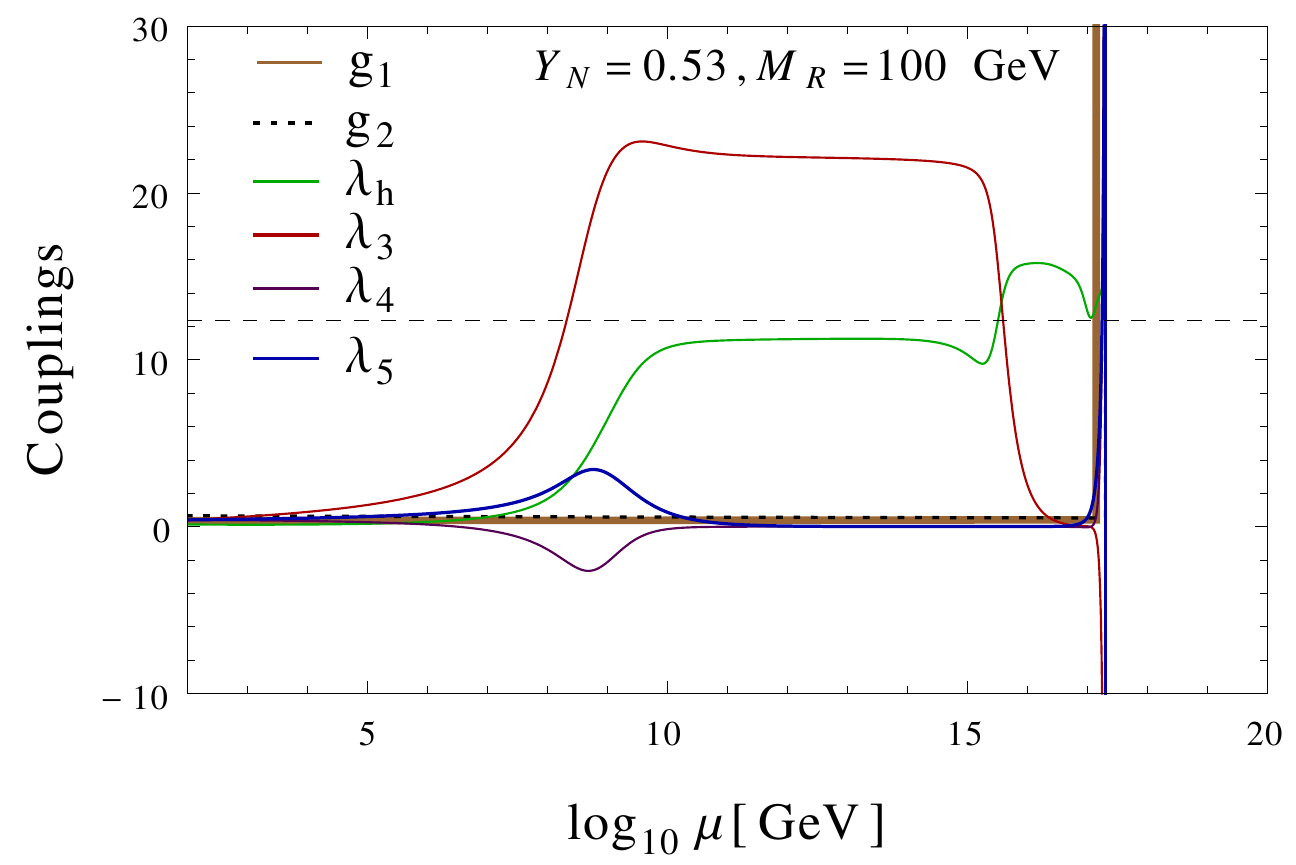}
			\caption{Two-loop RG evolution of dimensionless couplings $g_1$, $g_2$, $\lambda_h$ and $\lambda_i$ (with $i=3,4,5$) as a function of the energy scale $\mu$ for benchmark values of $Y_N=0.53,\, y_t=0.9369$, $M_R=100$ GeV and initial condition for $\lambda_i=0.4$ at the EW scale. The horizontal dashed line shows the perturbativity limit for scalar quartic and gauge couplings.}\label{fig4l}
		\end{center}
	\end{figure}

Figure \ref{fig4l} describes the variations of different dimensionless couplings with the energy scale $\mu$. Here we have shown the two-loop RG evolution of $g_1$ (yellow), $g_2$ (dotted blue), $\lambda_h$ (green), $\lambda_3$ (red), $\lambda_4$ (purple) and $\lambda_5$ (blue) as a function of the energy scale $\mu$ for benchmark values of $Y_N=0.53$ and $M_R=100$ GeV and with the initial conditions $g_1$=0.3583, $g_2$=0.6478, $y_t=0.9369$, $\lambda_h$=0.1264, and $\lambda_i=0.4$ (for $i=3,4,5$) at the EW scale. The important feature to be noted from this plot is that the theory becomes non-perturbative around $10^{8.5}$ GeV, as the $\lambda_3$ coupling overshoots the perturbativity limit, mainly driven by $ \rm \lambda_3Tr(Y^\dagger Y_N)$ (see Appendix ~\ref{betaf}) for the large Yukawa coupling $Y_N=0.53$ chosen here. 
	This is to illustrate that the perturbativity of the couplings up to the Planck scale is an additional constraint we have to take into account along with the vacuum stability constraint, while doing the RG-analysis.	
	
	%%%%%%%%%%%%%%%%%%%%%%%%%Perturbativity bounds %%%%%%%%%%%%%%
%%%%%%%%%%%%%%%%%%%%%%%%%%%%%%%%%%%%%%%%%%%%%%%%%%%%%%%%%%%%

\begin{figure}[!t]
		\hspace*{-0.9cm}
		\mbox{\subfigure[$Y_N=0.1$]{\includegraphics[width=0.32\linewidth,angle=-0]{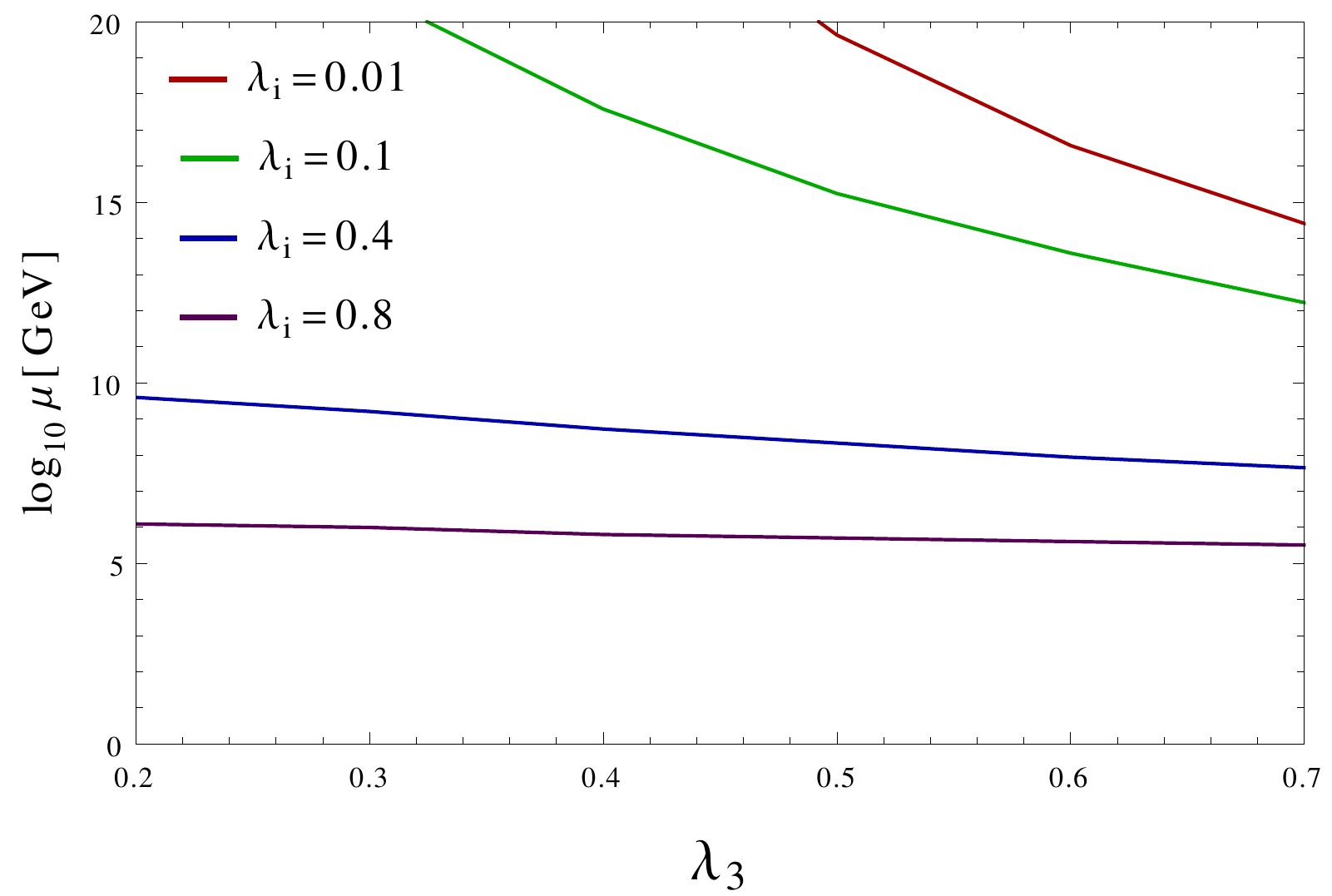}\label{f1}}
			\hspace*{+2mm}
			\subfigure[$Y_N=0.4$]{\includegraphics[width=0.32\linewidth,angle=-0]{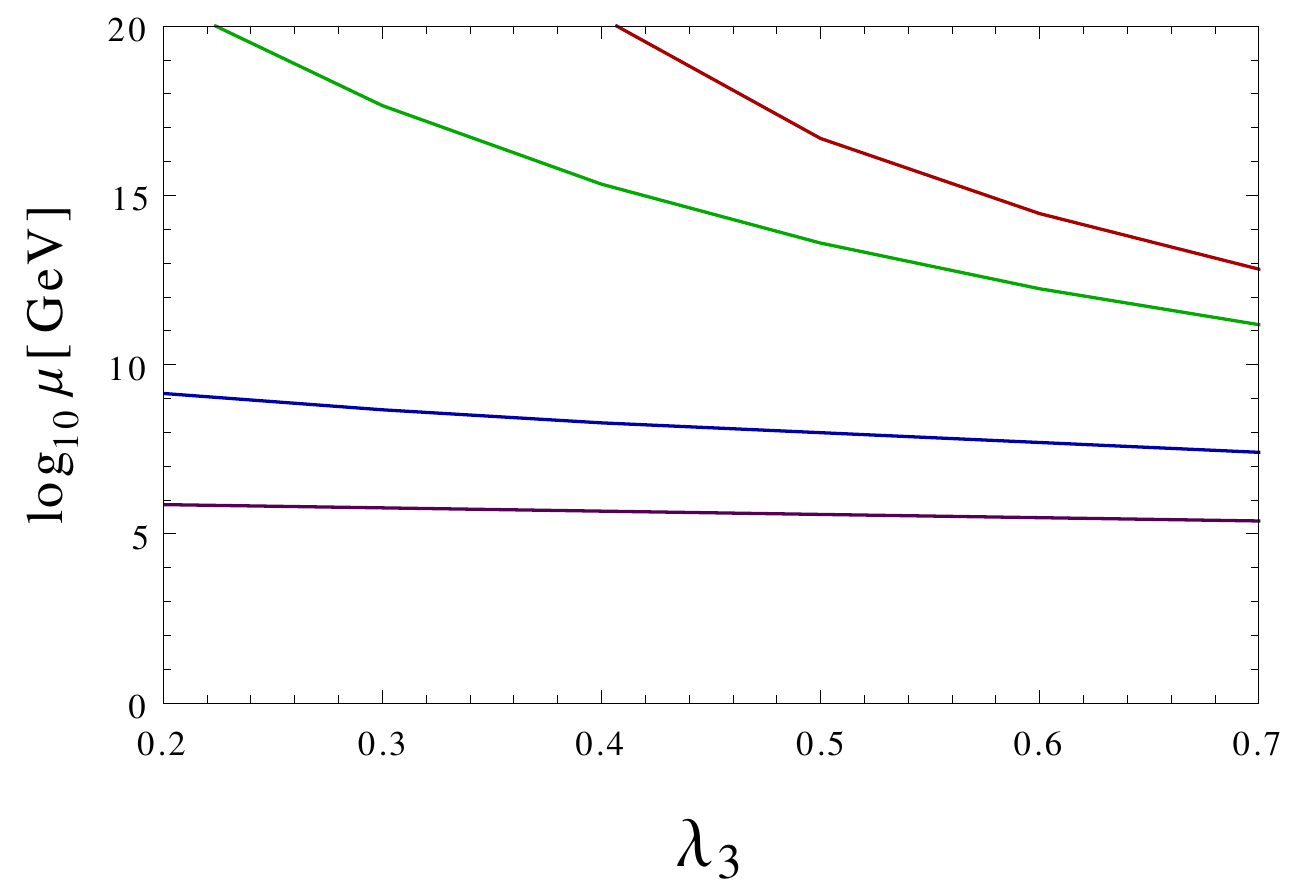}\label{f2}}
			\hspace{+2mm}
			\subfigure[$Y_N=0.9$]{\includegraphics[width=0.32\linewidth,angle=-0]{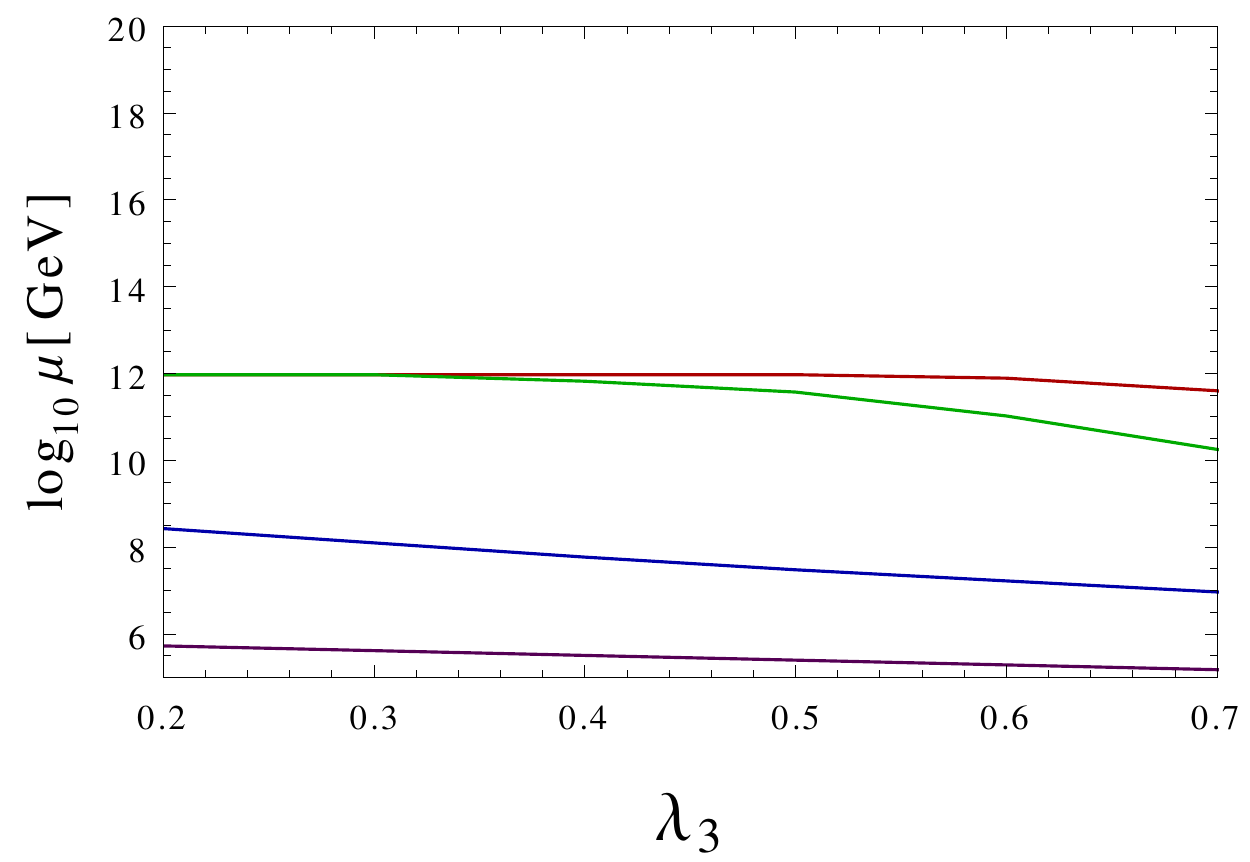}\label{f3}}}
		\caption{Two-loop running of the scalar quartic coupling $\lambda_3$ as a function of the perturbativity scale (scale where the perturbativity limit is violated) for three benchmark values of the Yukawa coupling $Y_N$ with $y_t=0.9369$. Here red, green, blue and purple curves in each plot correspond to different initial conditions for $\lambda_i$ (with $i=2,4,5$) at the EW scale, representative of very weak ($\lambda_i = 0.01$), weak ($\lambda_i = 0.1$), moderate ($\lambda_i = 0.4$) and strong ($\lambda_i = 0.8$) coupling limits respectively.}\label{fig5l}
	\end{figure}

\begin{figure}[t!]
		\hspace*{-0.9cm}
		\mbox
		{\subfigure[$Y_N=0.1$]{\includegraphics[width=0.33\linewidth,angle=-0]{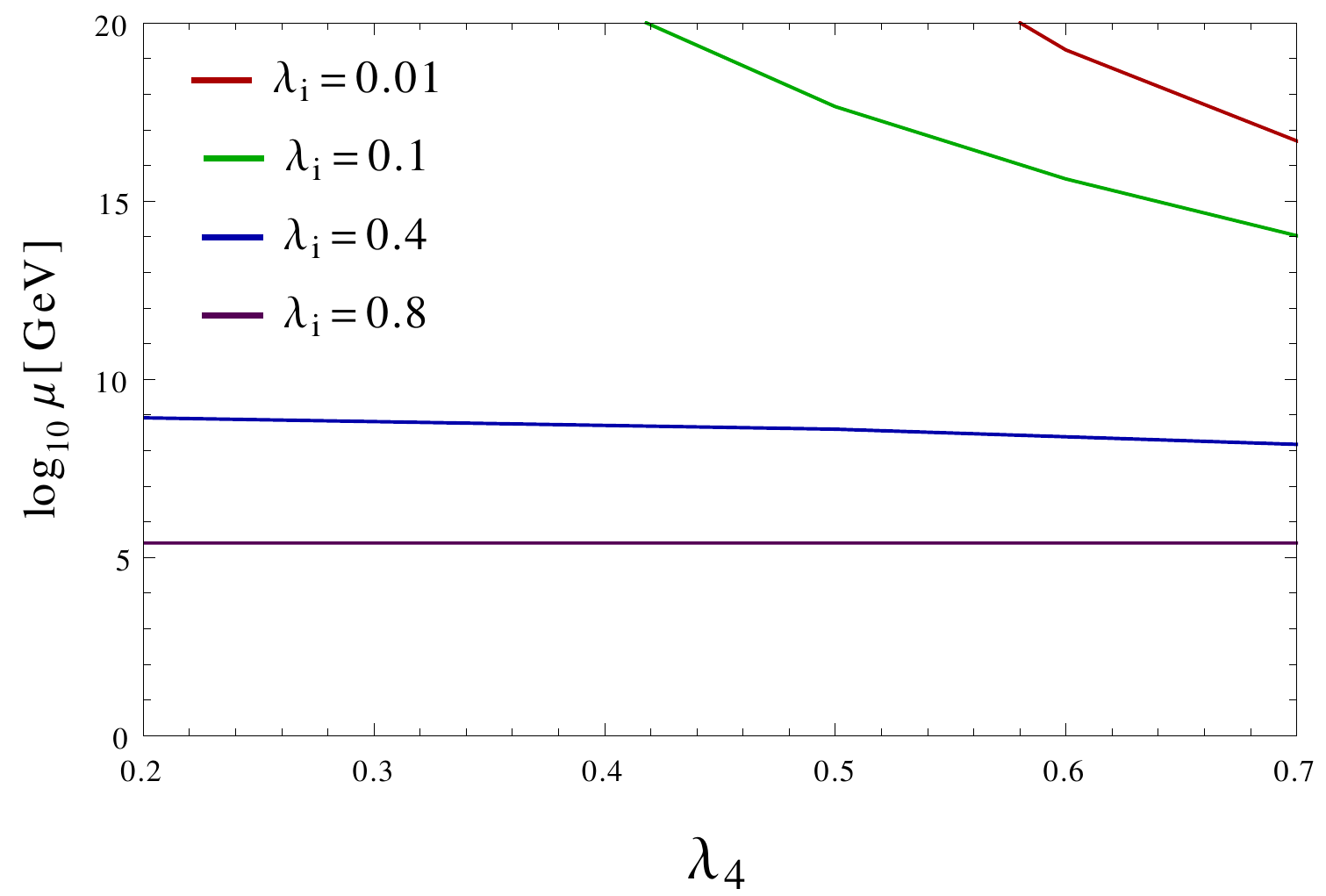}\label{f4}}
	\subfigure[$Y_N=0.4$]{\includegraphics[width=0.33\linewidth,angle=-0]{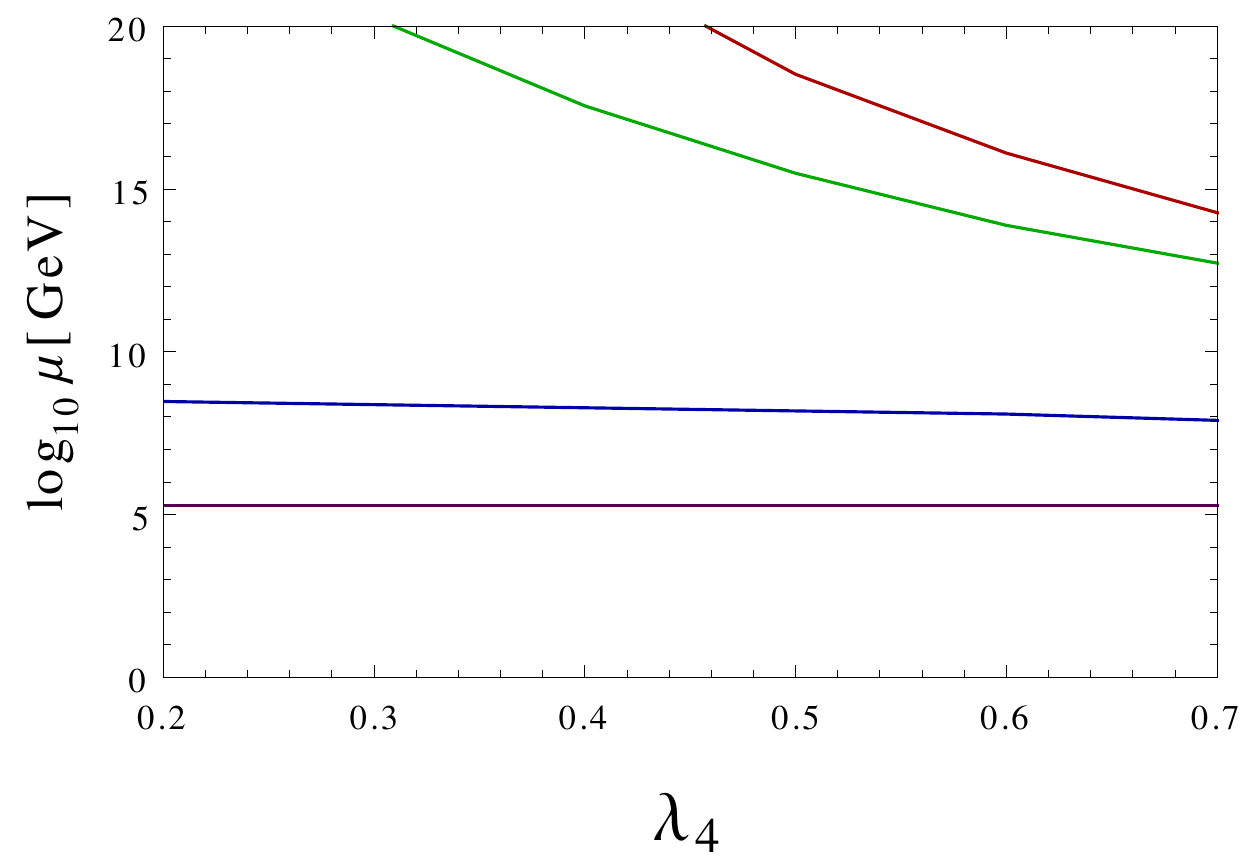}\label{f5}}
\subfigure[$Y_N=0.9$]{\includegraphics[width=0.33\linewidth,angle=-0]{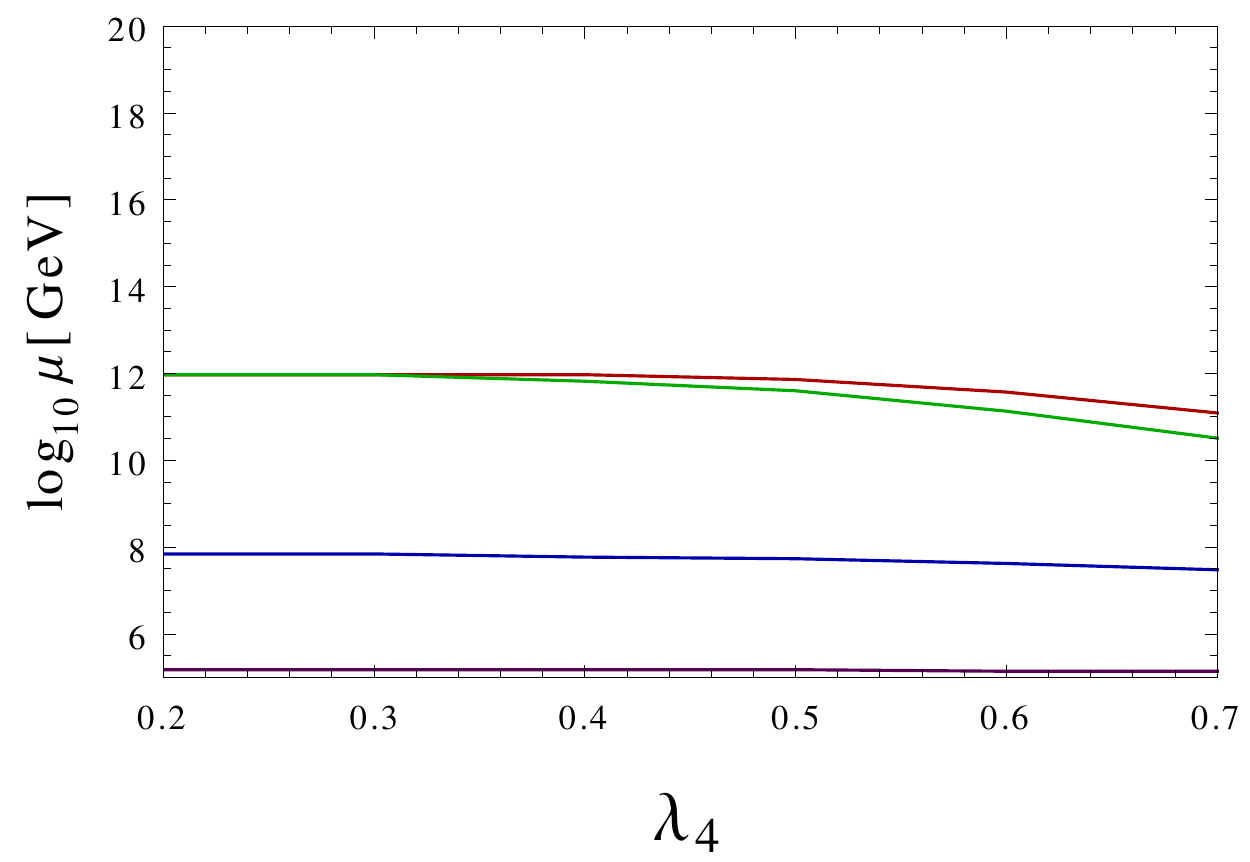}\label{f6}}}
		\caption{Two-loop running of the scalar quartic coupling $\lambda_4$ as a function of the perturbativity scale for three benchmark values of the Yukawa coupling $Y_N$ with $y_t=0.9369$. Here red, green, blue and purple curves in each plot correspond to different initial conditions for $\lambda_i$ (with $i=2,3,5$) at the EW scale, representative of very weak ($\lambda_i = 0.01$), weak ($\lambda_i = 0.1$), moderate ($\lambda_i = 0.4$) and strong ($\lambda_i = 0.8$) coupling limits respectively.}\label{fig6l}	
	\end{figure}

\begin{center}
\begin{figure}[t!]
\hspace*{-0.7cm}
\mbox{\subfigure[$Y_N=0.1$]{\includegraphics[width=0.33\linewidth,angle=-0]{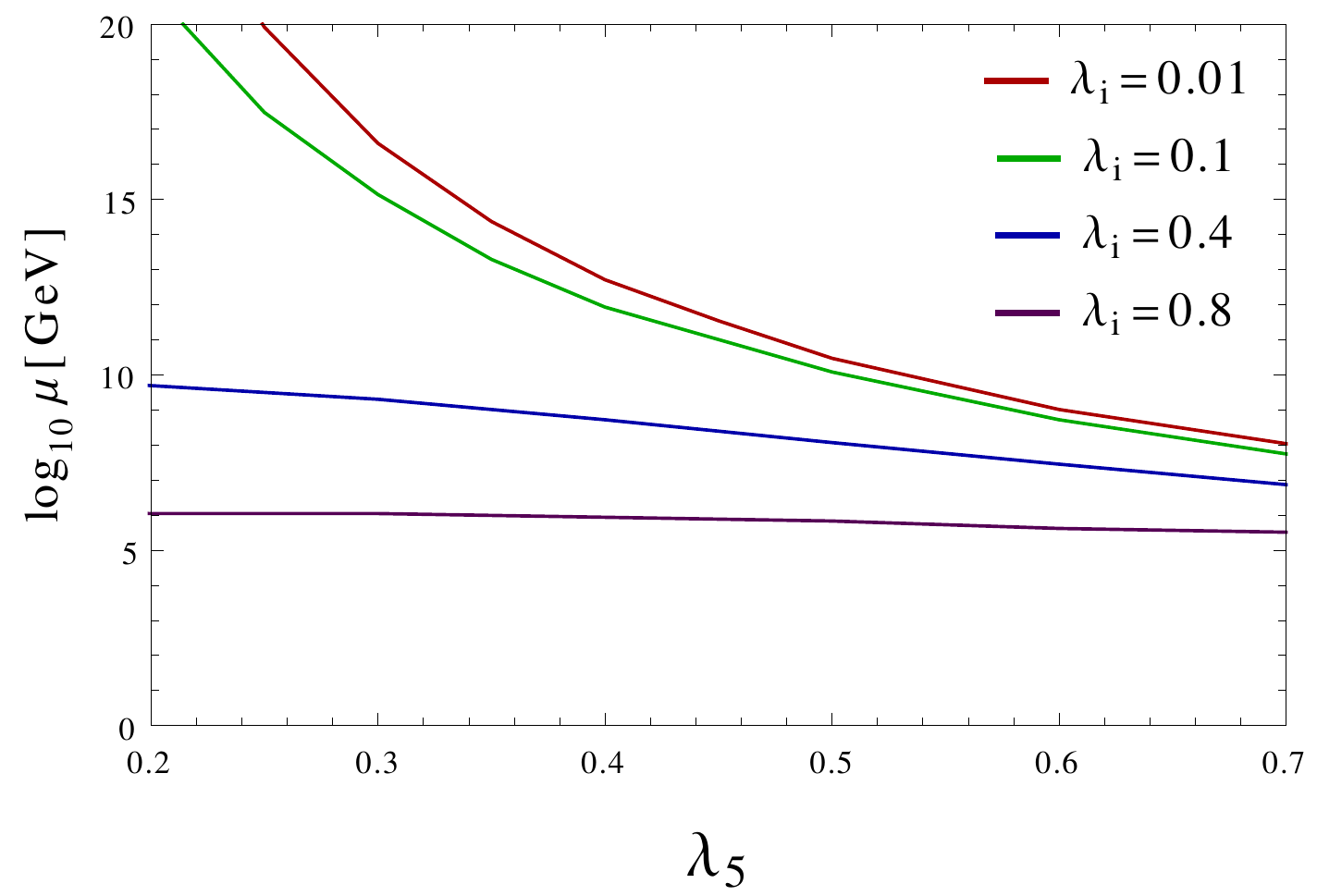}\label{f7}}	\subfigure[$Y_N=0.4$]{\includegraphics[width=0.33\linewidth,angle=-0]{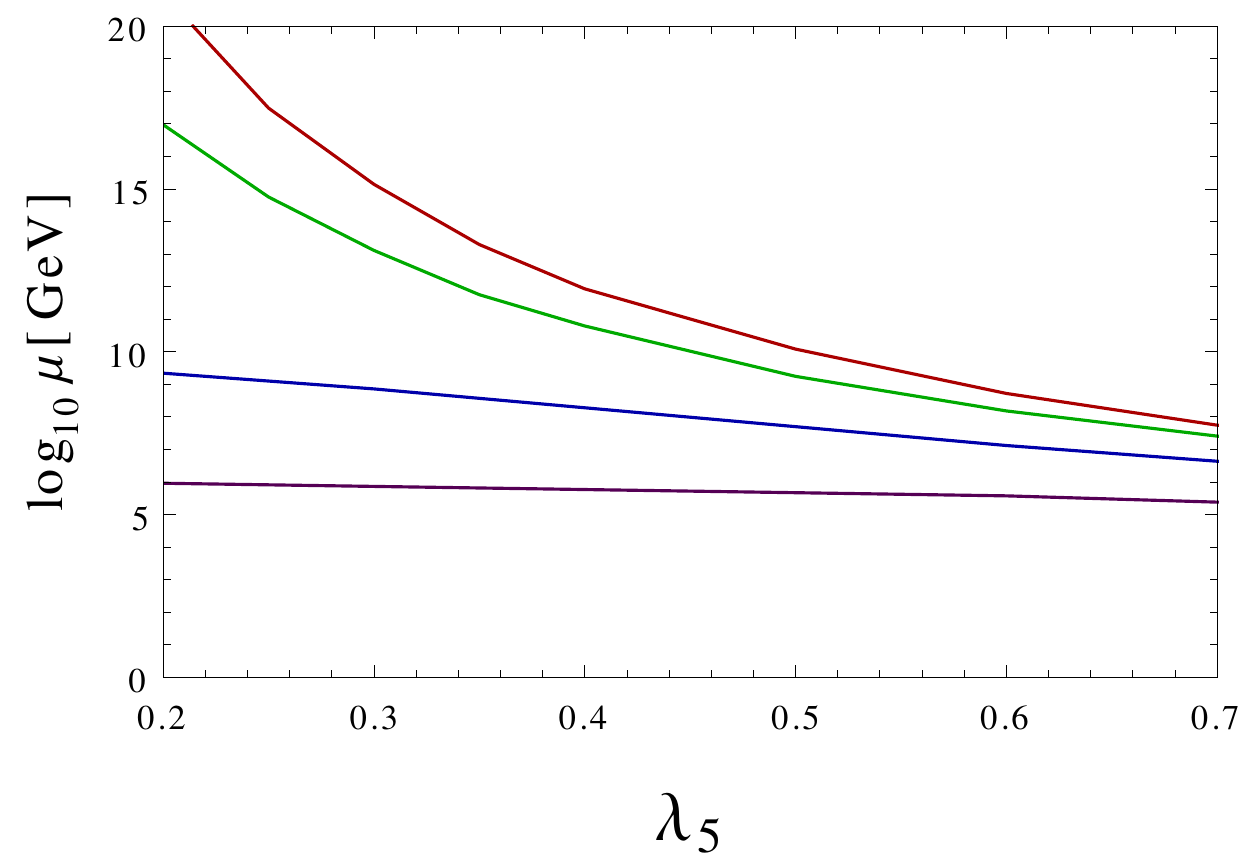}\label{f8}}
\subfigure[$Y_N=0.9$]{\includegraphics[width=0.33\linewidth,angle=-0]{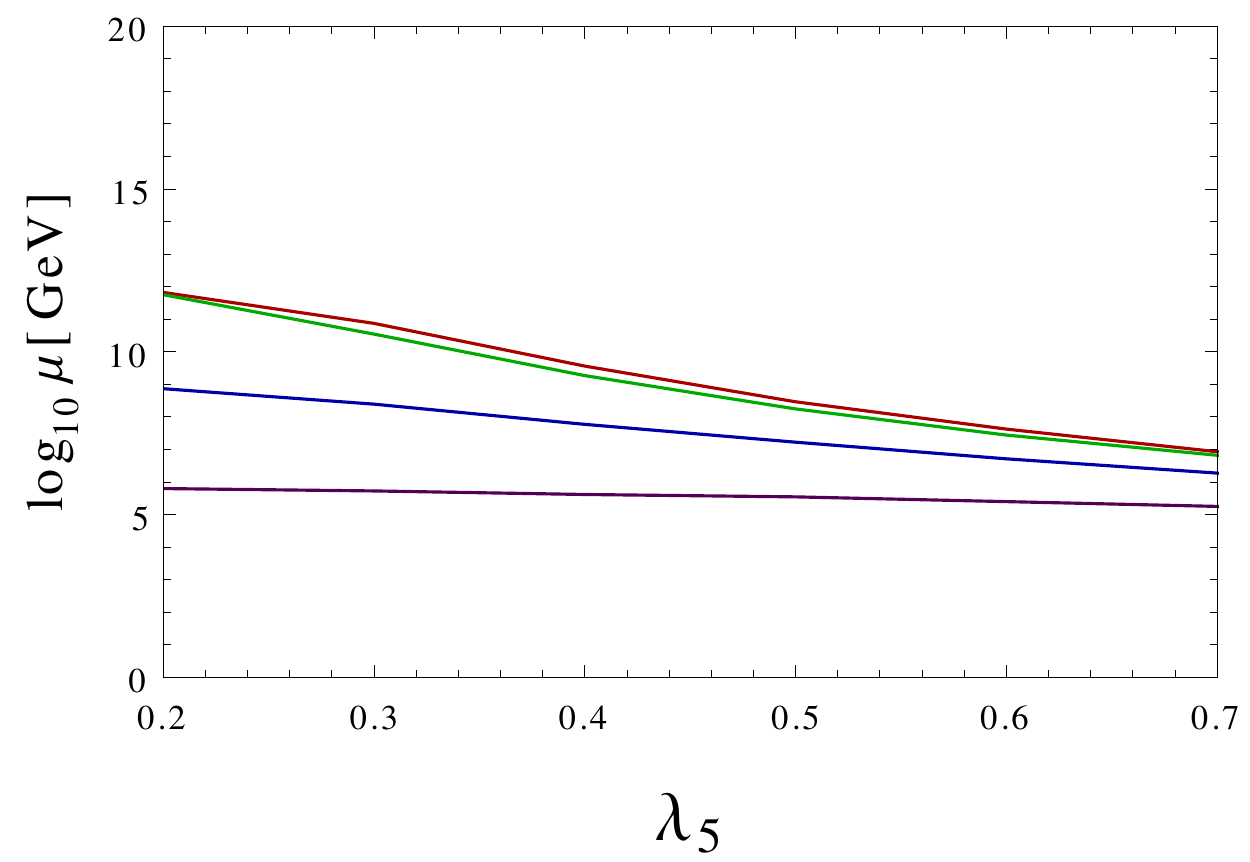}\label{f9}}}
\caption{Two-loop running of the scalar quartic coupling $\lambda_5$ as a function of the perturbativity scale for three benchmark values of the Yukawa coupling $Y_N$ with $y_t=0.9369$. Here red, green, blue and purple curves in each plot correspond to different initial conditions for $\lambda_i$ (with $i=2,3,4$) at the EW scale, representative of very weak ($\lambda_i = 0.01$), weak ($\lambda_i = 0.1$), moderate ($\lambda_i = 0.4$) and strong ($\lambda_i = 0.8$) coupling limits respectively.}\label{fig7l}
\end{figure}
\end{center}

The perturbativity behaviuor of the scalar quartic couplings $\lambda_{3,4,5}$ is studied in Figures~\ref{fig5l}-\ref{fig7l} respectively. In each case, we consider three benchmark values for the Yukawa coupling $Y_N=0.1$ (left), 0.4 (middle) and 0.9 (right). In each subplot, the various curves correspond to different benchmark initial values for the remaining unknown quartic couplings at the EW scale: red, green, blue and purple respectively for very weak coupling ($\lambda_i=0.01$), weak coupling ($\lambda_i=0.1$), moderate coupling ($\lambda_i=0.4$) and strong coupling ($\lambda_i=0.8$), while the SM Higgs quartic coupling is fixed at $\lambda_h=0.126$ for $y_t=0.9369$ and one of the quartic coupling value is varied (as shown along the $x$-axis) at the EW scale.  From Figure~\ref{fig5l}, we see that for a given $Y_N$ value, the scale at which  $\lambda_3$ hits the perturbative limit decreases as the scalar effect is increased. For example, in the strong coupling limit (with $\lambda_{2,4,5}=0.8$ at the EW scale), $\lambda_3$ hits the Landau pole at $ \mu \sim 10^{6}$ GeV making the theory non-perturbative much below the Planck scale. As we increase the $Y_N$ value (going from left to right panel), the perturbative limit is reached even for smaller values of $\lambda_i$. For instance, for $Y_N=0.9$ (right panel of Figure~\ref{fig5l}), $\lambda_3$ hits the Landau pole even in the very weak coupling limit (with $\lambda_i=0.01$) at $\mu\sim 10^{12}$ GeV. The results for $\lambda_4$ (cf.~Figure~\ref{fig6l}) and $\lambda_5$ (cf.~Figure~\ref{fig7l}) are   very similar to those of $\lambda_3$ discussed above.

% Figure~\ref{f2} and Figure~\ref{f3} show such variations of $\lambda_3$ for $Y_N$=0.4 and $Y_N$=0.9 respectively. As $Y_N$ increases the corresponding $\lambda_3$ coupling hits the Landau pole even before for all scenarios: very weak, weak, moderate and strong. Specially for $Y_N$=0.9, $\lambda_3$ coupling hits the Landau pole before Planck scale ($\mu \simeq 10^{12}$) even for weak and very week scenarios. For the case of  strong coupling limit $\lambda_{3}$ hits the Landau pole at $ \sim 10^{5.8}$ GeV. Similarly, Figure~\ref{fig6l} and Figure~\ref{fig7l}  describe the variation of $\lambda_4$ and $\lambda_5$ for very weak, weak, moderate  and  strong  $\lambda_i$  for $Y_N$=0.1, $Y_N=0.4$ and $Y_N=0.8$ respectively. The results are very similar to the case of $\lambda_3$. 

 %to get constraints from perturbativity. Figure~\ref{f1} shows the variation of $\lambda_3$ with the scale (in log10) in GeV for very weak, weak, moderate  and  strong  where $ i=2,4,5$. The other couplings are chose as  $Y_N$=0.1, 0.4, 0.9  and $\lambda_h=0.126$ at two-loop level. 

%T

%%%%%%%%%%%%%%%%%%%%%%%%%Runing of Lambdas %%%%%%%%%%%%%%%%%%%%%%%%
%%%%%%%%%%%%%%%%%%%%%%%%%%%%%%%%%%%%%%%%%%%%%%%%%%%%%%%%%%%%%
	
%%%%%%%%%%%%%%%%%%%%%%%%%%%%%%%%%%%%%%%%%%%%%%%%%%%%%%%%%%%%%	

%%%%%%%%%%%%%%%%%%%%% perturbativity lambda5 %%%%%%%%%%%%%%%%%%%%%%%
%%%%%%%%%%%%%%%%%%%%%%%%%%%%%%%%%%%%%%%%%%%%%%%%%%%%%%%%%%%

\begin{center} 
		\begin{figure}[t!]
			\mbox{\hspace{3cm}\includegraphics[width=0.5\linewidth,angle=-0]{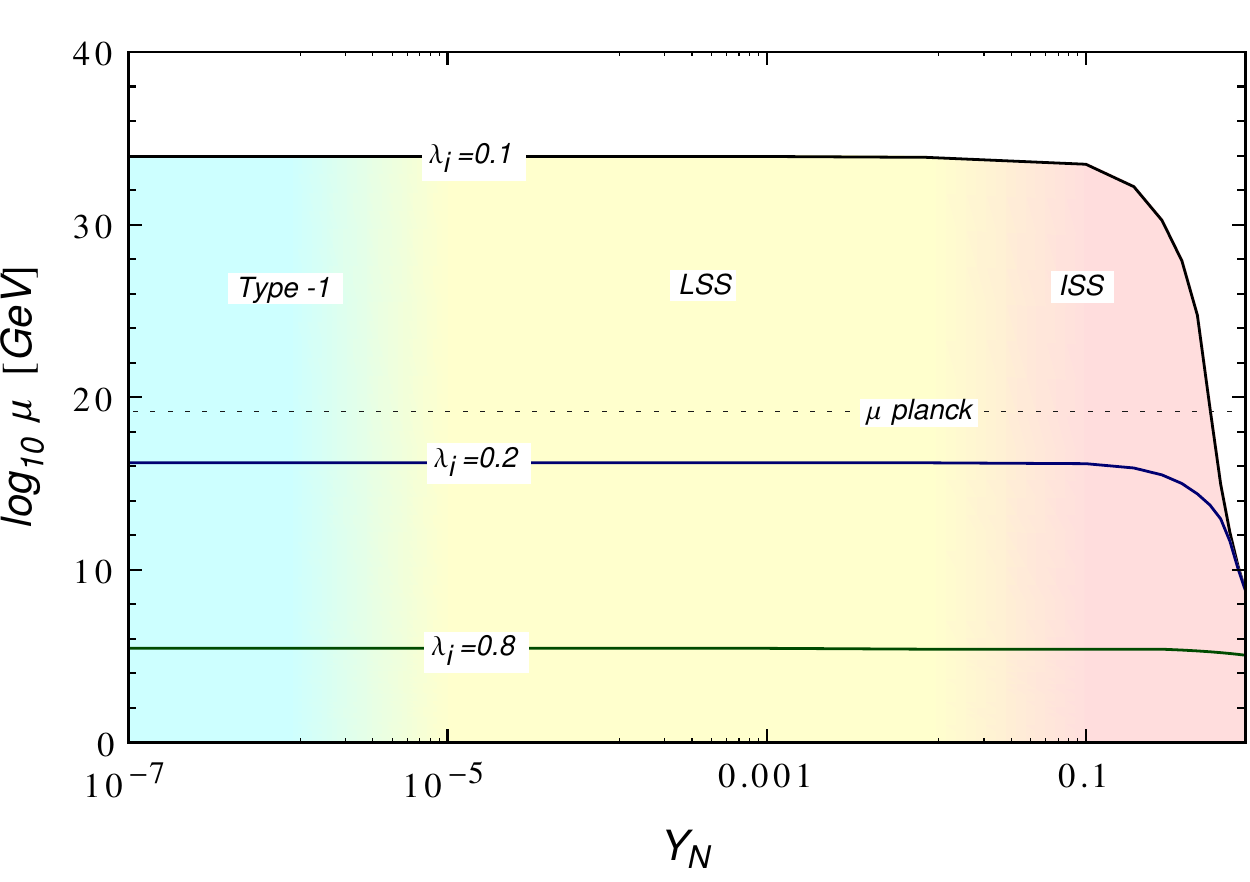}}
			\caption{Bounds from perturbativity on $Y_N$  as a function of the perturbativity scale for different values of $\lambda_i$ with $y_t=0.9369$, $M_R=100$ GeV.  The color coding refers to the size of Yukawa coupling, with sky-blue, yellow and red-colored regions roughly corresponding to the canonical type-I seesaw, low-scale seesaw (with fine-tuning) and inverse seesaw scenarios. }\label{fig8l}
		\end{figure}
	\end{center}

Figure~\ref{fig8l}  shows the bounds on Yukawa coupling  $Y_N$ from perturbativity of $\lambda_i$ for different initial $\lambda_i$ values for the choice of $y_t=0.9369$ at the EW scale. Here the color coding refers to the size of the Yukawa coupling. For small $Y_N$ $\sim$ $10^{-7}$ corresponding to the canonical type-I seesaw limit (sky-blue region), no significant effect of RHN is noticed on the perturbativity bound. Even if we allow for $Y_N$ values up to $10^{-2}$ as in low-scale seesaw models with cancellation in the seesaw matrix (yellow region), the effect of RHN on the perturbativity of $\lambda_i$ is hardly noticeable. However, as we increase $Y_N$ to the level of 0.1 and above, the perturbativity scale decreases quickly due to the positive effect of RHNs via $\rm \lambda_i Tr(Y^\dagger_N T_N)$ in the RG equations. The exact value of $Y_N$ where this starts to happen depends on the initial value of $\lambda_i$. For $\lambda_i=0.1$, the perturbativity scale occurs below the Planck scale and  the effect of RHN starts showing up for $Y_N>0.15$. For $\lambda_i=0.2$, the perturbativity limit is constant  $\sim$ $10^{16}$ GeV  and the effect of RHN starts becoming important for a larger $Y_N>0.3$ or so. On the other hand, for $\lambda_i$ =0.8, the perturbativity limit is constant at $\sim$ $10^{6}$ GeV  and the effect of RHN comes much later for $Y_N>0.8$. Thus as $\lambda_i$ increases, it can accommodate higher values of $Y_N$ for vacuum stability, but on the contrary, it makes the theory non-perturbative at much lower scale. We infer from Figure~\ref{fig8l} that an upper bound comes from perturbativity on $\lambda_i$ and $Y_N$ values, i.e. $\lambda_i\leq 0.15$ and $Y_N\leq 0.3$ for the given theory to remain perturbative till the Planck scale. For comparison, it is worth noting that the perturbativity limit on $Y_N$ derived here is a factor of few weaker than those coming from EW precision data, which vary between 0.02 to 0.07, depending on the lepton flavor, for the minimal seesaw case (i.e. without the inert doublet)~\cite{delAguila:2008pw, deBlas:2013gla, Akhmedov:2013hec, Antusch:2014woa, Flieger:2019eor}. 

%This  region  corresponding to the blue colour and represents the Type-I seesaw. For $Y_N$  $\sim$ $10^{-2}$ to $10^{-1}$ limit the yellow colour  as low-scale seesaw and  Red colour corresponds to the  Inverse seesaw limit. There is significant effect of RHN  in Inverse seesaw limit, i.e.  for $Y_N$ $\sim$ $10^{-1}$ on wards. As $\lambda_i$ increases, the corresponding perturbative scale decreases. 

%%%%%%%%%%%%%%%%%%%% Yn and perturbativity limit %%%%%%%%%%%%%%%%%%%
%%%%%%%%%%%%%%%%%%%%%%%%%%%%%%%%%%%%%%%%%%%%%%%%%%%%%%%%%	

%\pagebreak
%%%%%%%%%%%%%%%%%From effective potential approach%%%%%%%%%%%%%

\section{Vacuum Stability from RG-improved potential}\label{stability}
In this section, we investigate the stability of the EW vacuum including the quantum corrections at one-loop level. Here we follow the RG-improved effective potential approach by Coleman and Weinberg~\cite{Coleman:1973jx}, and calculate the effective potential at one-loop for our model. The parameter space of the model is then scanned for the stability, metastability and instability of the potential by calculating the effective Higgs quartic coupling and demanding appropriate limits. We then translate it into constraints on the model parameter space.

Considering the running of couplings with the energy scale in the SM, we know that the  Higgs quartic coupling $\lambda_h$ gets a negative contribution from top Yukawa coupling $y_t$, which makes it negative around $10^{9-10}$ GeV and we expect a second deeper minimum for the high field values of $\Phi_1$ as it couples to top quark. It has been shown that other direction almost remains flat as it is unlikely to get quantum corrections which generates much deeper minima, especially for the inert doublet which does not couple to top quark and RHNs\cite{AK, Chakrabarty:2016smc, Khan:2015ipa}. 
 Since the other minimum exists at much higher scale  than the EW minimum in $h$ direction, we can safely consider the effective potential in the $h$-direction to be 
\begin{align}
V_{\rm eff}(h,\mu) \ \simeq \ \lambda_{\rm eff}(h,\mu)\frac{h^4}{4},\quad {\rm with}~h\gg v \, ,
\label{eq:4.2}
\end{align}
where $\lambda_{\rm eff}(h,\mu)$ is the effective quartic coupling which can be calculated from the RG-improved potential. The stability of the vacuum can then be guaranteed at a given scale $\mu$ by demanding that $\lambda_{\rm eff}(h,\mu)\geq 0$. This approach gives us the RG-improved stability condition at the one-loop level, which supersedes the tree-level condition given in Eq.~\eqref{stabTHDM1}. We follow the same strategy as in the SM in order to calculate $\lambda_{\rm eff}(h,\mu)$ in our model, as described below.

\subsection{Effective Potential}
The one-loop RG-improved effective potential at high field values ( keeping the form of Eq.~\ref{eq:4.2}) in our model can be written as 
\begin{align}
 V_{\rm eff} \ \simeq \ V_0+V_1^{\rm SM}+V_{1}^{\rm inert}+V_{1}^{\rm RHN} \, ,
 \label{eq:4.3}
 \end{align}
 where contributions at high Higgs field values come from $V_0$, the tree-level potential; $V_1^{\rm SM}$, the SM one-loop potential at zero temperature with vanishing momenta;  $V_{1}^{\rm inert}$ and $V_{1}^{\rm RHN}$, the one-loop potentials for the inert scalar doublet and the RHN loops in the model. In general, $V_1$ can be written as 
\begin{align}\label{qc}
V_1(h, \mu) \ = \ \frac{1}{64\pi^2}\sum_{i} (-1)^F n_iM_i^4(h) \Bigg[\log\frac{M_i^2(h)}{\mu^2}-c_i\Bigg],
\end{align}
where the sum runs over all the particles that couple to the $h$-field, $F=1$ for fermions in the loop and 0 for bosons, $n_i$ is the number of degrees of freedom of each particle, $M_i^2$ are the tree-level field-dependent masses given by 
\begin{align}
M_i^2(h) \ = \ \kappa_i h^2-\kappa'_i \, ,
\label{eq:4.5}
\end{align}
with the coefficients given in Table~\ref{table:1}. In the last column, $m^2$ corresponds to the tree-level Higgs mass parameter. Note that the massless particles do not contribute to Eq.~\eqref{eq:4.5}, and hence, neither to Eq.~\eqref{qc}. Therefore, for the SM fermions, we only include the dominant contribution from top quarks, and neglect the other quarks. It is also important to note that the RHN contributions come after each threshold value of $M_{R_i}$. 

%For scalars and gauge bosons, $n_i$ is positive, while for fermions $n_i$ is negative.
%%%%%%%%%%%%%%%%%% DOF of SM %%%%%%%%%%%%%%%%%%%%%%%%%%%%
\begin{table}[h!]
\begin{center}
\begin{tabular}{||c|c|c|c|c|c|c||}\hline\hline
Particles & $i$ & $F$ & $n_i$ & $c_i$ & $\kappa_i$ & $\kappa'_i$ \\ \hline\hline
& $W^\pm$ & 0 & 6 & 5/6 & $g_2^2/4$ & 0\\
& $Z$ & 0 & 3 & 5/6 & $(g_1^2+g_2^2)/4$ & 0\\
SM & $t$ & 1 & 12 & 3/2 & $Y_t^2$ & 0\\
& $h$ & 0 & 1 & 3/2 & $\lambda_h$ & $m^2$\\
& $G^\pm$ & 0 & 2 & 3/2 & $\lambda_h$ & $m^2$\\
& $G^0$ & 0 & 1 & 3/2 & $\lambda_h$ & $m^2$\\ \hline
& $H^\pm$ & 0 & 2 & 3/2 & $\lambda_3/2$ & 0\\
Inert & $H$ & 0 & 1 & 3/2 & $(\lambda_3+\lambda_4+2\lambda_5)/2$ & 0\\
& $A$ & 0 & 1 & 3/2 & $(\lambda_3+\lambda_4-2\lambda_5)/2$ & 0\\ \hline
RHN & $N_i$ & 1 & 2 & 3/2 & $Y_N^2/2$ & 0\\ \hline\hline
\end{tabular}
\end{center}
\caption{Coefficients entering in the Coleman-Weinberg effective potential, cf.~Eq.~\eqref{qc}.}	\label{table:1}
\end{table}

Using Eq.~\eqref{qc} for the one-loop potentials, the effective potential in Eq.~\eqref{eq:4.3} can be written in terms of an effective quartic coupling as in Eq.~\eqref{eq:4.2}. This effective coupling can be written as follows: 
%\begin{align}
% \lambda_{\rm eff}\left(h,\mu\right) \ \equiv \  \lambda_h\left(\mu\right)+\Delta \lambda_{eff}^{SM}+\Delta \lambda_{eff}^{IDM}+\Delta \lambda_{eff}^{\nu}
% \end{align} 
	\begin{align} \label{totalL}
		\lambda_{\rm eff}\left(h,\mu\right) & \ \simeq \  \underbrace{\lambda_1\left(\mu\right)}_{\text{tree-level}}+\frac{1}{16\pi^2}\Bigg\{\underbrace{\sum_{\substack{i=W^\pm, Z, t, \\ h, G^\pm, G^0}} n_i\kappa_i^2 \Big[\log\frac{\kappa_i h^2}{\mu^2}-c_i\Big]}_{\text{Contribution from SM}}   
		\nonumber \\ 
		& \qquad +\underbrace{\sum_{i = H,A,H^\pm}n_i\kappa_i^2 \Big[\log\frac{\kappa_i h^2}{\mu^2}-c_i\Big]}_{\text{ Contribution from inert doublet }}
		+ \underbrace{2\sum_{i = 1,2,3}  n_i  \kappa_i^2 \Big[\log\frac{\kappa_i h^2}{\mu^2}-c_i\Big]}_{\text{ Contribution from RHN }}\Bigg\}.
		\end{align}
Note that in the inverse seesaw case and in the limit $\mu_S\to 0$, each of the RHN mass eigenvalue is double-degenerate, and therefore, we have an extra factor of two for each RHN contribution in Eq.~\eqref{totalL}. The nature of $\lambda_{\rm eff}(h,\mu)$ in our model thus guides us to identify the possible instability and metastability regions, as discussed below. We take the field value $h=\mu$ for the numerical analysis as at that scale the potential remains scale-invariant~\cite{Casas:1994us}.

\subsection{Stable, Metastable and Unstable Regions}
The parameter space where $\lambda_{\rm eff }\geq 0$ is termed as the {\it stable} region, since the EW vacuum is the global minimum in this region. For $\lambda_{\rm eff}<0$, there exists a second minimum deeper than the EW vacuum. In this case, the EW vacuum could be either unstable or metastable, depending on the tunneling probability from the EW vacuum to the true vacuum. The parameter space with $\lambda_{\rm eff}<0$, but with the tunneling lifetime longer than the age of the universe is termed as the {\it metastable} region. The expression for the tunneling probability to the deeper vacuum at zero temperature is given by
\begin{equation}
P \ = \ T_0^4 {\mu}^4 \exp\left[\frac{-8 {\pi}^2}{3 \lambda_{\rm eff}(\mu)}\right] \, ,
\label{eq:P}
\end{equation}
where $T_0$ is the age of the universe and $\mu$ denotes the scale where the probability is maximized, i.e. $\frac{\partial P}{\partial \mu}=0$. This gives us a relation between the $\lambda$ values at different scales: 
\begin{align}
\lambda_{\rm eff}(\mu) \ = \ \frac{\lambda_{\rm eff}(v)}{1-\frac{3}{2\pi^2}\log\left(\frac{v}{\mu}\right)\lambda_{\rm eff}(v)} \, ,
\label{eq:lamb}
\end{align}
where $v\simeq 246$ GeV is the EW VEV. Setting $P=1$, $T= 10^{10}$ years and $\mu=v$ in Eq.~\eqref{eq:P}, we find $\lambda_{\rm eff}(v)$ =0.0623. The condition $P< 1$, for a universe about  $T= 10^{10}$ years old is equivalent to the requirement that the tunneling lifetime from the EW vacuum to the deeper one is larger than $T_0$ and we obtain the following condition for metastability  \cite{Isidori:2001bm}: 
\begin{align}\label{meta}
0 \ > \ \lambda_{\rm eff}(\mu) \ \gtrsim \ \frac{-0.065}{1-0.01 \log\left(\frac{v}{\mu}\right)} \, .
\end{align}
The remaining parameter space with $\lambda_{\rm eff}<0$, where the condition~\eqref{meta} is not satisfied is termed as the {\it unstable} region. As can be seen from Eq.~\eqref{totalL}, these regions depend on the energy scale $\mu$, as well as the model parameters, including the RHN mass and the gauge, scalar quartic and Yukawa couplings (see also Ref.~\cite{Khan:2015ipa}).

%%%%%%%%%%%%%%%%%%% Metastability %%%%%%%%%%%%%%%%%%%%
\begin{figure}[t!]
	%	\vspace*{-2cm}
	{\mbox{\subfigure[$\lambda_i = 0.01$, $M_R=10^{3}$ GeV]{\includegraphics[width=0.45\linewidth,angle=-0]{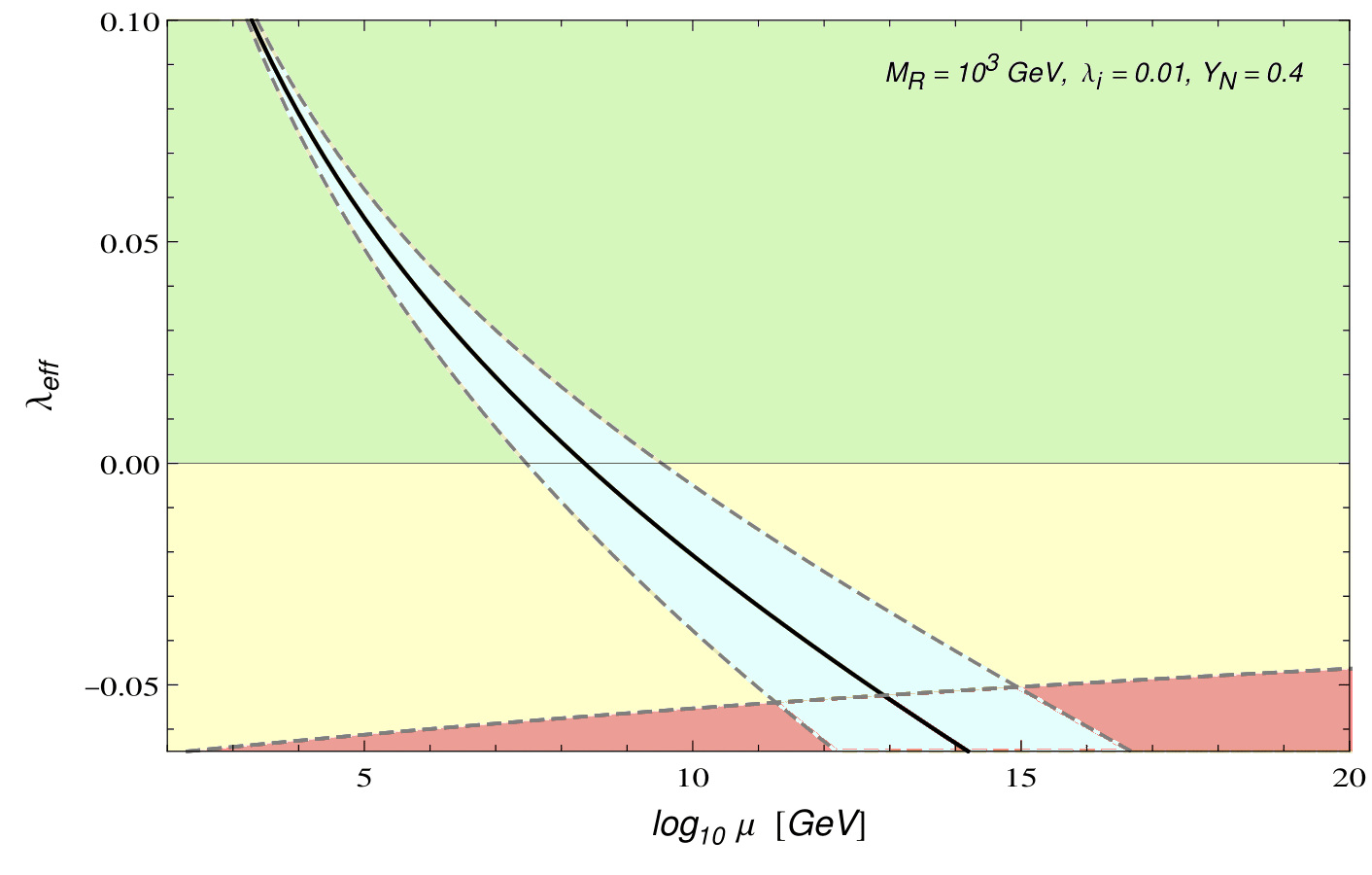}\label{f11}}
			\subfigure[$\lambda_i = 0.1$, $M_R=10^{3}$ GeV]{\includegraphics[width=0.45\linewidth,angle=-0]{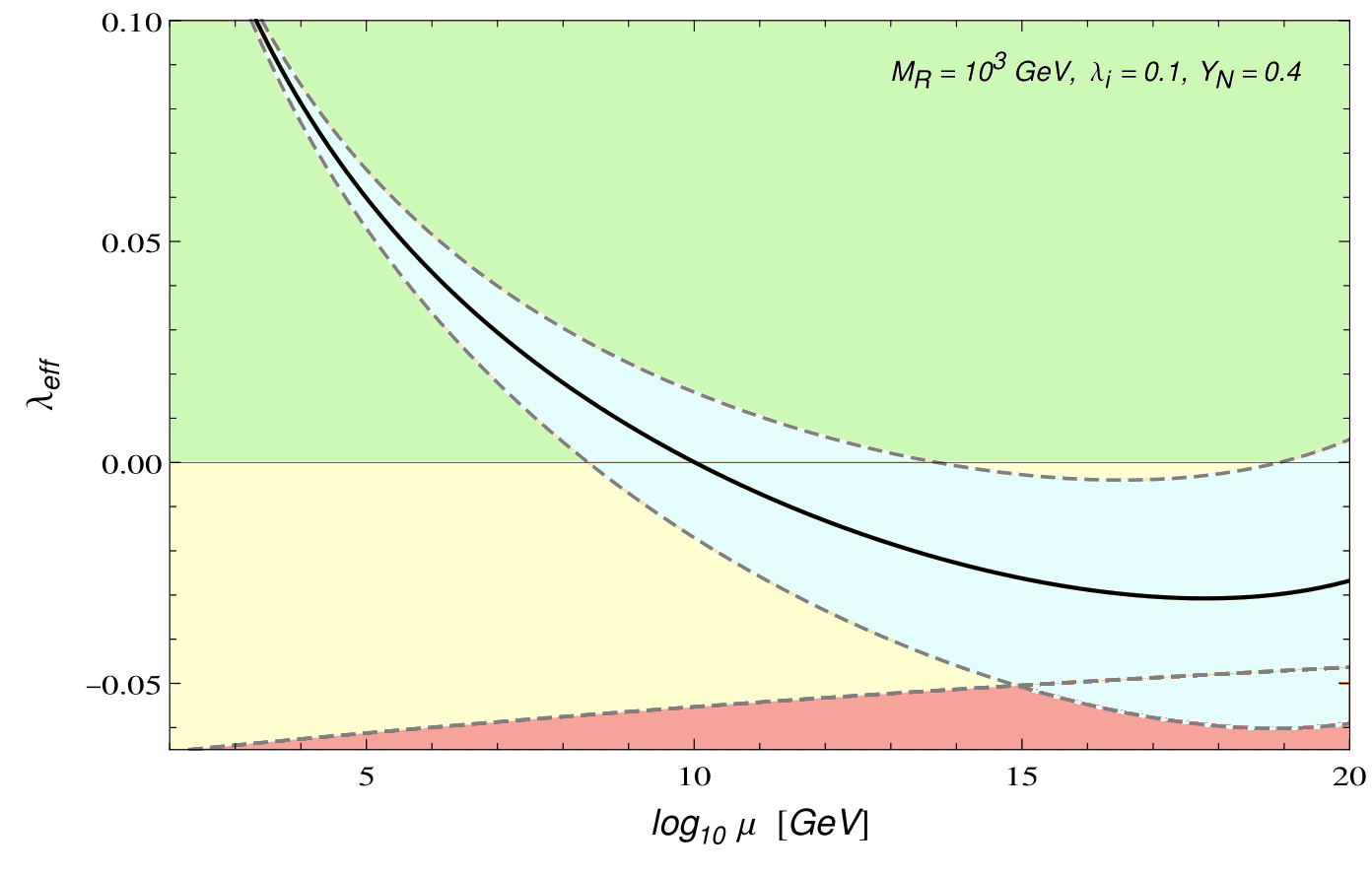}\label{f12}}}
		%				\hspace*{-2cm}
		\mbox{\subfigure[$\lambda_i = 0.01$, $M_R=10^{4}$ GeV]{\includegraphics[width=0.45\linewidth,angle=-0]{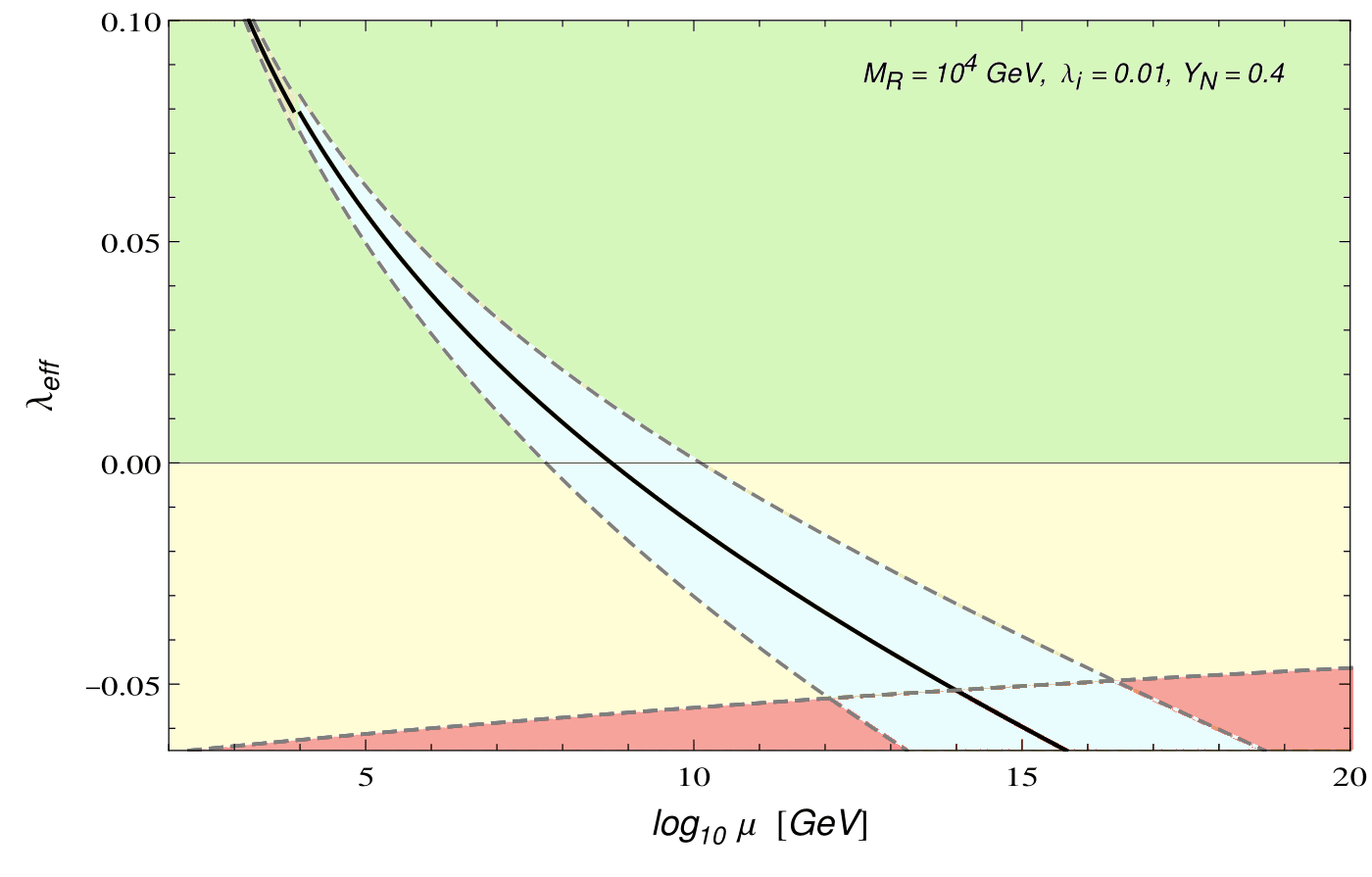}\label{f13}}
			\subfigure[$\lambda_i = 0.1$, $M_R=10^{4}$ GeV]{\includegraphics[width=0.45\linewidth,angle=-0]{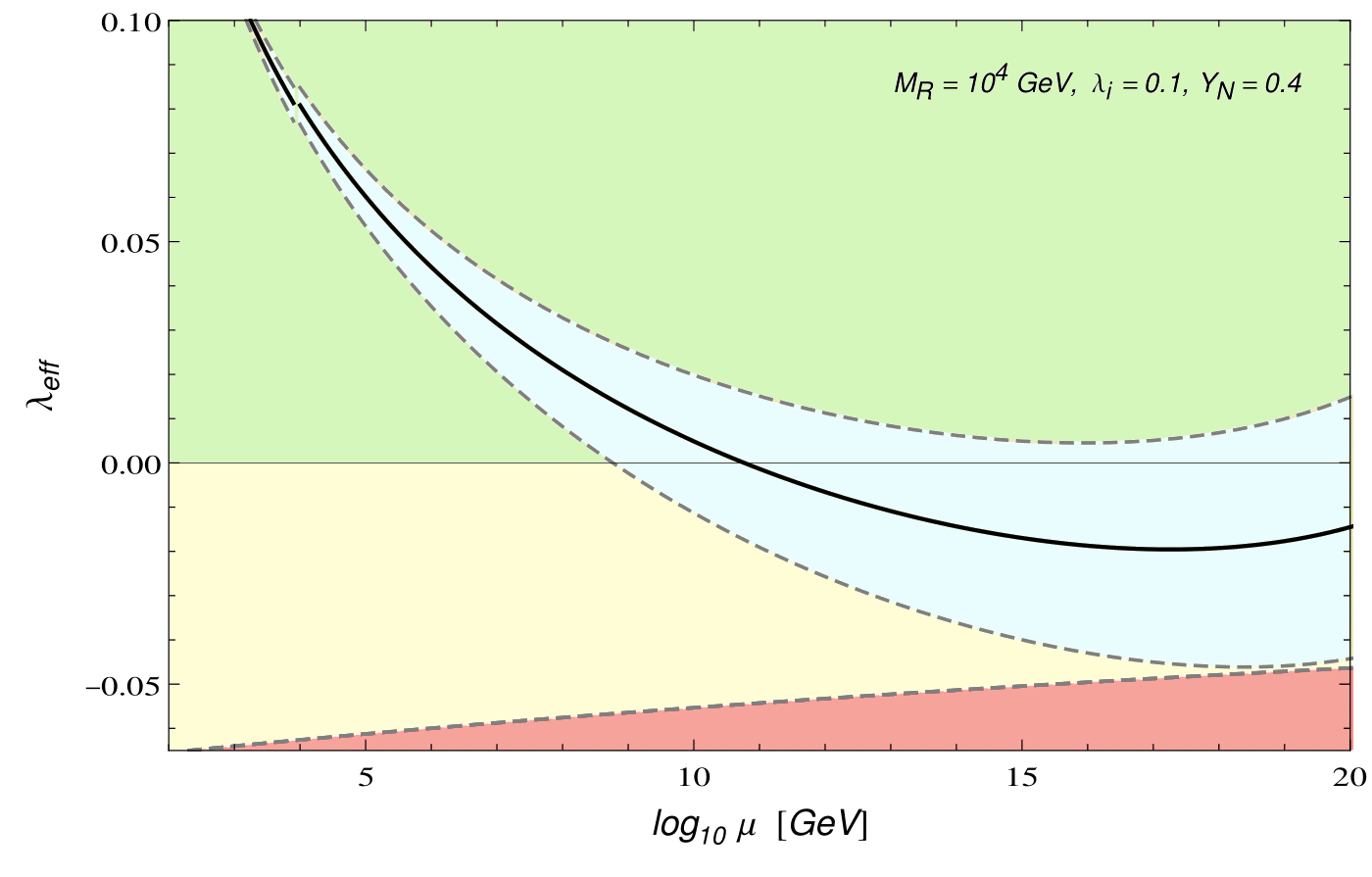}\label{f14}}}
		%				\hspace*{-2cm}
		\mbox{\subfigure[$\lambda_i = 0.01$, $M_R=10^{8}$ GeV]{\includegraphics[width=0.45\linewidth,angle=-0]{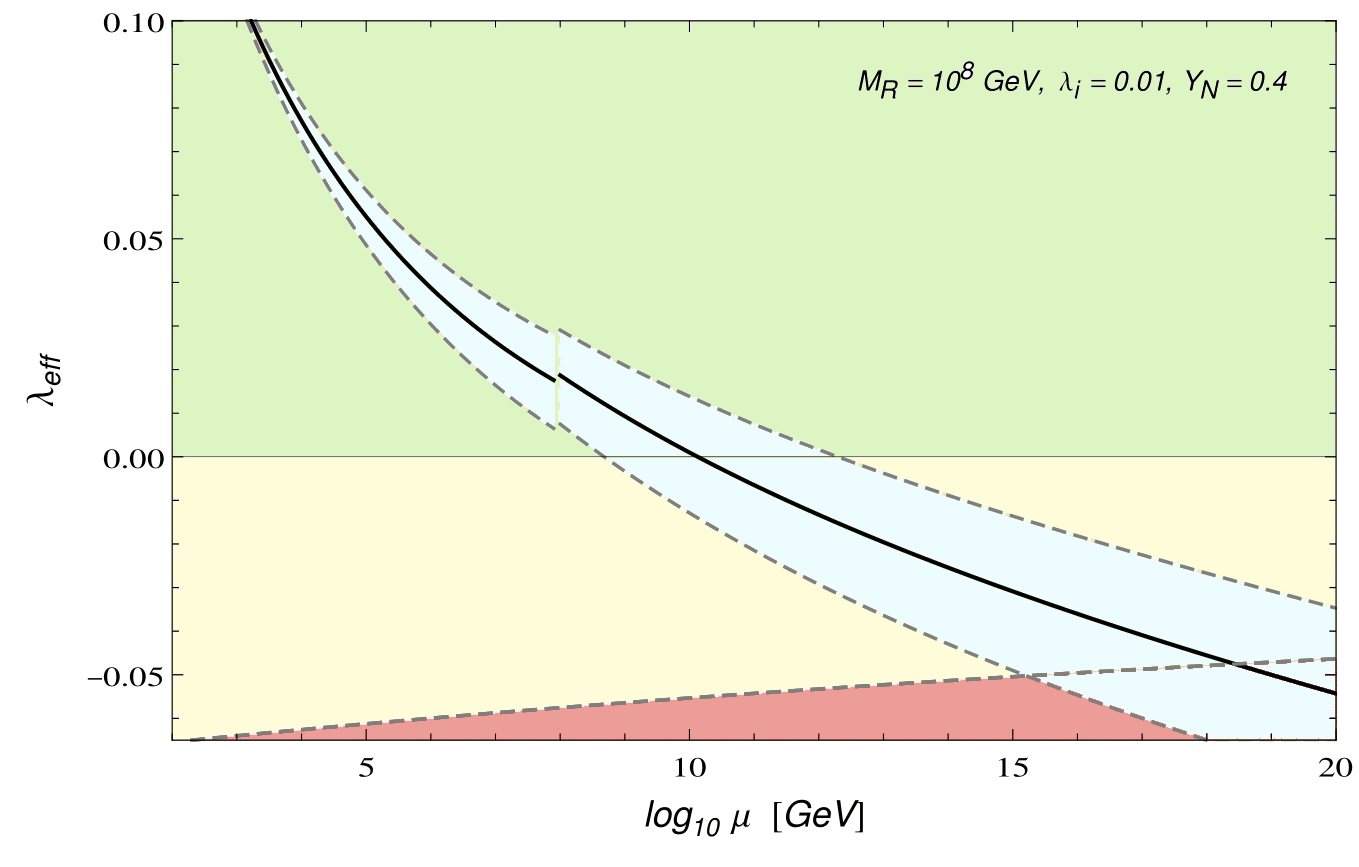}\label{f15}}
			\subfigure[$\lambda_i = 0.1$, $M_R=10^{8}$ GeV]{\includegraphics[width=0.45\linewidth,angle=-0]{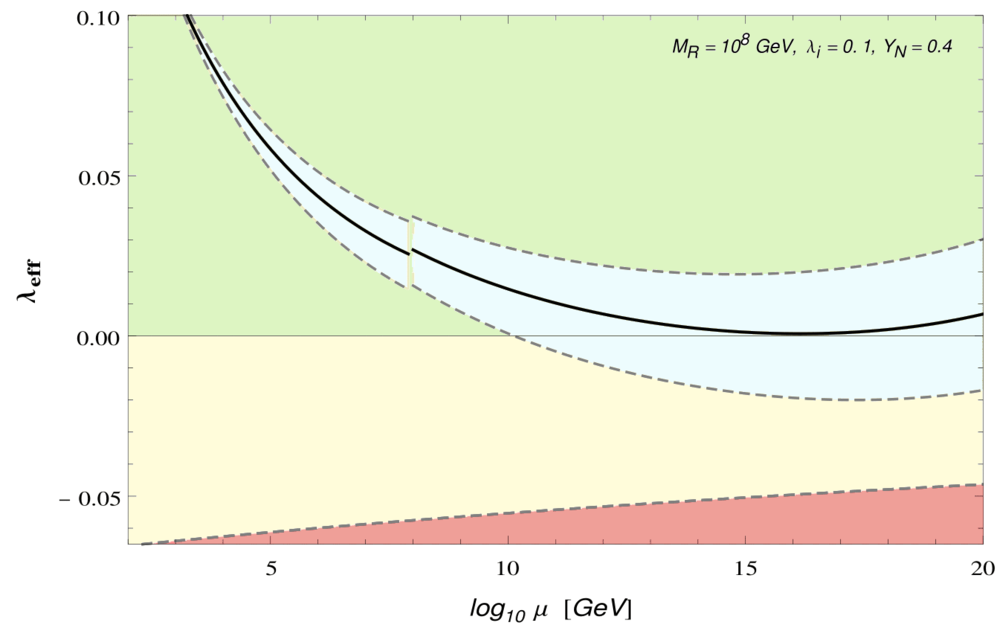}\label{f16}}}
	\caption{Running of $\lambda_{\rm{eff}}$ with energy scale for six different scenarios: $\lambda_i=0.01$ (left) and 0.1 (right); $M_R=10^3$ GeV (top), $10^4$ GeV (middle) and $10^8$ GeV (bottom). We have fixed $Y_N=0.4$ in all the subplots. The three different lines for $\lambda_{\rm eff}$ correspond to different values of the top Yukawa coupling obtained by varying the top mass from 170 GeV (upper dashed line) to 176 GeV (lower dashed line) with the median value of 173 GeV (middle solid line). The red, yellow and green regions correspond to the unstable, metastable and stable regions, respectively.}\label{fig9l} }
	\end{figure}
	%%%%%%%%%%%%%%%%%%%%%%%%%%%%%%%%%%%%%%%%%%%%%%%%%%	
Figure~\ref{fig9l} shows the variation of $\lambda_{\rm eff}$ in our model with the energy scale for different values of $\lambda_i$ (with $i=2,3,4,5$) and $M_R$ values with a fixed $Y_N=0.4$. The three different lines correspond to different values of the top Yukawa coupling by varying the top mass from $170$ to $176$ GeV with median value at $173$ GeV~\cite{Degrassi:2012ry}. The red region in Figure~\ref{fig9l} corresponds to the instability region and the yellow region below the horizontal line $\lambda_{\rm eff}=0$ corresponds to the metastable region, whereas the green region above $\lambda_{\rm eff}=0$ is the stability region. Figure~\ref{f11} and Figure~\ref{f12} show that as the values of $\lambda_i$ are increased from 0.01 to 0.1 for the same value of $Y_N=0.4$ and $M_R=10^{3}$, $\lambda_{\rm eff}$ becomes unstable at $10^{15}$ GeV instead of $10^{11}$ GeV (with higher end of the top mass). Figure~\ref{f11}, Figure~\ref{f13} and Figure~\ref{f15} [or Figure~\ref{f12}, Figure~\ref{f14} and Figure~\ref{f16}] show that for fixed $\lambda_i$ and $Y_N$, the stability scale also gets enhanced as we increase RHN mass $M_R$, because the RHNs contribute to the $\beta$-function only at scales $\mu\geq M_R$. This is the reason for the discontinuity at $M_R$ value, which is obvious in Figure~\ref{f15} and Figure~\ref{f16}. 

\begin{figure}[t!]
	\hspace*{-0.4cm}
	{\mbox{\subfigure[$Y_N=0.1$, $M_R=10^{3}$ GeV]{\includegraphics[width=0.32\linewidth,angle=-0]{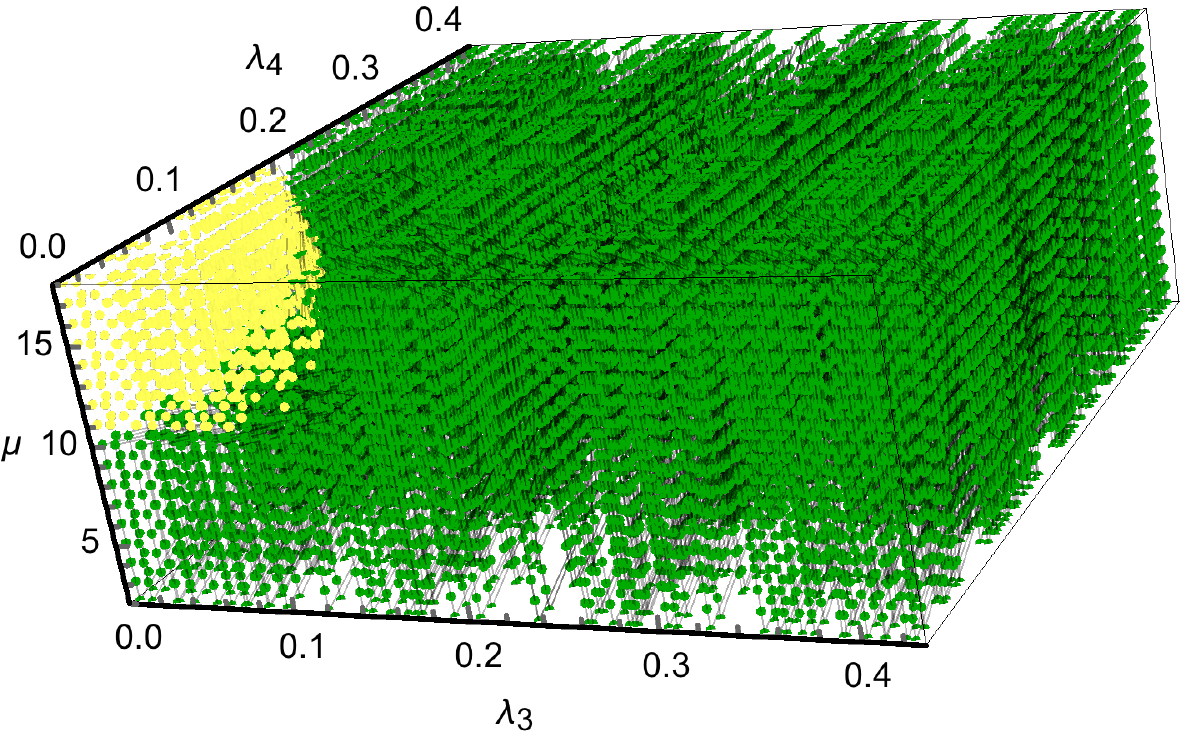}\label{f17}}	
\subfigure[$Y_N=0.4$, $M_R=10^{3}$ GeV]{\includegraphics[width=0.32\linewidth,angle=-0]{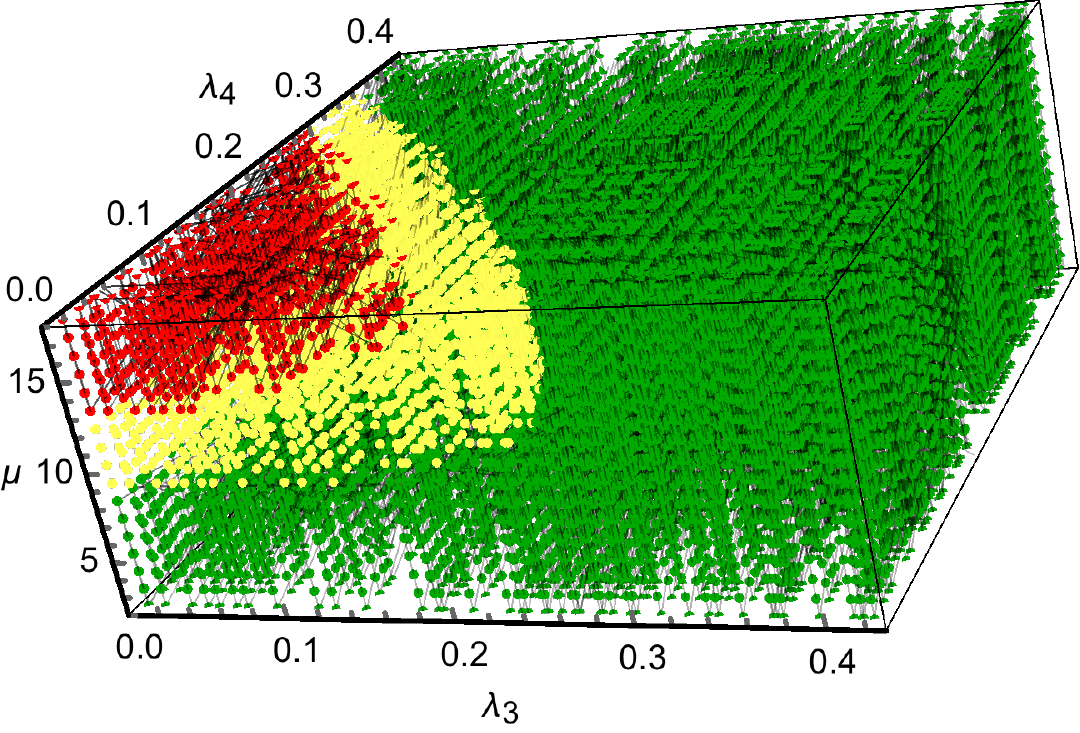}\label{f18}}
\subfigure[$Y_N =0.5$, $M_R=10^{3}$ GeV]{\includegraphics[width=0.38\linewidth,angle=-0]{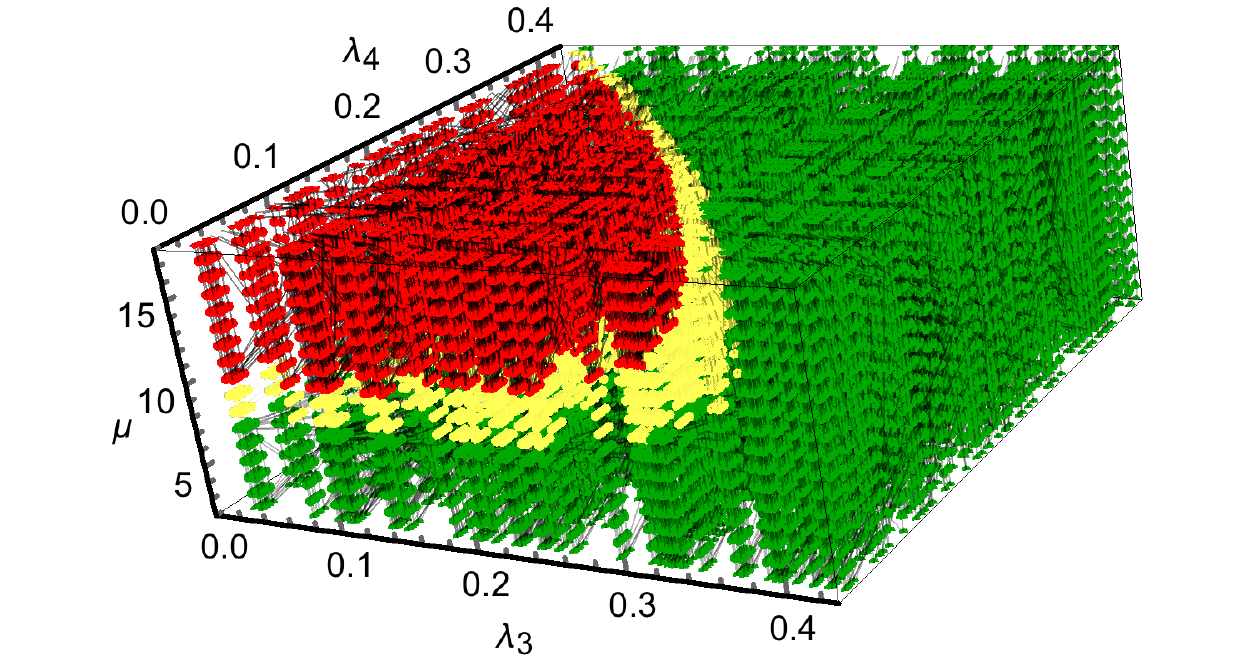}\label{f19}}}
		\hspace*{-0.4cm}
		\mbox{\subfigure[$Y_N =0.4$, $M_R=10^{2}$ GeV]	{\includegraphics[width=0.33\linewidth,angle=-0]{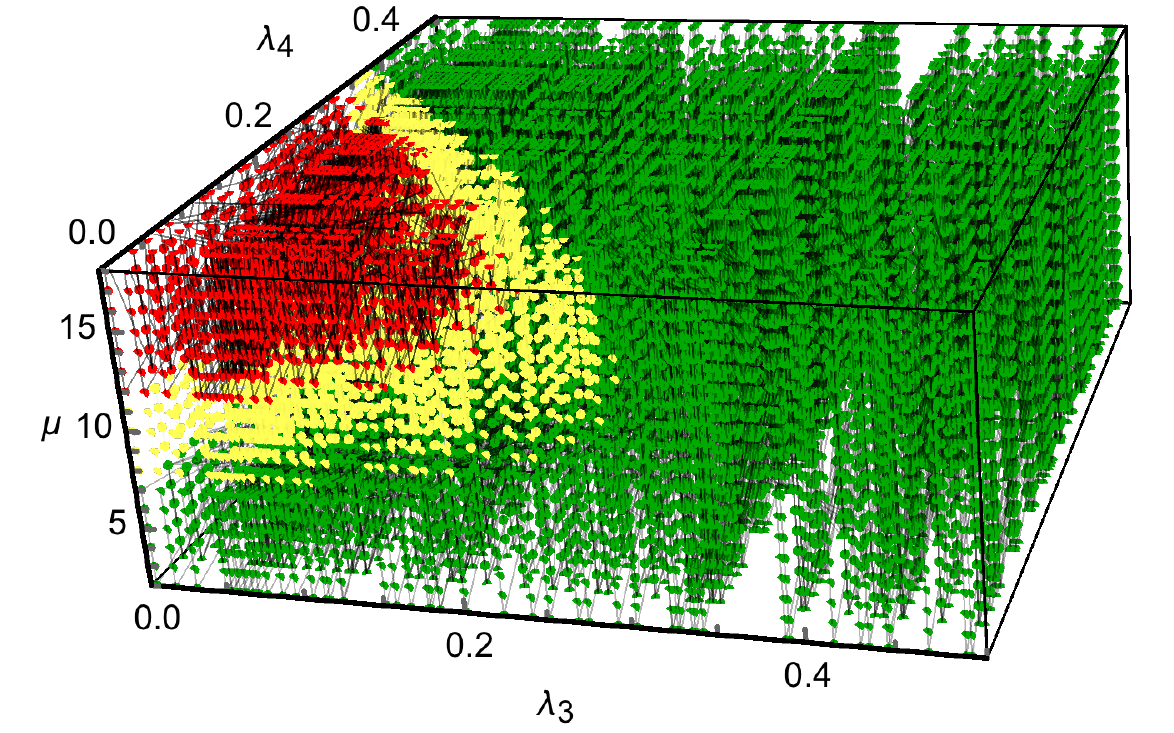}\label{f20}}}\mbox{\subfigure[$Y_N =0.4$, $M_R=10^{5}$ GeV]{\includegraphics[width=0.35\linewidth,angle=-0]{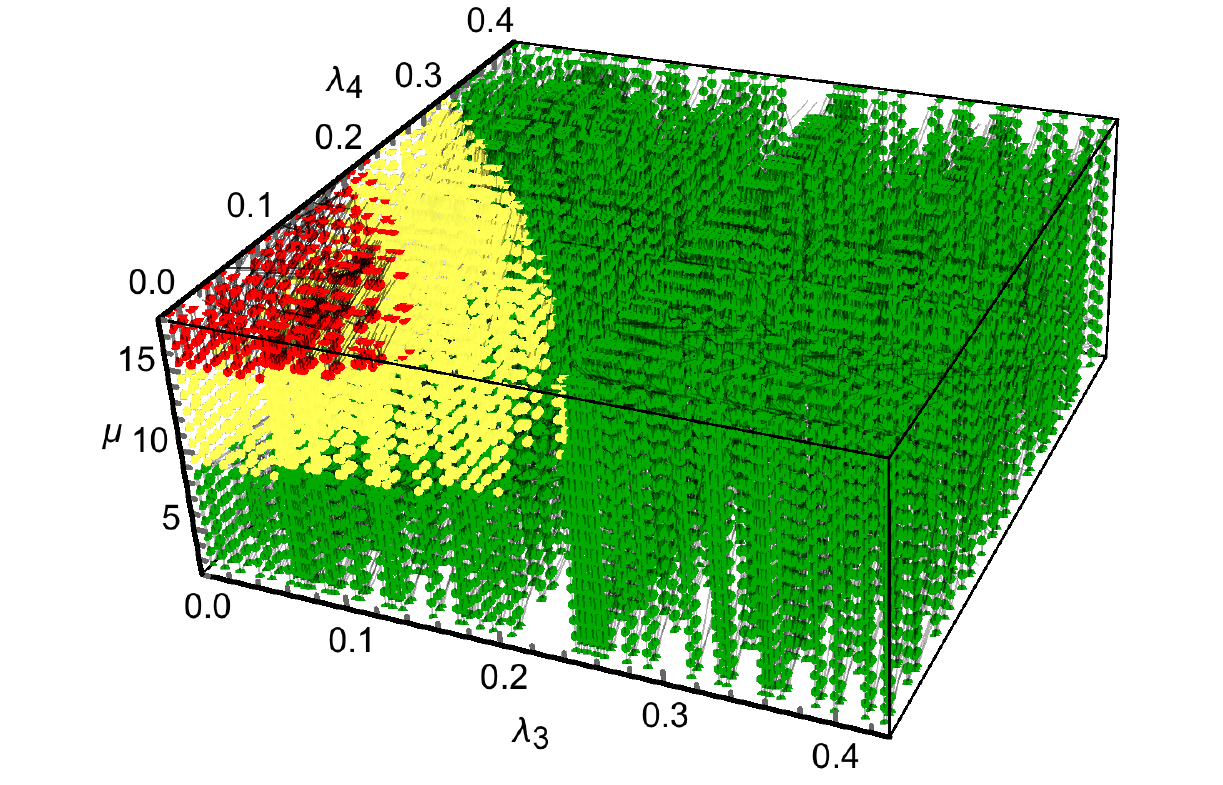}\label{f21}}\subfigure[$Y_N =0.4$, $M_R=10^{8}$ GeV]{\includegraphics[width=0.36\linewidth,angle=-0]{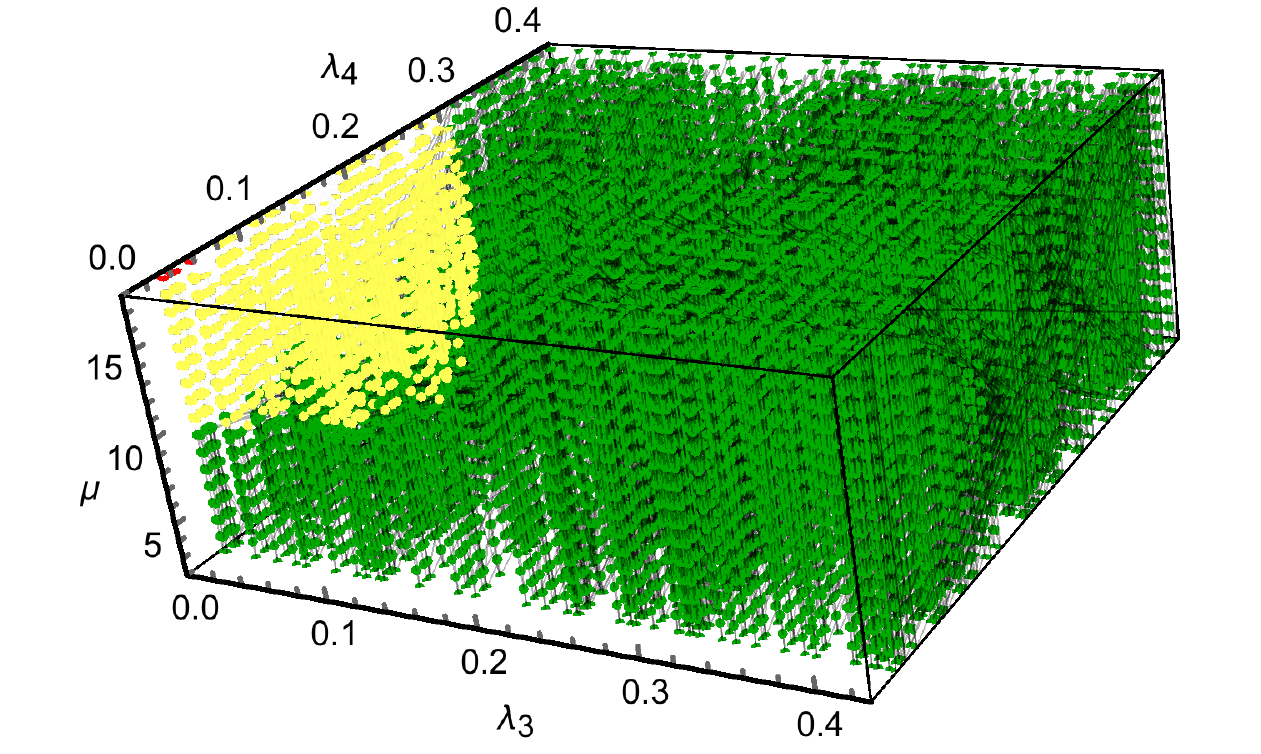}\label{f22}}}
		\caption{Three-dimensional correlation plot for $\lambda_3$ versus $\lambda_4$ with energy scale [log(10) in GeV] in  six different scenarios. In the top three panels, we fix $M_R=10^3$ GeV, $y_t= 0.93693$  and vary $Y_N$ from 0.1 (left) to 0.4 (middle) and 0.5 (right). In the bottom three panels, we fix $Y_N=0.4$ and vary $M_R$ from $10^2$ GeV (left) to $10^5$ GeV (middle) and $10^8$ GeV (right). In all the subplots, we have fixed $\lambda_2=\lambda_5=0.01$. The red, yellow and green regions correspond to the unstable, metastable and stable regions, respectively.} \label{fig10l}} 
	\end{figure}

To see the individual effects of the scalar quartic couplings $\lambda_{2,3,4,5}$ on the stability scale, we show in Figure~\ref{fig10l} the three-dimensional correlation plots for $\lambda_3$ versus $\lambda_4$ with energy scale $\mu$ for different values of $Y_N$ and $M_R$ with a fixed $\lambda_{2}=\lambda_5=0.01$. As in Figure~\ref{fig9l}, the red, yellow and green regions correspond to the unstable, metastable and stable regions respectively. Figure~\ref{f17}, Figure~\ref{f18} and Figure~\ref{f19} show the effect of the RHN Yukawa coupling on the stability scale. For smaller $Y_N$ =0.1, there is no unstable region. As the value of $Y_N$ is increased to 0.4 and 0.5 the stability and metastability regions decrease, while the unstable region increases. Similarly, Figure~\ref{f20}. Figure~\ref{f21} and Figure~\ref{f22}  describe the dependence on the $M_R$ scale. Here the metastable and stable regions increase as we increase the value of $M_R$ from $10^{2}$ to $10^{8}$ GeV. 
%\newpage

%%%%%%%%%%%%%%%%%%%%%%%%%%%%%%%%%%%%%%%%%%%%%%%%	

%	\subsection*{Instability Bound:} The scalar potential becomes unstable for large negative values of $\lambda_{eff}$. In this case there exists a second minima other than EW minima and the EW vacuum decays to the second minima within the lifetime of our universe. The region $0<\lambda_{eff}<- \frac{-0.065}{1-0.01 ln \frac{v}{\mu}}$ is defined as the instability region. 

\begin{figure}
\hspace*{-1cm}
{\mbox{\subfigure[$Y_N =10^{-7}$]{\includegraphics[width=0.50\linewidth,angle=-0]{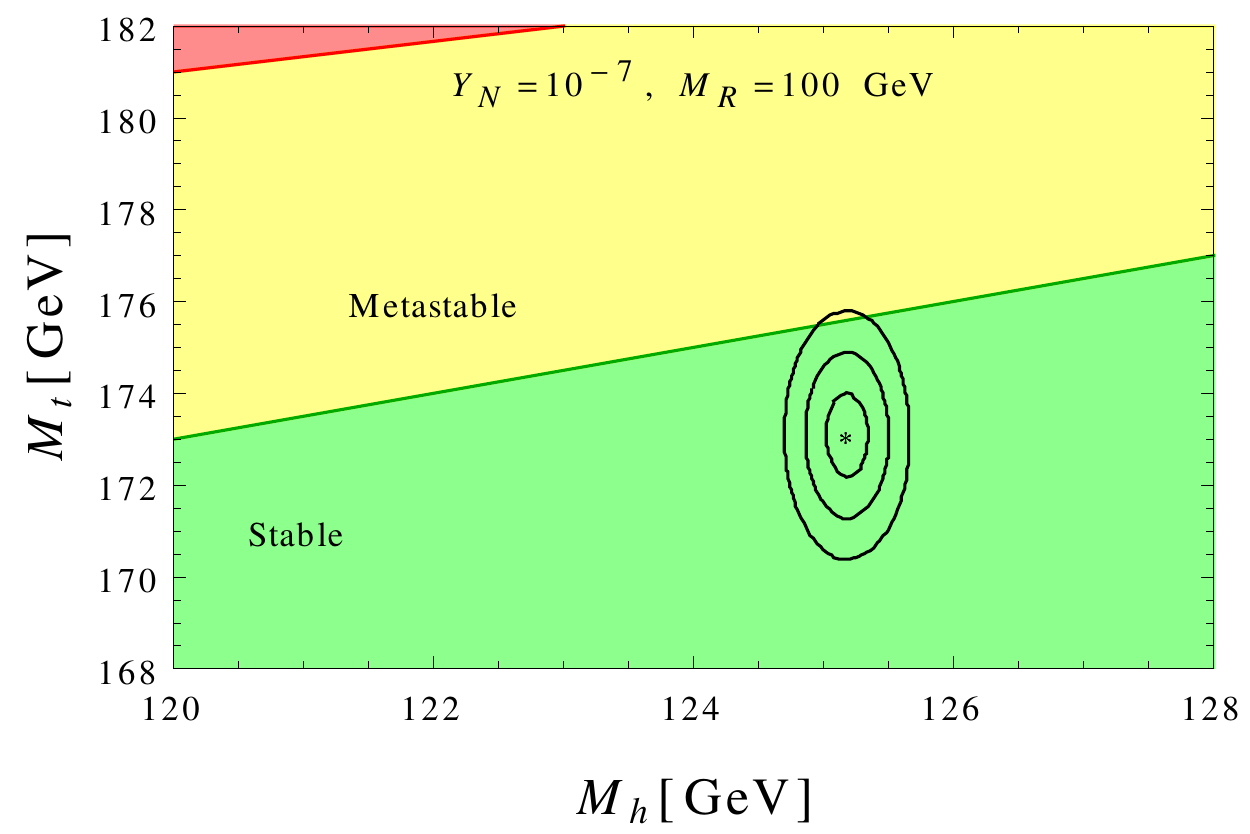}\label{f23}}
\subfigure[$Y_N = 0.38$]{\includegraphics[width=0.50\linewidth,angle=-0]{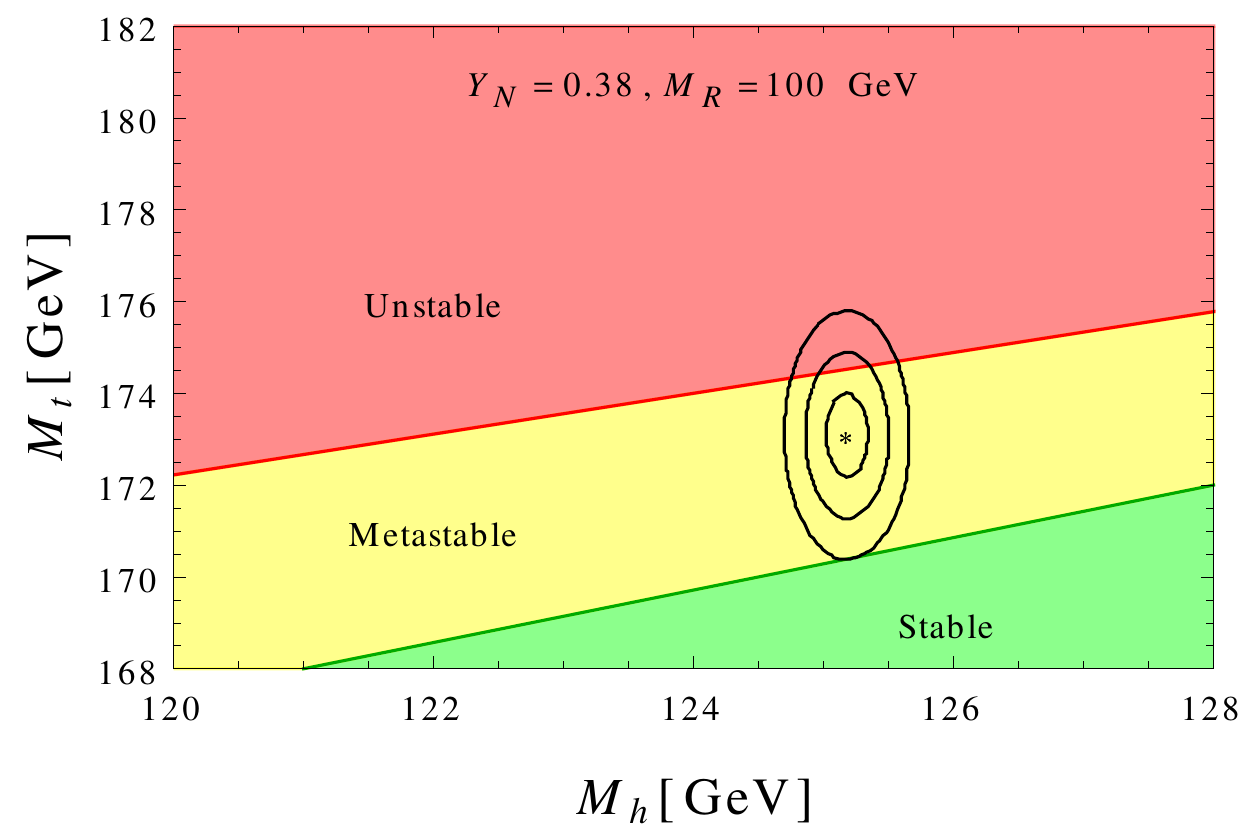}\label{f24}}}
\caption{Stability phase diagram in terms of the SM Higgs boson and top-quark pole masses. Here we have fixed $\lambda_i=0.1$ and $M_R=100$ GeV, while $Y_N$ is varied from $10^{-7}$ (left) to $Y_N = 0.38$ (right).  The red, yellow and green regions correspond to the unstable, metastable and stable regions respectively, which change depending on the model parameters. The contours and the dot show the current experimental $1\sigma,2\sigma,3\sigma$ regions and central value in the $(M_h,M_t)$ plane.}\label{fig11l}}
\end{figure}

As can be seen from Figure~\ref{fig9l}, the stability scale crucially depends on the top Yukawa coupling. The running of $\lambda_{\rm eff}$ also depends on the initial value of $\lambda_h$, which comes from the experimental value of the SM Higgs mass. Figure~\ref{fig11l} shows the stability phase diagram in terms of Higgs boson mass and top pole mass for two different choices of $Y_N=10^{-7}$ and 0.38 while keeping $M_R$ fixed at 100 GeV. The contours show the current experimental $1\sigma,2\sigma,3\sigma$ regions in the $(M_h,M_t)$ plane, while the dot represents the central value~\cite{Tanabashi:2018oca}. 
Figure~\ref{f23} describes that  for small $Y_N=10^{-7}$, the current $3\sigma$ values for the Higgs boson  mass and top mass mostly lie in the stable region. However, as $Y_N$ is increased to a large value of $0.38$ in  Figure~\ref{f24}, the Higgs boson mass value lies in  the stable region but the top mass value lies in the unstable/metastable region.  The bound that comes on $Y_N$ from stability for which both Higgs boson mass and the top mass lie in the stability region is  $Y_N \lesssim 0.32$ for $M_R=100$ GeV and $\lambda_i=0.1$. Although this turns out to be weaker than the existing experimental constraints~\cite{deGouvea:2015euy, Bolton:2019pcu}, this provides an independent, purely theoretical constraint on the model. 
%%%%%%%%%%%%%%%%%%%%%%%%%%%%%%%%%%%%%%%%%%%%%%%%
%%%%%%%%%%%%%%%%%%%%%%%%%%%%%%%%%%%%%%%%%%%%%%%

\section{LHC Phenomenology}\label{pheno}
The collider phenomenology of inert Higgs doublet with RHN is quite interesting as some decay modes involving RHNs are not allowed due to the $Z_2$ symmetry and this feature can be used to distinguish it from other scenarios. The pseudoscalar boson, the heavy CP-even Higgs boson and the charged Higgs boson ($A, H, H^\pm$) are all from  the inert doublet $\Phi_2$, which is $Z_2$ odd and their mass splittings  are mostly $\lesssim M_W$ [cf.~Eq.~\eqref{mass}]. However, mass splittings around $\gsim M_{W^\pm, Z}$ are also possible some parameter space. The $Z_2$ symmetry prohibits any kind of mass-mixing of these inert Higgs bosons with the SM-like Higgs boson, which is coming from $Z_2$-even $\Phi_1$. The couplings of $\Phi_2$ with fermions are also prohibited, leaving only the gauge and scalar couplings. Nevertheless, as shown above, the inert Higgs doublet $\Phi_2$ plays a crucial role in determining the stability and perturbativity conditions, and therefore, it is important to study their potential signatures at colliders. In Table~\ref{bps} we present ten benchmark points for the future collider study which are allowed by the vacuum stability and perturbativity bounds.  The scenario with the lightest charged Higgs bosons ($H^\pm$)  causes an electromagnetically-charged DM candidate and such points are phenomenologically disallowed. This leaves us with two kind of scenarios with either $H$ or $A$ as the lightest heavy scalar, to be identified as the DM candidate.

%%%%%%%%%%%%%%%%%%%%%%%%%%%%%%%%

\begin{table}[t]
	%\begin{ruledtabular}
	\begin{center}
		\renewcommand{\arraystretch}{1.4}
		\begin{tabular}{||c ||c|c|c|c|c|c|c||}
			\hline\hline
			BP &$\lambda_3$&$\lambda_4$&$\lambda_5$&$m_{22}$&$M_H$&$M_A$&$M_{H^{\pm}}$\\
			\hline\hline
			BP1&0.10&0.10&0.10&200&228.26&200.00&207.42\\
			\hline
			BP2&0.10&0.10&0.10&300&319.53&300.00&305.00\\
			\hline
				BP3&0.20&0.20&0.20&250&294.53&250.00&261.84\\
				\hline
			BP4&0.11&0.11&$-0.20$&200&185.88&242.40&208.15\\
			\hline
			BP5&0.22&0.22&$-0.16$&300&305.99&336.14&310.89\\
			\hline
			BP6&0.32&$-0.10$&$-0.01$&300&309.92&311.86&315.72\\
			\hline
			BP7&0.32&$-0.20$&$-0.08$&250&247.56&266.40&268.66\\
			\hline
			BP8&0.29&0.31&0.31&2200&2208.38&2199.86&2201.99\\
			\hline
			BP9&0.23&0.11&0.12&1200&1207.30&1201.26&1202.90\\
			\hline
			BP10&0.20&0.23&0.28&2000&2007.48&1999.01&2001.51\\
			\hline
			\hline
		\end{tabular}
		\caption{Benchmark points allowed by the vacuum stability, perturbativity and DM constraints. Here we have chosen $ Y_N=0.4$ and $M_R =$ 1 TeV.
		}\label{bps}
	\end{center}
\end{table}
%%%%%%%%%%%%%%%%%%%%%%%%%%%%%%%%%%%%%%%%%%%%%%%%%

%%%%%%%%%%%%%%%%%%%%%%%DECAY MODES FEYNMAN DIAGRAMS%%%%%%%
\begin{figure}[thb]
%	\hspace*{-0.9cm}
	\mbox{\subfigure[]{\includegraphics[width=0.35\linewidth,angle=-0]{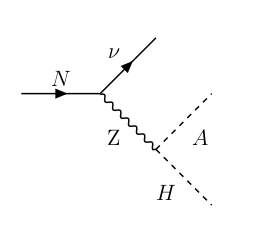}\label{f25}}
		\subfigure[]{\includegraphics[width=0.35\linewidth,angle=-0]{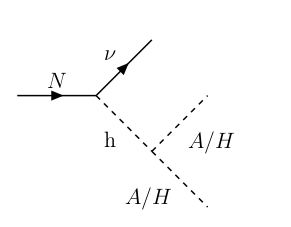}\label{f26}}
		\subfigure[]{\includegraphics[width=0.35\linewidth,angle=-0]{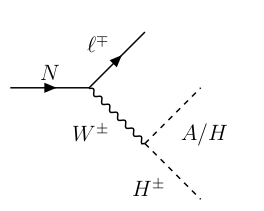}\label{f27}}}
		\mbox{\subfigure[]{\includegraphics[width=0.35\linewidth,angle=-0]{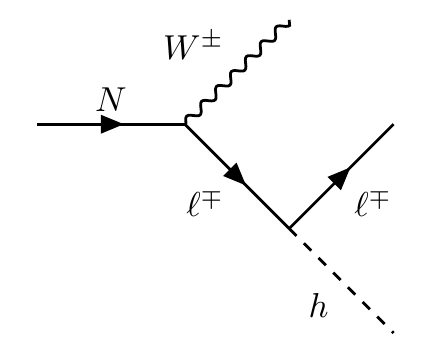}\label{f27a}}
		\subfigure[]{\includegraphics[width=0.35\linewidth,angle=-0]{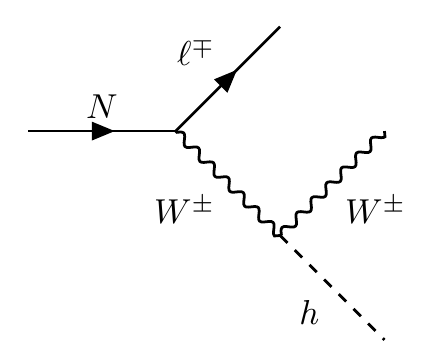}\label{f27b}}
		\subfigure[]{\includegraphics[width=0.35\linewidth,angle=-0]{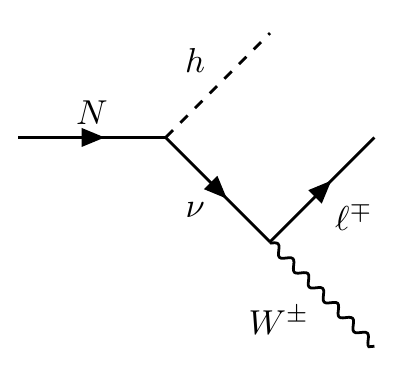}\label{f27c}}}
	\caption{Various three-body decays of RHNs involving heavy Higgs bosons in the final state: (a) Decay to light neutrinos and $H/A$ via an off-shell $Z$ boson; (b) decay to light neutrinos and $H/A$ pairs via an off-shell Higgs boson; (c) decay to a charged-lepton and charged Higgs boson in association with $H/A$ via an off-shell $W$ boson; (d)-(f) decay to a charged lepton and SM $W$ and Higgs bosons.}\label{fig12l}
\end{figure}
%%%%%%%%%%%%%%%%%%%% Decay branching fraction %%%%%%%%%%%%%%%
\begin{table}[t]
	%\begin{ruledtabular}
	\begin{center}
		\renewcommand{\arraystretch}{1.4}
		\begin{tabular}{||c |c||}
			\hline\hline
			Decay Modes &  BR \\
			& in percentage \\
			\hline
			$N_i\to h W^\pm \ell^\mp$ & 0.36\\
			\hline
			$N_i\to H H^\pm \ell^\mp$ & $2.4\times 10^{-4}$ \\
			\hline
			$N_i\to A H^\pm \ell^\mp$ & $5.2\times 10^{-5}$ \\
			\hline
		\end{tabular}
		\caption{Dominant three-body decay BRs of RHN involving Higgs bosons in the final states for a benchmark  point allowed by the vacuum stability and perturbativity  with $M_R=$ 1 TeV. Note that these BRs are independent of the choice of $Y_N$. 
		}\label{br}
	\end{center}
\end{table}

The RHNs on the other hand only couple to $\Phi_1$, leaving the Yukawa interactions with the SM-like Higgs boson.  Via their mixing with the light neutrinos, the RHNs also couple to the SM $W$ and $Z$ gauge bosons after EW symmetry breaking, which are proportional to the VEV of $\Phi_1$ and decay dominantly to $W^\pm\ell^\mp,\, Z\nu$, and  $h\nu$. In principle, the RHN sector and the inert scalar sector do not talk to each other. However, couplings with the gauge sectors open up a window to the inert Higgs sector from the RHN decay. This is possible via the three-body decays of the RHNs with heavy Higgs bosons in the final states that can be seen from  Figure~\ref{fig12l}. The RHNs can decay to light neutrinos and $H, A$ via an off-shell $Z$ boson [cf.~Figure~\ref{f25}], to light neutrinos and $H/A$ pairs via a off-shell $h$ [cf.~Figure~\ref{f26}], to a charged lepton and charged Higgs boson in association with $H/A$ [cf.~Figure~\ref{f27}], and to a charged lepton and SM Higgs boson in association with $W^\pm$ [cf.~Figures~\ref{f27a}-\ref{f27c}].  For a RHN with mass 1 TeV, though the two-body decay modes (with on-shell $W^\pm, Z$ and $h$) dominate, but the three-body decay modes involving the heavy Higgs sector can still be explored at the LHC. The highest three-body decay mode is $N_i\to h W^\pm \ell^\mp$ [cf.~Figure~\ref{f27a}] with branching ratio (BR) $\sim 0.36 \%$ and other modes are with  ${\rm BR}(N_i\to H H^\pm \ell^\mp)\sim 2.4\times 10^{-4}\%$ and ${\rm BR}(N_i\to A H^\pm \ell^\mp)\sim 5.2 \times 10^{-5}\%$ respectively, as given in Table~\ref{br} for $ Y_N=0.01$ and $M_R=$ 1 TeV. 
%For $N$ pair production the cross-sections  are $1.8 \times 10^{-4}$ and $1.2\times 10^{-3}$ respectively at the LHC with center of mass energy 14 TeV and 100 TeV. Certainly the cross-sections are not that encouraging and we need to go for higher luminosity at the 100 TeV LHC. 

%%%%%%%%%%%%%%%%% Production modes feynman diagrams%%%%%

\begin{figure}[thb]
%	\hspace*{-0.7cm}
	{\mbox{\subfigure[]{\includegraphics[width=0.32\linewidth,angle=-0]{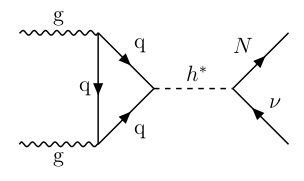}\label{f28}}\quad	\subfigure[]{\includegraphics[width=0.32\linewidth,angle=-0]{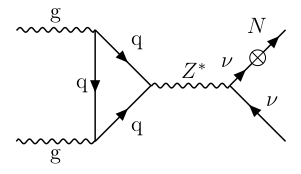}\label{f29}}
			\quad\subfigure[]{\includegraphics[width=0.32\linewidth,angle=-0]{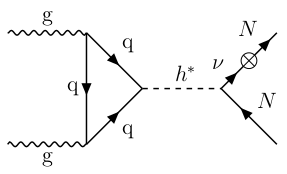}\label{f30}}}
%		\hspace*{-0.25cm}
		\mbox{\subfigure[]	{\includegraphics[width=0.33\linewidth,angle=-0]{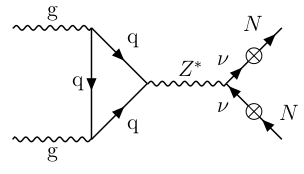}\label{f31}}}\quad
		\mbox{\subfigure[]{\includegraphics[width=0.29\linewidth,angle=-0]{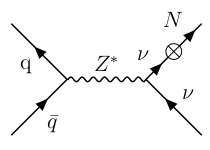}\label{f32}}\quad\subfigure[]{\includegraphics[width=0.3\linewidth,angle=-0]{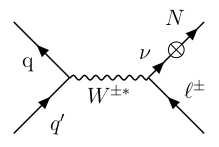}\label{f33}}}
		\caption{Feynman diagrams for RHN production via either gluon-gluon fusion [(a) to (d)] or Drell-Yan process [(e) and (f)]. The cross $\otimes$ indicates light-heavy neutrino mixing.} \label{fig13l}} 
\end{figure}
%%%%%%%%%%%%%%%%%%%%%%%%%%%%%%%%%%%%%%%%%%%%%%%%%%%%%%

%%%%%%%%%%%%%%%%%%%% Production Table %%%%%%%%%%%%%%%
\begin{table}[t]
	%\begin{ruledtabular}
	\begin{center}
		\renewcommand{\arraystretch}{1.4}
		\begin{tabular}{||c|c|c|c|c|c|c|c||}
			\hline\hline
			\multicolumn{2}{||c|}{Parameters} & \multicolumn{6}{|c||}{Processes}\\
			\hline
	& &  \multicolumn{2}{|c|}{$\sigma(gg \to \sum_i N_i \nu_i)$ }&\multicolumn{2}{|c|}{$\sigma(gg \to \sum_i N_i N_i)$ }&\multicolumn{2}{|c||}{$\sigma_{\rm{DY}}(pp \to \sum_i N_i +X)$ } \\
		$Y_N$ 	&$M_R$ &\multicolumn{2}{|c|}{in fb}&\multicolumn{2}{|c|}{in fb}&\multicolumn{2}{|c||}{in fb}\\
		\cline{3-8}
			             &in GeV & 14 TeV & 100  TeV &14 TeV &100 TeV &14 TeV &100 TeV\\
			\hline
			0.1&500&0.15&9.70 &$1.8\times 10^{-4}$&$1.2\times 10^{-2}$&0.34&6.90 \\
			 \hline
			 0.1&1000&$1.6\times 10^{-3}$&0.36 &$5.0\times 10^{-7}$&$1.1\times 10^{-4}$&$4.5\times10^{-3}$&0.18\\
			 \hline
			 0.4&500&2.40& 155.40 &0.30&0.50 &5.00&95.60\\
			 \hline
			  0.4&1000&0.03& 5.83&$1.2\times 10^{-4}$&0.03&0.06&2.55\\
			 \hline
			 \hline
		\end{tabular}
		\caption{NLO production cross-sections of the RHNs at the LHC for 14 TeV and 100 TeV center of mass energy. Here the other parameters are as in BP3 of Table~\ref{bps}. }\label{cross}
	\end{center}
\end{table}
%%%%%%%%%%%%%%%%%%%%%%%%%%%%%%%%%%%%%%%%%%%%%%%%%%%

As for the RHN production at the LHC, being SM gauge-singlets, they can only be produced via their mixing with active neutrinos in the minimal seesaw model. The dominant production modes are shown in Figure~\ref{fig13l}. There are two types of processes: (a)-(d) involve RHN production \cite {vonBuddenbrock:2017gvy, Das:2016hof} via off-shell Higgs boson from gluon-gluon fusion, whereas (e)-(f) involve production via off-shell $W^\pm/Z$ from Drell-Yan processes. 
% and this makes probing such  right-handed neutrinos at the LHC  challenging. The right-handed neutrino in this case only talks to SM gauge bosons via the mixing angle ($\frac{Y_N v}{\sqrt{2}M_R}$). However, in $U(1)'$ extensions of SM  the right-handed neutrinos they can be pair produced at the LHC via $U(1)'$ gauge boson. Such Type-I right-handed neutrino productions and their phenomenologies have been studied in the literature \cite{RHNU1, RHNLFV, RHNLFV2,  RHNBML}. In the case of Type-III Seesaw, the right-handed neutrinos have charged partner and couple to $W^\pm$ boson \cite{Foot:1988aq}. The LHC phenomenologies of such extensions with and without additional Higgs doublet have been looked into \cite{TypeIII2HDM, Franceschini:2008pz, TypeIII}.  The inverse-seesaw phenomenologies 
 %probing the right-handed neutrinos at the LHC along with heavier Higgs bosons are examined \cite{ISS, ISS2}. 
% Particularly in this case of Type-I Seesaw with inert Higgs doublet extension of SM Production of these right-handed neutrinos at colliders again depends on the mixing angle  ($\frac{Y_N v}{\sqrt{2}M_R}$) with SM  gauge bosons. The right-handed neutrino sector neither talk to the other heavy Higgs bosons ($H/A/H^\pm$) directly nor these heavy Higgs bosons couple to SM fermions. 
%The right-handed neutrinos can be produced via off-shell SM like Higgs boson in association with light neutrinos or in pair as shown in Figure~\ref{fig13l} either by gluon gluon fusion (see, Figure~\ref{f28} to Figure~\ref{f31}) and/or normal Drell-Yan processes (see, Figure~\ref{f32} and Figure~\ref{f33}). 
The next-to-leading order (NLO) cross-sections for $Y_N=0.1,0.4$ and $M_R=500$ GeV, 1 TeV are given in Table~\ref{cross} where other parameters are kept as in BP3 of Table~\ref{bps}. For the process $N\nu$ [cf.~Figure~\ref{f28}], the production cross-section at NLO for $Y_N=0.1$ and $M_R=500$ GeV is: $\sigma (gg \to \sum_{i = 1,2,3} N_i \nu_i)$ is  $\sim 0.15$ and $9.7$ fb  respectively at the LHC with 14 TeV and 100 TeV center of mass energy \cite{cross1}. For pair production the cross-sections are $1.8\times 10^{-4}$
  and $1.2\times 10^{-3}$  respectively at the LHC with 14 TeV and 100 TeV center of mass energy. Here we have used {\tt CalcHEP 3.7.5}~\cite{calchep} for calculating the tree-level cross sections and decay branching fraction and have chosen {\tt NNPDF 3.0 QED NLO} \cite{pdf} and $\sqrt{\hat{s}}$ (parton-level center of mass energy) as the energy scale for the cross-section calculations.  The third column of Table~\ref{cross} also give NLO Drell-Yan cross-sections for the same scale and PDF. We can see that for $\sqrt{s}=14$ TeV at the LHC Drell-Yan processes are more dominant than gluon gluon fusion, whereas at $\sqrt{s}=100$ TeV gluon gluon fusion processes surpass Drell-Yan ones.  Though the overall cross-sections are small, but higher luminosity LHC can probe these three-body decays. The maximum cross-section comes for $Y_N=0.4$ and $M_R=500$ GeV and for $\sqrt{s}=100$ TeV and these are 155.40 fb, 95.60 fb, 0.50 fb respectively for $(gg \to \sum_i N_i \nu_i)$, $(pp \to \sum_i N_i +X)_{\rm DY}$ and $(gg \to \sum_i N_i N_i)$. Note that although such large values of $Y_N$ might have been excluded from indirect constraints such as EW precision data, it is still useful to get an independent direct constraint from the collider searches. 

Coming to the inert Higgs boson signatures we have to rely on the mass spectrum of the Higgs bosons which depend on the couplings $\lambda_{3,4,5}$ as shown in Eq.~\eqref{mass}. Table~\ref{bps} shows benchmark points with the $\lambda_{3,4,5}$  that are allowed by the vacuum stability and perturbativity conditions.  Depending on the phase space available, the charged Higgs boson in this model can decay into $A W^\pm$ and/or $HW^\pm$ mostly via off-shell $W$ boson as the heavy Higgs bosons stay degenerate.  The lighter of $A$ and $H$ is the DM candidate and thus can give rise to the signature of mono-lepton plus missing energy or dijet plus missing energy. However, because of the $Z_2$-odd nature of $H,A,H^\pm$ we can only produce the charged Higgs bosons as pair or in association with $H/A$. The heavier of $A/H$ in that case decays to dilepton plus missing energy via off-shell $Z$ boson.  The production of $H^\pm$ pair gives rise to dilepton plus missing energy and $H^\pm A/H$ give rise to trilepton or mono-lepton plus missing energy signatures, which can be searched for at the LHC and FCC-hh~\cite{Benedikt:2018csr}. The inert Higgs boson productions in association with the DM candidate leaving to jet plus lepton and missing energy signatures  are studied in Ref.~\cite{Belyaev:2016lok,Chakrabarty:2017qkh}. The inert doublet signatures along with the three-body decays of RHNs with Higgs boson in the final state can shed light on this model at the LHC with higher luminosity. 

The LHC phenomenology discussed here is different from $U(1)'$ extensions where the RHNs can be pair-produced at the LHC via the $U(1)'$ gauge boson~\cite{Basso:2008iv, Kang:2015uoc, Cox:2017eme, Das:2017deo, Das:2019fee}. Phenomenological signatures of such RHN decays in the type-I seesaw in presence of extra scalars have been studied in the literature~\cite{RHNU1, RHNLFV, RHNLFV2,  RHNBML, Ko:2013zsa}. Similarly, in the case of type-III seesaw, the RHNs have charged partner and couple to $W^\pm$ bosons~\cite{Foot:1988aq}. The LHC phenomenology of such extensions with and without additional Higgs doublet has also been looked into~\cite{TypeIII2HDM, Franceschini:2008pz, TypeIII}.  The inverse-seesaw phenomenologies probing the RHNs at the LHC along with heavier Higgs bosons were also examined~\cite{ISS, ISS2}.

\section{Conclusion}\label{concl}
We have considered a simple extension of the SM with a $Z_2$-odd inert Higgs doublet, supplemented by right-handed neutrinos with potentially large Dirac Yukawa couplings. The neutral part of the inert-Higgs doublet is a suitable DM candidate, while the RHNs are responsible for the correct light neutrino masses via seesaw mechanism. We have studied the effect of these new scalars and fermions on the stability of the EW vacuum by performing an RG analysis for the scalar quartic couplings. 

We find that the additional scalars enhance the EW stability bound with respect to the SM case, as expected. Although the introduction of RHNs with relatively larger Yukawa couplings can be a spoiler for vacuum stability, the inert doublet comes to a rescue by contributing positively to the $\beta$-functions.  On the other hand, the scalar quartic couplings cannot take arbitrarily large values at the EW scale due to perturbativity considerations at higher scales. In particular, we find upper bounds on the scalar quartic couplings $\lambda_i$ (with $i=2,3,4,5$) and the Dirac Yukawa couplings $Y_N$, depending on the RHN mass scale $M_R$, to satisfy both stability and perturbativity constraints.  

We also analyzed the RG-improved effective potential to identify the regions of parameter space giving rise to stable, metastable and unstable vacua. For fixed values of $\lambda_i$, increasing $Y_N$ enlarges the unstable vacuum region, whereas decreasing $Y_N$ and/or increasing the RHN mass scale $M_R$ enhances the stability prospects. The effect of the RHNs on vacuum stability is only relevant in the low-scale seesaw scenarios with relatively large Dirac Yukawa couplings, which can be realized either via cancellations in the type-I seesaw matrix or via some form of inverse seesaw mechanism.

We also studied the phenomenological signatures of the heavy Higgs bosons along with RHNs at the LHC and future 100 TeV collider. Since the heavy Higgs bosons in this model come from the $Z_2$-odd doublet, they are relatively non-interacting with the SM particles and are almost mass-degenerate, thus making their collider searches rather difficult. We have identified some new three-body decay modes of the RHNs to heavy Higgs bosons (assuming that the RHNs are heavier than the Higgs bosons) which can be used to distinguish this model from other vanilla RHN models.

\acknowledgments
PB wants to thank Washington University in St. Louis for a visit during the project and SERB CORE Grant CRG/2018/004971 and Anomalies 2019-IUSSTF for the support.  BD would like to thank the organizers of FPCP 2018 at University of Hyderabad and IIT Hyderabad for warm hospitality during which part of this work was done. The work BD is supported in part by the U.S. Department of Energy under Grant No. DE-SC0017987 and in part by the MCSS funds. SJ thanks DST/INSPIRES/03/2018/001207 for the financial support towards the PhD program. AK thanks DST/INSPIRES/03/2018/000344. SJ thanks Anirban Karan and Saunak Dutta for help in Mathematica. SJ wants to thank Dr. Gajendranath Chaudhury for giving office space during this work. 

%\vspace{-0.1cm}
%\samepage
\appendix
\section{Two-loop $\beta$-functions} \label{betaf}
\subsection{Scalar Quartic Couplings}
\footnotesize{
\begingroup
\allowdisplaybreaks
\begin{align*}
	\beta_{\lambda_h} \ =  \ &
\frac{1}{16\pi^2} \Bigg[	\frac{27}{200} g_{1}^{4} +\frac{9}{20} g_{1}^{2} g_{2}^{2} +\frac{9}{8} g_{2}^{4} -\frac{9}{5} g_{1}^{2} \lambda_1 -9 g_{2}^{2} \lambda_1 +24 \lambda_{1}^{2} +2 \lambda_{3}^{2} +2 \lambda_3 \lambda_4 +\lambda_{4}^{2}+4 \lambda_{5}^{2} \nonumber \\ 
	&\qquad \quad +12 \lambda_1 \mbox{Tr}\Big({Y_d  Y_{d}^{\dagger}}\Big) +4 \lambda_1 \mbox{Tr}\Big({Y_e  Y_{e}^{\dagger}}\Big) +4 \lambda_1 \mbox{Tr}\Big({Y_N  Y_{N}^{\dagger}}\Big) +12 \lambda_1 \mbox{Tr}\Big({Y_u  Y_{u}^{\dagger}}\Big)  \nonumber \\ 
	&\qquad \quad -6 \mbox{Tr}\Big({Y_d  Y_{d}^{\dagger}  Y_d  Y_{d}^{\dagger}}\Big) -2 \mbox{Tr}\Big({Y_e  Y_{e}^{\dagger}  Y_e  Y_{e}^{\dagger}}\Big)  -6 \mbox{Tr}\Big({Y_u  Y_{u}^{\dagger}  Y_u  Y_{u}^{\dagger}}\Big)-2 \mbox{Tr}\Big({Y_N  Y_{N}^{\dagger}  Y_N  Y_{N}^{\dagger}}\Big)\Bigg] \nonumber \\
&+\frac{1}{(16\pi^2)^2}\Bigg[-\frac{3537}{2000} g_{1}^{6} -\frac{1719}{400} g_{1}^{4} g_{2}^{2} -\frac{303}{80} g_{1}^{2} g_{2}^{4} +\frac{291}{16} g_{2}^{6} +\frac{1953}{200} g_{1}^{4} \lambda_1 +\frac{117}{20} g_{1}^{2} g_{2}^{2} \lambda_1 \nonumber \\ 
& \qquad \qquad -\frac{51}{8} g_{2}^{4} \lambda_1 +\frac{108}{5} g_{1}^{2} \lambda_{1}^{2} 
+108 g_{2}^{2} \lambda_{1}^{2} -312 \lambda_{1}^{3} +\frac{9}{10} g_{1}^{4} \lambda_3 +\frac{15}{2} g_{2}^{4} \lambda_3 +\frac{12}{5} g_{1}^{2} \lambda_{3}^{2} +12 g_{2}^{2} \lambda_{3}^{2} \nonumber \\ 
&\qquad \qquad 
-20 \lambda_1 \lambda_{3}^{2} -8 \lambda_{3}^{3} +\frac{9}{20} g_{1}^{4} \lambda_4 
-\frac{3}{2} g_{1}^{2} g_{2}^{2} \lambda_4 +\frac{15}{4} g_{2}^{4} \lambda_4 +\frac{12}{5} g_{1}^{2} \lambda_3 \lambda_4 +12 g_{2}^{2} \lambda_3 \lambda_4 \nonumber \\ 
&\qquad \qquad 
-20 \lambda_1 \lambda_3 \lambda_4 -12 \lambda_{3}^{2} \lambda_4 +\frac{6}{5} g_{1}^{2} \lambda_{4}^{2} +3 g_{2}^{2} \lambda_{4}^{2} -12 \lambda_1 \lambda_{4}^{2} -16 \lambda_3 \lambda_{4}^{2} -6 \lambda_{4}^{3} -\frac{12}{5} g_{1}^{2} \lambda_{5}^{2} \nonumber \\ 
&\qquad \qquad 
-56 \lambda_1 \lambda_{5}^{2} -80 \lambda_3 \lambda_{5}^{2} +8 \lambda_4 \lambda_{5}^{2} +\frac{1}{20} \Big(-5 \Big(64 \lambda_1 \Big(-5 g_{3}^{2}  + 9 \lambda_1 \Big) -90 g_{2}^{2} \lambda_1  + 9 g_{2}^{4} \Big) 
\nonumber \\ 
&\qquad \qquad 
+ 9 g_{1}^{4}  + g_{1}^{2} \Big(50 \lambda_1  + 54 g_{2}^{2} \Big)\Big)\mbox{Tr}\Big({Y_d  Y_{d}^{\dagger}}\Big) -\frac{3}{20} \Big(15 g_{1}^{4}  -2 g_{1}^{2} \Big(11 g_{2}^{2}  + 25 \lambda_1 \Big) \nonumber \\ 
&\qquad \qquad 
+ 5 \Big(-10 g_{2}^{2} \lambda_1  + 64 \lambda_{1}^{2}  + g_{2}^{4}\Big)\Big)\mbox{Tr}\Big({Y_e  Y_{e}^{\dagger}}\Big) -\frac{9}{100} g_{1}^{4} \mbox{Tr}\Big({Y_N  Y_{N}^{\dagger}}\Big) \nonumber \\ 
&\qquad \qquad 
-\frac{3}{10} g_{1}^{2} g_{2}^{2} \mbox{Tr}\Big({Y_N  Y_{N}^{\dagger}}\Big) -\frac{3}{4} g_{2}^{4} \mbox{Tr}\Big({Y_N  Y_{N}^{\dagger}}\Big) +\frac{3}{2} g_{1}^{2} \lambda_1 \mbox{Tr}\Big({Y_N  Y_{N}^{\dagger}}\Big) +\frac{15}{2} g_{2}^{2} \lambda_1 \mbox{Tr}\Big({Y_N  Y_{N}^{\dagger}}\Big)\nonumber \\ 
&\qquad \qquad 
-48 \lambda_{1}^{2} \mbox{Tr}\Big({Y_N  Y_{N}^{\dagger}}\Big) -\frac{171}{100} g_{1}^{4} \mbox{Tr}\Big({Y_u  Y_{u}^{\dagger}}\Big) +\frac{63}{10} g_{1}^{2} g_{2}^{2} \mbox{Tr}\Big({Y_u  Y_{u}^{\dagger}}\Big) -\frac{9}{4} g_{2}^{4} \mbox{Tr}\Big({Y_u  Y_{u}^{\dagger}}\Big) \nonumber \\ 
&\qquad \qquad +\frac{17}{2} g_{1}^{2} \lambda_1 \mbox{Tr}\Big({Y_u  Y_{u}^{\dagger}}\Big) +\frac{45}{2} g_{2}^{2} \lambda_1 \mbox{Tr}\Big({Y_u  Y_{u}^{\dagger}}\Big) +80 g_{3}^{2} \lambda_1 \mbox{Tr}\Big({Y_u  Y_{u}^{\dagger}}\Big) -144 \lambda_{1}^{2} \mbox{Tr}\Big({Y_u  Y_{u}^{\dagger}}\Big) \nonumber \\ 
&\qquad \qquad +\frac{4}{5} g_{1}^{2} \mbox{Tr}\Big({Y_d  Y_{d}^{\dagger}  Y_d  Y_{d}^{\dagger}}\Big) -32 g_{3}^{2} \mbox{Tr}\Big({Y_d  Y_{d}^{\dagger}  Y_d  Y_{d}^{\dagger}}\Big) -3 \lambda_1 \mbox{Tr}\Big({Y_d  Y_{d}^{\dagger}  Y_d  Y_{d}^{\dagger}}\Big) \nonumber \\ 
&\qquad \qquad -\frac{12}{5} g_{1}^{2} \mbox{Tr}\Big({Y_e  Y_{e}^{\dagger}  Y_e  Y_{e}^{\dagger}}\Big)  - \lambda_1 \mbox{Tr}\Big({Y_e  Y_{e}^{\dagger}  Y_e  Y_{e}^{\dagger}}\Big) -14 \lambda_1 \mbox{Tr}\Big({Y_e  Y_{N}^{\dagger}  Y_N  Y_{e}^{\dagger}}\Big) \nonumber \\ 
&\qquad \qquad - \lambda_1 \mbox{Tr}\Big({Y_N  Y_{N}^{\dagger}  Y_N  Y_{N}^{\dagger}}\Big) -\frac{8}{5} g_{1}^{2} \mbox{Tr}\Big({Y_u  Y_{u}^{\dagger}  Y_u  Y_{u}^{\dagger}}\Big) -32 g_{3}^{2} \mbox{Tr}\Big({Y_u  Y_{u}^{\dagger}  Y_u  Y_{u}^{\dagger}}\Big) \nonumber \\ 
&\qquad \qquad  -3 \lambda_1 \mbox{Tr}\Big({Y_u  Y_{u}^{\dagger}  Y_u  Y_{u}^{\dagger}}\Big) -42 \lambda_1 \mbox{Tr}\Big({Y_u  Y_{u}^{\dagger}  Y_{d}^\intercal  Y_d^*}\Big) +30 \mbox{Tr}\Big({Y_d  Y_{d}^{\dagger}  Y_d  Y_{d}^{\dagger}  Y_d  Y_{d}^{\dagger}}\Big) \nonumber \\ 
&\qquad \qquad +10 \mbox{Tr}\Big({Y_e  Y_{e}^{\dagger}  Y_e  Y_{e}^{\dagger}  Y_e  Y_{e}^{\dagger}}\Big) -4 \mbox{Tr}\Big({Y_e  Y_{e}^{\dagger}  Y_e  Y_{N}^{\dagger}  Y_N  Y_{e}^{\dagger}}\Big) +2 \mbox{Tr}\Big({Y_e  Y_{N}^{\dagger}  Y_N  Y_{e}^{\dagger}  Y_e  Y_{e}^{\dagger}}\Big) \nonumber \\ 
&\qquad \qquad +10 \mbox{Tr}\Big({Y_N  Y_{N}^{\dagger}  Y_N  Y_{N}^{\dagger}  Y_N  Y_{N}^{\dagger}}\Big) +30 \mbox{Tr}\Big({Y_u  Y_{u}^{\dagger}  Y_u  Y_{u}^{\dagger}  Y_u  Y_{u}^{\dagger}}\Big) -6 \mbox{Tr}\Big({Y_u  Y_{u}^{\dagger}  Y_u  Y_{u}^{\dagger}  Y_{d}^\intercal  Y_d^*}\Big) \nonumber\\
&\qquad \qquad -6 \mbox{Tr}\Big({Y_u  Y_{u}^{\dagger}  Y_{d}^\intercal  Y_d^*  Y_{d}^\intercal  Y_d^*}\Big) -2 \mbox{Tr}\Big({Y_e  Y_{N}^{\dagger}  Y_N  Y_{N}^{\dagger}  Y_N  Y_{e}^{\dagger}}\Big)\Bigg] \, . \\
%%%%%%%%%%%%%%%%%
\beta_{\lambda_2} \  = \ &
\frac{1}{16\pi^2}\Bigg[24 \lambda_{2}^{2}  + 2 \lambda_{3}^{2}  + 2 \lambda_3 \lambda_4  + 4 \lambda_{5}^{2}  -9 g_{2}^{2} \lambda_2  + \frac{27}{200} g_{1}^{4}  + \frac{9}{20} g_{1}^{2} \Big(-4 \lambda_2  + g_{2}^{2}\Big) + \frac{9}{8} g_{2}^{4}  + \lambda_{4}^{2}\Bigg] \nonumber \\
&+\frac{1}{(16\pi^2)^2}\Bigg[-\frac{3537}{2000} g_{1}^{6} -\frac{1719}{400} g_{1}^{4} g_{2}^{2} -\frac{303}{80} g_{1}^{2} g_{2}^{4} +\frac{291}{16} g_{2}^{6} +\frac{1953}{200} g_{1}^{4} \lambda_2 +\frac{117}{20} g_{1}^{2} g_{2}^{2} \lambda_2 -\frac{51}{8} g_{2}^{4} \lambda_2  \nonumber \\ 
&\qquad \qquad+\frac{108}{5} g_{1}^{2} \lambda_{2}^{2}+108 g_{2}^{2} \lambda_{2}^{2} -312 \lambda_{2}^{3} +\frac{9}{10} g_{1}^{4} \lambda_3 +\frac{15}{2} g_{2}^{4} \lambda_3 +\frac{12}{5} g_{1}^{2} \lambda_{3}^{2} +12 g_{2}^{2} \lambda_{3}^{2} -20 \lambda_2 \lambda_{3}^{2} \nonumber \\ &\qquad \qquad
-8 \lambda_{3}^{3} +\frac{9}{20} g_{1}^{4} \lambda_4 -\frac{3}{2} g_{1}^{2} g_{2}^{2} \lambda_4 +\frac{15}{4} g_{2}^{4} \lambda_4 +\frac{12}{5} g_{1}^{2} \lambda_3 \lambda_4 +12 g_{2}^{2} \lambda_3 \lambda_4 -20 \lambda_2 \lambda_3 \lambda_4 \nonumber \\
&\qquad \qquad
-12 \lambda_{3}^{2} \lambda_4 +\frac{6}{5} g_{1}^{2} \lambda_{4}^{2} +3 g_{2}^{2} \lambda_{4}^{2} -12 \lambda_2 \lambda_{4}^{2} -16 \lambda_3 \lambda_{4}^{2} -6 \lambda_{4}^{3} -\frac{12}{5} g_{1}^{2} \lambda_{5}^{2}-56 \lambda_2 \lambda_{5}^{2} \nonumber \\
&\qquad \qquad
 -80 \lambda_3 \lambda_{5}^{2} +8 \lambda_4 \lambda_{5}^{2} -6 \Big(2 \lambda_{3}^{2}  + 2 \lambda_3 \lambda_4  + 4 \lambda_{5}^{2}  + \lambda_{4}^{2}\Big)\mbox{Tr}\Big({Y_d  Y_{d}^{\dagger}}\Big) \nonumber \\
&\qquad \qquad
-2 \Big(2 \lambda_{3}^{2}  + 2 \lambda_3 \lambda_4  + 4 \lambda_{5}^{2}  + \lambda_{4}^{2}\Big)\mbox{Tr}\Big({Y_e  Y_{e}^{\dagger}}\Big) -4 \lambda_{3}^{2} \mbox{Tr}\Big({Y_N  Y_{N}^{\dagger}}\Big) -4 \lambda_3 \lambda_4 \mbox{Tr}\Big({Y_N  Y_{N}^{\dagger}}\Big) 
\nonumber \\ 
&\qquad \qquad
-2 \lambda_{4}^{2} \mbox{Tr}\Big({Y_N  Y_{N}^{\dagger}}\Big) -8 \lambda_{5}^{2} \mbox{Tr}\Big({Y_N  Y_{N}^{\dagger}}\Big) -12 \lambda_{3}^{2} \mbox{Tr}\Big({Y_u  Y_{u}^{\dagger}}\Big) -12 \lambda_3 \lambda_4 \mbox{Tr}\Big({Y_u  Y_{u}^{\dagger}}\Big) 
\nonumber \\ 
&\qquad \qquad
-6 \lambda_{4}^{2} \mbox{Tr}\Big({Y_u  Y_{u}^{\dagger}}\Big) -24 \lambda_{5}^{2} \mbox{Tr}\Big({Y_u  Y_{u}^{\dagger}}\Big)\Bigg] \, .  \\
%%%%%%%%%
	\beta_{\lambda_3} \ =  \ &
	\frac{1}{16\pi^2}\Bigg[\frac{27}{100} g_{1}^{4} +\frac{9}{10} g_{1}^{2} g_{2}^{2} +\frac{9}{4} g_{2}^{4} -\frac{9}{5} g_{1}^{2} \lambda_3 -9 g_{2}^{2} \lambda_3 +12 \lambda_1 \lambda_3 +12 \lambda_2 \lambda_3 +4 \lambda_{3}^{2} +4 \lambda_1 \lambda_4 +4 \lambda_2 \lambda_4  \nonumber \\ 
	&\qquad \quad +2 \lambda_{4}^{2} +40 \lambda_{5}^{2} +6 \lambda_3 \mbox{Tr}\Big({Y_d  Y_{d}^{\dagger}}\Big) +2 \lambda_3 \mbox{Tr}\Big({Y_e  Y_{e}^{\dagger}}\Big) +2 \lambda_3 \mbox{Tr}\Big({Y_N  Y_{N}^{\dagger}}\Big) +6 \lambda_3 \mbox{Tr}\Big({Y_u  Y_{u}^{\dagger}}\Big)\Bigg]\nonumber \\
&+\frac{1}{(16\pi^2)^2}\Bigg[-\frac{3537}{1000} g_{1}^{6} -\frac{1719}{200} g_{1}^{4} g_{2}^{2} -\frac{303}{40} g_{1}^{2} g_{2}^{4} +\frac{291}{8} g_{2}^{6} +\frac{27}{10} g_{1}^{4} \lambda_1 +3 g_{1}^{2} g_{2}^{2} \lambda_1 +\frac{45}{2} g_{2}^{4} \lambda_1 \nonumber \\ 
&\qquad \qquad +\frac{27}{10} g_{1}^{4} \lambda_2 +3 g_{1}^{2} g_{2}^{2} \lambda_2 +\frac{45}{2} g_{2}^{4} \lambda_2 +\frac{1773}{200} g_{1}^{4} \lambda_3 +\frac{57}{20} g_{1}^{2} g_{2}^{2} \lambda_3 -\frac{111}{8} g_{2}^{4} \lambda_3 +\frac{72}{5} g_{1}^{2} \lambda_1 \lambda_3  \nonumber \\ 
&\qquad \qquad+72 g_{2}^{2} \lambda_1 \lambda_3-60 \lambda_{1}^{2} \lambda_3 +\frac{72}{5} g_{1}^{2} \lambda_2 \lambda_3 +72 g_{2}^{2} \lambda_2 \lambda_3 -60 \lambda_{2}^{2} \lambda_3 +\frac{6}{5} g_{1}^{2} \lambda_{3}^{2} +6 g_{2}^{2} \lambda_{3}^{2}  \nonumber \\ 
&\qquad \qquad-72 \lambda_1 \lambda_{3}^{2} -72 \lambda_2 \lambda_{3}^{2}-12 \lambda_{3}^{3} +\frac{9}{10} g_{1}^{4} \lambda_4 -3 g_{1}^{2} g_{2}^{2} \lambda_4 +\frac{15}{2} g_{2}^{4} \lambda_4 +\frac{24}{5} g_{1}^{2} \lambda_1 \lambda_4 +36 g_{2}^{2} \lambda_1 \lambda_4  \nonumber \\ 
&\qquad \qquad -16 \lambda_{1}^{2} \lambda_4 +\frac{24}{5} g_{1}^{2} \lambda_2 \lambda_4+36 g_{2}^{2} \lambda_2 \lambda_4 -16 \lambda_{2}^{2} \lambda_4 -12 g_{2}^{2} \lambda_3 \lambda_4 -32 \lambda_1 \lambda_3 \lambda_4 -32 \lambda_2 \lambda_3 \lambda_4  \nonumber \\ 
&\qquad \qquad -4 \lambda_{3}^{2} \lambda_4 +\frac{12}{5} g_{1}^{2} \lambda_{4}^{2}+6 g_{2}^{2} \lambda_{4}^{2} -28 \lambda_1 \lambda_{4}^{2} -28 \lambda_2 \lambda_{4}^{2} -16 \lambda_3 \lambda_{4}^{2} -12 \lambda_{4}^{3} +48 g_{1}^{2} \lambda_{5}^{2}  \nonumber \\ 
&\qquad \qquad +216 g_{2}^{2} \lambda_{5}^{2} -336 \lambda_1 \lambda_{5}^{2}-336 \lambda_2 \lambda_{5}^{2} -264 \lambda_3 \lambda_{5}^{2} +16 \lambda_4 \lambda_{5}^{2} -\frac{3}{4} g_{2}^{4} \mbox{Tr}\Big({Y_N  Y_{N}^{\dagger}}\Big)  \nonumber \\ 
&\qquad \qquad +\frac{3}{4} g_{1}^{2} \lambda_3 \mbox{Tr}\Big({Y_N  Y_{N}^{\dagger}}\Big)+\frac{1}{20} \Big(-5 \Big(-45 g_{2}^{2} \lambda_3  + 8 \Big(-20 g_{3}^{2} \lambda_3  + 3 \Big(20 \lambda_{5}^{2}  + 2 \lambda_{3}^{2}    \nonumber \\ 
& \qquad \qquad + 4 \lambda_1 \Big(3 \lambda_3  + \lambda_4\Big) + \lambda_{4}^{2}\Big)\Big) + 9 g_{2}^{4} \Big)+ 9 g_{1}^{4} + g_{1}^{2} \Big(25 \lambda_3  + 54 g_{2}^{2} \Big)\Big)\mbox{Tr}\Big({Y_d  Y_{d}^{\dagger}}\Big)  \nonumber \\ 
& \qquad \qquad -\frac{1}{20} \Big(-3 g_{1}^{2} \Big(22 g_{2}^{2}  + 25 \lambda_3 \Big) + 45 g_{1}^{4}  + 5 \Big(-15 g_{2}^{2} \lambda_3  + 3 g_{2}^{4}+ 8 \Big(20 \lambda_{5}^{2}  + 2 \lambda_{3}^{2}   \nonumber \\ 
&\qquad \qquad + 4 \lambda_1 \Big(3 \lambda_3  + \lambda_4\Big) + \lambda_{4}^{2}\Big)\Big)\Big)\mbox{Tr}\Big({Y_e  Y_{e}^{\dagger}}\Big)-\frac{9}{100} g_{1}^{4} \mbox{Tr}\Big({Y_N  Y_{N}^{\dagger}}\Big) -\frac{3}{10} g_{1}^{2} g_{2}^{2} \mbox{Tr}\Big({Y_N  Y_{N}^{\dagger}}\Big) \nonumber \\ 
&\qquad \qquad +\frac{15}{4} g_{2}^{2} \lambda_3 \mbox{Tr}\Big({Y_N  Y_{N}^{\dagger}}\Big) -24 \lambda_1 \lambda_3 \mbox{Tr}\Big({Y_N  Y_{N}^{\dagger}}\Big) -4 \lambda_{3}^{2} \mbox{Tr}\Big({Y_N  Y_{N}^{\dagger}}\Big) -8 \lambda_1 \lambda_4 \mbox{Tr}\Big({Y_N  Y_{N}^{\dagger}}\Big) \nonumber \\ 
& \qquad \qquad 
-2 \lambda_{4}^{2} \mbox{Tr}\Big({Y_N  Y_{N}^{\dagger}}\Big) -40 \lambda_{5}^{2} \mbox{Tr}\Big({Y_N  Y_{N}^{\dagger}}\Big) -\frac{171}{100} g_{1}^{4} \mbox{Tr}\Big({Y_u  Y_{u}^{\dagger}}\Big) +\frac{63}{10} g_{1}^{2} g_{2}^{2} \mbox{Tr}\Big({Y_u  Y_{u}^{\dagger}}\Big) \nonumber \\ 
&
\qquad \qquad -\frac{9}{4} g_{2}^{4} \mbox{Tr}\Big({Y_u  Y_{u}^{\dagger}}\Big) +\frac{17}{4} g_{1}^{2} \lambda_3 \mbox{Tr}\Big({Y_u  Y_{u}^{\dagger}}\Big) +\frac{45}{4} g_{2}^{2} \lambda_3 \mbox{Tr}\Big({Y_u  Y_{u}^{\dagger}}\Big) +40 g_{3}^{2} \lambda_3 \mbox{Tr}\Big({Y_u  Y_{u}^{\dagger}}\Big) \nonumber \\ 
&\qquad \qquad -72 \lambda_1 \lambda_3 \mbox{Tr}\Big({Y_u  Y_{u}^{\dagger}}\Big) -12 \lambda_{3}^{2} \mbox{Tr}\Big({Y_u  Y_{u}^{\dagger}}\Big) -24 \lambda_1 \lambda_4 \mbox{Tr}\Big({Y_u  Y_{u}^{\dagger}}\Big) -6 \lambda_{4}^{2} \mbox{Tr}\Big({Y_u  Y_{u}^{\dagger}}\Big) \nonumber \\ 
&\qquad \qquad -120 \lambda_{5}^{2} \mbox{Tr}\Big({Y_u  Y_{u}^{\dagger}}\Big) -\frac{27}{2} \lambda_3 \mbox{Tr}\Big({Y_d  Y_{d}^{\dagger}  Y_d  Y_{d}^{\dagger}}\Big) -\frac{9}{2} \lambda_3 \mbox{Tr}\Big({Y_e  Y_{e}^{\dagger}  Y_e  Y_{e}^{\dagger}}\Big)  \nonumber \\ 
&\qquad \qquad -7 \lambda_3 \mbox{Tr}\Big({Y_e  Y_{N}^{\dagger}  Y_N  Y_{e}^{\dagger}}\Big)-8 \lambda_4 \mbox{Tr}\Big({Y_e  Y_{N}^{\dagger}  Y_N  Y_{e}^{\dagger}}\Big) -\frac{9}{2} \lambda_3 \mbox{Tr}\Big({Y_N  Y_{N}^{\dagger}  Y_N  Y_{N}^{\dagger}}\Big) \nonumber \\ 
&\qquad \qquad -\frac{27}{2} \lambda_3 \mbox{Tr}\Big({Y_u  Y_{u}^{\dagger}  Y_u  Y_{u}^{\dagger}}\Big) -21 \lambda_3 \mbox{Tr}\Big({Y_u  Y_{u}^{\dagger}  Y_{d}^\intercal  Y_d^*}\Big) -24 \lambda_4 \mbox{Tr}\Big({Y_u  Y_{u}^{\dagger}  Y_{d}^\intercal  Y_d^*}\Big)\Bigg] \, . \\
%%%%%%%%%%%%%%%%%%
\beta_{\lambda_4} \ = \  &
\frac{1}{16\pi^2}\Bigg[-\frac{9}{5} g_{1}^{2} g_{2}^{2} -\frac{9}{5} g_{1}^{2} \lambda_4 -9 g_{2}^{2} \lambda_4 +4 \lambda_1 \lambda_4 +4 \lambda_2 \lambda_4 +8 \lambda_3 \lambda_4 +4 \lambda_{4}^{2} -32 \lambda_{5}^{2} +6 \lambda_4 \mbox{Tr}\Big({Y_d  Y_{d}^{\dagger}}\Big) \nonumber \\ 
&+2 \lambda_4 \mbox{Tr}\Big({Y_e  Y_{e}^{\dagger}}\Big) +2 \lambda_4 \mbox{Tr}\Big({Y_N  Y_{N}^{\dagger}}\Big) +6 \lambda_4 \mbox{Tr}\Big({Y_u  Y_{u}^{\dagger}}\Big)\Bigg]\nonumber \\
&+\frac{1}{(16\pi^2)^2} \Bigg[+\frac{657}{50} g_{1}^{4} g_{2}^{2} +\frac{42}{5} g_{1}^{2} g_{2}^{4} -6 g_{1}^{2} g_{2}^{2} \lambda_1 -6 g_{1}^{2} g_{2}^{2} \lambda_2 -\frac{6}{5} g_{1}^{2} g_{2}^{2} \lambda_3 +\frac{1413}{200} g_{1}^{4} \lambda_4 +\frac{129}{20} g_{1}^{2} g_{2}^{2} \lambda_4 \nonumber \\ 
&-\frac{231}{8} g_{2}^{4} \lambda_4 +\frac{24}{5} g_{1}^{2} \lambda_1 \lambda_4 -28 \lambda_{1}^{2} \lambda_4 +\frac{24}{5} g_{1}^{2} \lambda_2 \lambda_4 -28 \lambda_{2}^{2} \lambda_4 +\frac{12}{5} g_{1}^{2} \lambda_3 \lambda_4 +36 g_{2}^{2} \lambda_3 \lambda_4 \nonumber \\ 
&-80 \lambda_1 \lambda_3 \lambda_4 -80 \lambda_2 \lambda_3 \lambda_4 -28 \lambda_{3}^{2} \lambda_4 -\frac{12}{5} g_{1}^{2} \lambda_{4}^{2} +18 g_{2}^{2} \lambda_{4}^{2} -40 \lambda_1 \lambda_{4}^{2} -40 \lambda_2 \lambda_{4}^{2} -28 \lambda_3 \lambda_{4}^{2} \nonumber \\ 
&-\frac{192}{5} g_{1}^{2} \lambda_{5}^{2} -216 g_{2}^{2} \lambda_{5}^{2} +192 \lambda_1 \lambda_{5}^{2} +192 \lambda_2 \lambda_{5}^{2} +192 \lambda_3 \lambda_{5}^{2} +88 \lambda_4 \lambda_{5}^{2} +27 \lambda_4 \mbox{Tr}\Big({Y_u  Y_{u}^{\dagger}  Y_{d}^\intercal  Y_d^*}\Big) \nonumber \\ 
&+\Big(4 \Big(10 g_{3}^{2} \lambda_4  -3 \Big(2 \lambda_1 \lambda_4  + 2 \lambda_3 \lambda_4  -8 \lambda_{5}^{2}  + \lambda_{4}^{2}\Big)\Big) + \frac{45}{4} g_{2}^{2} \lambda_4  + g_{1}^{2} \Big(-\frac{27}{5} g_{2}^{2}  + \frac{5}{4} \lambda_4 \Big)\Big)\mbox{Tr}\Big({Y_d  Y_{d}^{\dagger}}\Big) \nonumber \\ 
&+\Big(-4 \Big(2 \lambda_1 \lambda_4  + 2 \lambda_3 \lambda_4  -8 \lambda_{5}^{2}  + \lambda_{4}^{2}\Big) + \frac{15}{4} g_{2}^{2} \lambda_4  + g_{1}^{2} \Big(\frac{15}{4} \lambda_4  -\frac{33}{5} g_{2}^{2} \Big)\Big)\mbox{Tr}\Big({Y_e  Y_{e}^{\dagger}}\Big) +\frac{3}{5} g_{1}^{2} g_{2}^{2} \mbox{Tr}\Big({Y_N  Y_{N}^{\dagger}}\Big) \nonumber \\ 
&+\frac{3}{4} g_{1}^{2} \lambda_4 \mbox{Tr}\Big({Y_N  Y_{N}^{\dagger}}\Big) +\frac{15}{4} g_{2}^{2} \lambda_4 \mbox{Tr}\Big({Y_N  Y_{N}^{\dagger}}\Big) -8 \lambda_1 \lambda_4 \mbox{Tr}\Big({Y_N  Y_{N}^{\dagger}}\Big) -8 \lambda_3 \lambda_4 \mbox{Tr}\Big({Y_N  Y_{N}^{\dagger}}\Big) \nonumber \\ 
&-4 \lambda_{4}^{2} \mbox{Tr}\Big({Y_N  Y_{N}^{\dagger}}\Big) +32 \lambda_{5}^{2} \mbox{Tr}\Big({Y_N  Y_{N}^{\dagger}}\Big) -\frac{63}{5} g_{1}^{2} g_{2}^{2} \mbox{Tr}\Big({Y_u  Y_{u}^{\dagger}}\Big) +\frac{17}{4} g_{1}^{2} \lambda_4 \mbox{Tr}\Big({Y_u  Y_{u}^{\dagger}}\Big) \nonumber \\ 
&+\frac{45}{4} g_{2}^{2} \lambda_4 \mbox{Tr}\Big({Y_u  Y_{u}^{\dagger}}\Big) +40 g_{3}^{2} \lambda_4 \mbox{Tr}\Big({Y_u  Y_{u}^{\dagger}}\Big) -24 \lambda_1 \lambda_4 \mbox{Tr}\Big({Y_u  Y_{u}^{\dagger}}\Big) -24 \lambda_3 \lambda_4 \mbox{Tr}\Big({Y_u  Y_{u}^{\dagger}}\Big) \nonumber \\ 
&-12 \lambda_{4}^{2} \mbox{Tr}\Big({Y_u  Y_{u}^{\dagger}}\Big) +96 \lambda_{5}^{2} \mbox{Tr}\Big({Y_u  Y_{u}^{\dagger}}\Big) -\frac{27}{2} \lambda_4 \mbox{Tr}\Big({Y_d  Y_{d}^{\dagger}  Y_d  Y_{d}^{\dagger}}\Big) -\frac{9}{2} \lambda_4 \mbox{Tr}\Big({Y_e  Y_{e}^{\dagger}  Y_e  Y_{e}^{\dagger}}\Big) \nonumber \\ 
&+9 \lambda_4 \mbox{Tr}\Big({Y_e  Y_{N}^{\dagger}  Y_N  Y_{e}^{\dagger}}\Big) -\frac{9}{2} \lambda_4 \mbox{Tr}\Big({Y_N  Y_{N}^{\dagger}  Y_N  Y_{N}^{\dagger}}\Big) -\frac{27}{2} \lambda_4 \mbox{Tr}\Big({Y_u  Y_{u}^{\dagger}  Y_u  Y_{u}^{\dagger}}\Big)\Bigg] \, . \\
%%%%%%%%%%%%%%%%%%%%%%%%
\beta_{\lambda_5} \ = \ &
\frac{1}{16\pi^2}\Bigg[-\frac{9}{5} g_{1}^{2} \lambda_5 -9 g_{2}^{2} \lambda_5 +4 \lambda_1 \lambda_5 +4 \lambda_2 \lambda_5 +8 \lambda_3 \lambda_5 -4 \lambda_4 \lambda_5 \nonumber \\ 
&\qquad \quad +6 \lambda_5 \mbox{Tr}\Big({Y_d  Y_{d}^{\dagger}}\Big) +2 \lambda_5 \mbox{Tr}\Big({Y_e  Y_{e}^{\dagger}}\Big) +2 \lambda_5 \mbox{Tr}\Big({Y_N  Y_{N}^{\dagger}}\Big) +6 \lambda_5 \mbox{Tr}\Big({Y_u  Y_{u}^{\dagger}}\Big)\Bigg] 
\nonumber \\
 &+\frac{1}{(16\pi^2)^2}\Bigg[\frac{1413}{200} g_{1}^{4} \lambda_5 +\frac{57}{20} g_{1}^{2} g_{2}^{2} \lambda_5 -\frac{231}{8} g_{2}^{4} \lambda_5 -\frac{12}{5} g_{1}^{2} \lambda_1 \lambda_5 -28 \lambda_{1}^{2} \lambda_5 -\frac{12}{5} g_{1}^{2} \lambda_2 \lambda_5 -28 \lambda_{2}^{2} \lambda_5 \nonumber \\ 
&\qquad \qquad +\frac{48}{5} g_{1}^{2} \lambda_3 \lambda_5 +36 g_{2}^{2} \lambda_3 \lambda_5 -80 \lambda_1 \lambda_3 \lambda_5 -80 \lambda_2 \lambda_3 \lambda_5 -28 \lambda_{3}^{2} \lambda_5 -\frac{24}{5} g_{1}^{2} \lambda_4 \lambda_5 -36 g_{2}^{2} \lambda_4 \lambda_5 \nonumber \\ 
&\qquad \qquad +8 \lambda_1 \lambda_4 \lambda_5 +8 \lambda_2 \lambda_4 \lambda_5 +20 \lambda_3 \lambda_4 \lambda_5 +16 \lambda_{4}^{2} \lambda_5 +24 \lambda_{5}^{3} +\frac{15}{4} g_{2}^{2} \lambda_5 \mbox{Tr}\Big({Y_N  Y_{N}^{\dagger}}\Big) \nonumber \\ 
&\qquad \qquad +\frac{1}{4} \Big(16 \Big(10 g_{3}^{2}  + 3 \lambda_4  -6 \lambda_1  -6 \lambda_3 \Big) + 45 g_{2}^{2}  + 5 g_{1}^{2} \Big)\lambda_5 \mbox{Tr}\Big({Y_d  Y_{d}^{\dagger}}\Big) \nonumber \\ 
&\qquad \qquad +\frac{1}{4} \Big(15 g_{1}^{2}  + 15 g_{2}^{2}  + 16 \Big(-2 \lambda_1  -2 \lambda_3  + \lambda_4\Big)\Big)\lambda_5 \mbox{Tr}\Big({Y_e  Y_{e}^{\dagger}}\Big) +\frac{3}{4} g_{1}^{2} \lambda_5 \mbox{Tr}\Big({Y_N  Y_{N}^{\dagger}}\Big) \nonumber \\ 
&\qquad \qquad -8 \lambda_1 \lambda_5 \mbox{Tr}\Big({Y_N  Y_{N}^{\dagger}}\Big) -8 \lambda_3 \lambda_5 \mbox{Tr}\Big({Y_N  Y_{N}^{\dagger}}\Big) +4 \lambda_4 \lambda_5 \mbox{Tr}\Big({Y_N  Y_{N}^{\dagger}}\Big) +\frac{17}{4} g_{1}^{2} \lambda_5 \mbox{Tr}\Big({Y_u  Y_{u}^{\dagger}}\Big) \nonumber \\ 
&\qquad \qquad +\frac{45}{4} g_{2}^{2} \lambda_5 \mbox{Tr}\Big({Y_u  Y_{u}^{\dagger}}\Big) +40 g_{3}^{2} \lambda_5 \mbox{Tr}\Big({Y_u  Y_{u}^{\dagger}}\Big) -24 \lambda_1 \lambda_5 \mbox{Tr}\Big({Y_u  Y_{u}^{\dagger}}\Big) -24 \lambda_3 \lambda_5 \mbox{Tr}\Big({Y_u  Y_{u}^{\dagger}}\Big) \nonumber \\ 
&\qquad \qquad +12 \lambda_4 \lambda_5 \mbox{Tr}\Big({Y_u  Y_{u}^{\dagger}}\Big) -\frac{3}{2} \lambda_5 \mbox{Tr}\Big({Y_d  Y_{d}^{\dagger}  Y_d  Y_{d}^{\dagger}}\Big) -\frac{1}{2} \lambda_5 \mbox{Tr}\Big({Y_e  Y_{e}^{\dagger}  Y_e  Y_{e}^{\dagger}}\Big) +\lambda_5 \mbox{Tr}\Big({Y_e  Y_{N}^{\dagger}  Y_N  Y_{e}^{\dagger}}\Big) \nonumber \\ 
&\qquad \qquad -\frac{1}{2} \lambda_5 \mbox{Tr}\Big({Y_N  Y_{N}^{\dagger}  Y_N  Y_{N}^{\dagger}}\Big) -\frac{3}{2} \lambda_5 \mbox{Tr}\Big({Y_u  Y_{u}^{\dagger}  Y_u  Y_{u}^{\dagger}}\Big) +3 \lambda_5 \mbox{Tr}\Big({Y_u  Y_{u}^{\dagger}  Y_{d}^\intercal  Y_d^*}\Big)\Bigg] \, .
\end{align*}
\endgroup

\subsection{Gauge Couplings }
\begingroup
\allowdisplaybreaks
\begin{align*}
\beta_{g_1} \ = \ & 
\frac{1}{16\pi^2}\Bigg[\frac{21}{5} g_{1}^{3}\Bigg]
+\frac{1}{(16\pi^2)^2}\Bigg[\frac{1}{50} g_{1}^{3} \Big(180 g_{2}^{2}  + 208 g_{1}^{2}  + 440 g_{3}^{2}-15 \mbox{Tr}\Big({Y_N  Y_{N}^{\dagger}}\Big)  -25 \mbox{Tr}\Big({Y_d  Y_{d}^{\dagger}}\Big)  \nonumber \\
& \qquad \qquad \qquad \qquad  \qquad \qquad -75 \mbox{Tr}\Big({Y_e  Y_{e}^{\dagger}}\Big)-85 \mbox{Tr}\Big({Y_u  Y_{u}^{\dagger}}\Big) \Big)\Bigg] \, . \\
\beta_{g_2}  \ =  \ &
\frac{1}{16\pi^2}\Bigg[-3 g_{2}^{3}\Bigg]+\frac{1}{(16\pi^2)^2}\Bigg[
\frac{1}{10} g_{2}^{3} \Big(120 g_{3}^{2}  + 12 g_{1}^{2} + 80 g_{2}^{2} -15 \mbox{Tr}\Big({Y_d  Y_{d}^{\dagger}}\Big)  -15 \mbox{Tr}\Big({Y_u  Y_{u}^{\dagger}}\Big) \nonumber \\
& \qquad \qquad \qquad \qquad  \qquad \qquad -5 \mbox{Tr}\Big({Y_e  Y_{e}^{\dagger}}\Big)  -5 \mbox{Tr}\Big({Y_N  Y_{N}^{\dagger}}\Big)\Big)\Bigg] \, .  \\
\beta_{g_3} \ = \ &  
\frac{1}{16\pi^2}\Bigg[-7 g_{3}^{3}\Bigg]+
\frac{1}{(16\pi^2)^2}\Bigg[-\frac{1}{10} g_{3}^{3} \Big(-11 g_{1}^{2}   + 260 g_{3}^{2} -45 g_{2}^{2}\nonumber \\
& \qquad \qquad \qquad \qquad  \qquad \qquad + 20 \mbox{Tr}\Big({Y_d  Y_{d}^{\dagger}}\Big)  + 20 \mbox{Tr}\Big({Y_u  Y_{u}^{\dagger}}\Big)  \Big)\Bigg]  \, .
\end{align*}
\endgroup

\subsection{Yukawa Coupling}
\begingroup
\allowdisplaybreaks
\begin{align*}
\beta_{Y_u} \  = \ & 
\frac{1}{16\pi^2}\Bigg[\frac{3}{2} \Big(- {Y_{d}^\intercal  Y_d^*  Y_u}  + {Y_u  Y_{u}^{\dagger}  Y_u}\Big)
+Y_u \Big(3 \mbox{Tr}\Big({Y_d  Y_{d}^{\dagger}}\Big)  + 3 \mbox{Tr}\Big({Y_u  Y_{u}^{\dagger}}\Big)  -8 g_{3}^{2}  -\frac{17}{20} g_{1}^{2}  \nonumber \\
&\quad \qquad -\frac{9}{4} g_{2}^{2}  + \mbox{Tr}\Big({Y_e  Y_{e}^{\dagger}}\Big) + \mbox{Tr}\Big({Y_N  Y_{N}^{\dagger}}\Big)\Big)\Bigg]\nonumber \\
&+
\frac{1}{(16\pi^2)^2}\Bigg[\frac{1}{80} \Big(20 \Big(11 {Y_{d}^\intercal  Y_d^*  Y_{d}^\intercal  Y_d^*  Y_u}  -4 {Y_{d}^\intercal  Y_d^*  Y_u  Y_{u}^{\dagger}  Y_u} + 6 {Y_u  Y_{u}^{\dagger}  Y_u  Y_{u}^{\dagger}  Y_u}  - {Y_u  Y_{u}^{\dagger}  Y_{d}^\intercal  Y_d^*  Y_u} \Big) \nonumber \\
& \qquad \qquad 
+{Y_u  Y_{u}^{\dagger}  Y_u} \Big(1280 g_{3}^{2}  -180 \mbox{Tr}\Big({Y_e  Y_{e}^{\dagger}}\Big)  -180 \mbox{Tr}\Big({Y_N  Y_{N}^{\dagger}}\Big)+ 223 g_{1}^{2}  -540 \mbox{Tr}\Big({Y_d  Y_{d}^{\dagger}}\Big) \nonumber \\
& \qquad \qquad   -540 \mbox{Tr}\Big({Y_u  Y_{u}^{\dagger}}\Big)
 + 675 g_{2}^{2}  -960 \lambda_1 \Big)+{Y_{d}^\intercal  Y_d^*  Y_u} \Big(100 \mbox{Tr}\Big({Y_e  Y_{e}^{\dagger}}\Big) + 100 \mbox{Tr}\Big({Y_N  Y_{N}^{\dagger}}\Big) \nonumber \\ 
 &\qquad \qquad  -1280 g_{3}^{2}  + 300 \mbox{Tr}\Big({Y_d  Y_{d}^{\dagger}}\Big) 
 + 300 \mbox{Tr}\Big({Y_u  Y_{u}^{\dagger}}\Big)  -43 g_{1}^{2}  + 45 g_{2}^{2} \Big)\Big)\nonumber \\
 &\qquad \qquad +Y_u \Big(\frac{1267}{600} g_{1}^{4} -\frac{9}{20} g_{1}^{2} g_{2}^{2} -\frac{21}{4} g_{2}^{4} +\frac{19}{15} g_{1}^{2} g_{3}^{2} +9 g_{2}^{2} g_{3}^{2} -108 g_{3}^{4}  
+6 \lambda_{1}^{2} +\lambda_{3}^{2}+\lambda_3 \lambda_4 \nonumber \\
&\qquad \qquad +\lambda_{4}^{2}+6 \lambda_{5}^{2}+\frac{5}{8} \Big(32 g_{3}^{2}  + 9 g_{2}^{2}  + g_{1}^{2}\Big)\mbox{Tr}\Big({Y_d  Y_{d}^{\dagger}}\Big) +\frac{15}{8} \Big(g_{1}^{2} + g_{2}^{2}\Big)\mbox{Tr}\Big({Y_e  Y_{e}^{\dagger}}\Big) 
\nonumber \\
&\qquad \qquad +\frac{3}{8} g_{1}^{2} \mbox{Tr}\Big({Y_N  Y_{N}^{\dagger}}\Big) +\frac{15}{8} g_{2}^{2} \mbox{Tr}\Big({Y_N  Y_{N}^{\dagger}}\Big)+\frac{17}{8} g_{1}^{2} \mbox{Tr}\Big({Y_u  Y_{u}^{\dagger}}\Big) +\frac{45}{8} g_{2}^{2} \mbox{Tr}\Big({Y_u  Y_{u}^{\dagger}}\Big) \nonumber \\
&\qquad \qquad +20 g_{3}^{2} \mbox{Tr}\Big({Y_u  Y_{u}^{\dagger}}\Big) 
-\frac{27}{4} \mbox{Tr}\Big({Y_d  Y_{d}^{\dagger}  Y_d  Y_{d}^{\dagger}}\Big) -\frac{9}{4} \mbox{Tr}\Big({Y_e  Y_{e}^{\dagger}  Y_e  Y_{e}^{\dagger}}\Big)+\frac{1}{2} \mbox{Tr}\Big({Y_e  Y_{N}^{\dagger}  Y_N  Y_{e}^{\dagger}}\Big)  \nonumber \\
&\qquad \qquad -\frac{9}{4} \mbox{Tr}\Big({Y_N  Y_{N}^{\dagger}  Y_N  Y_{N}^{\dagger}}\Big) -\frac{27}{4} \mbox{Tr}\Big({Y_u  Y_{u}^{\dagger}  Y_u  Y_{u}^{\dagger}}\Big)+\frac{3}{2} \mbox{Tr}\Big({Y_u  Y_{u}^{\dagger}  Y_{d}^\intercal  Y_d^*}\Big) \Big)\Bigg] \, .
\end{align*}
\endgroup

\end{document}